\documentclass[prb,twocolumn,amsfonts,amssymb,amsmath,floatfix,tightenlines]{revtex4-1}

\usepackage{amsfonts,amssymb}
\usepackage{amsmath,times}
\usepackage{graphicx}
\usepackage{hyperref}    % links
\usepackage{color}
\usepackage{cleveref}
\usepackage[lofdepth,lotdepth,caption=false]{subfig}
\renewcommand{\vec}[1]{\mathbf{#1}}
\renewcommand{\mathrm}[1]{\text{#1}}

\begin{document}
%%%%%%%%%%%%%%%%%%%%%%%%%% Title and Abstract %%%%%%%%%%%%%%%%%%%%%%%%%%
\title{Quantum Geometric Contributions to the BKT Transition: Beyond Mean Field Theory}
\author{Zhiqiang Wang}
\affiliation{James Franck Institute, University of Chicago, Chicago, Illinois 60637, USA}
\author{Gaurav Chaudhary}
\affiliation{James Franck Institute, University of Chicago, Chicago, Illinois
60637, USA}
\author{Qijin Chen}
 \affiliation{Shanghai Branch, National Laboratory for Physical Sciences at Microscale and Department of Modern Physics, University
  of Science and Technology of China, Shanghai 201315, China}
\author{K. Levin}
\affiliation{James Franck Institute, University of Chicago, Chicago, Illinois 60637, USA}

\begin{abstract}
  We study quantum geometric contributions to the
  Berezinskii-Kosterlitz-Thouless (BKT) transition temperature,
  $T_{\mathrm{BKT}}$, in the presence of fluctuations beyond BCS
  theory.  Because quantum geometric effects become
  progressively more important with
  stronger pairing attraction, a full understanding of 
  2D multi-orbital superconductivity
  requires the incorporation of preformed pairs.
  We find it is through the effective mass of these pairs that
  quantum geometry enters the theory and this suggests that the quantum geometric
  effects are present in the non-superconducting pseudogap phase as well. Increasing these
  geometric contributions tends to raise
  $T_{\mathrm{BKT}}$ which then competes with fluctuation effects that
  generally depress it. 
  We argue that a way to physically quantify the magnitude
of these geometric terms is 
  in terms of the ratio of the pairing onset temperature $T^*$
  to $T_{\mathrm{BKT}}$. Our paper calls attention to an
experimental study demonstrating how both temperatures 
and, thus, their ratio may be currently accessible. They can be extracted from
  the same voltage-current measurements which are generally used
to establish BKT physics. 
  We use these observations to provide rough preliminary estimates of the
  magnitude of the geometric contributions in, for example, magic angle twisted bilayer graphene.
\end{abstract}

\maketitle

\section{Introduction} 

The recent discovery of superconducting phases
in twisted bilayer graphene (TBLG) at the first magic angle has
attracted much attention
\cite{Cao2018,Cao2018a,Bistritzer2011,Wu2018,Dodaro2018,Kang2018,Yankowitz2019,Yuan2018,Po2018,Xu2018,Roy2019,Isobe2018,Po2019,Tarnopolsky2019, Arora2020}.
The excitement surrounding this material is driven
largely by the flatness of the energy bands, which effectively enhances 
the importance of electron-electron interactions. This 
stronger interaction effect is consistent with the observed high
superconducting transition temperatures~\cite{Cao2018a} and has been
speculated to place TBLG somewhere in the crossover between the BCS and the
Bose-Einstein condensation (BEC) regimes
\cite{Uemura2004,Cao2018a,Wang2020}.
Because of its two dimensionality (2D) this superconductivity is
associated with a BKT instability, in which the transition temperature
$T_{\mathrm{BKT}}$ is directly proportional to the superfluid phase
stiffness ~\cite{Berezinskii1971,Kosterlitz1973,Benfatto2007}.  In
a single flat band this stiffness vanishes; however, in multi-orbital band
models, it was shown that the inclusion of quantum geometric effects
may reinstate a finite transition temperature
\cite{Peotta2015,Liang2017,Hu2019,Julku2020,Xie2020}.

This physical picture of flat-band superconductivity has been
established within BCS mean field (MF) theory, which is
known to be problematic in 2D.  Moreover, quantum geometric effects
become most apparent outside the BCS regime, where non-condensed pairs,
neglected in MF theory, play an important role in the phase stiffness.

In this paper we present a theory which addresses these
shortcomings through studies of the interplay of
preformed pairs with quantum geometric effects. We determine
$T_{\mathrm{BKT}}$, in 2D superconductors
using a simple two-band tight-binding model
\cite{Hofmann2019,Neupert2011} that captures some key ingredients in
common with its TBLG counterpart, including potentially nontrivial
band topology.  The model has some formal similarities to a spin-orbit
coupled Fermi gas Hamiltonian, where the nature of (albeit, three
dimensional) pairing fluctuations within the BCS-BEC crossover is well
studied \cite{He2013, Zhang2014, Fu2013, Zheng2014, Wu2015a}.  Built
on the BCS-Leggett ground state \cite{Leggett1980}, our approach yields
results for $T_{\mathrm{BKT}}$ that are consistent with the mean field
literature at weak attraction, precisely where the MF theory is expected to work.

A major contribution of this paper is to establish the important competition:
bosonic excitations lead to a
decrease in the effective phase stiffness,
whereas, geometric effects generally cause an
increase. These latter become more appreciable as the bands
become flatter. As a result, $T_{\mathrm{BKT}}$
remains substantial, even though it is reduced by beyond mean-field fluctuations.
An important finding is that geometric contributions appear through the
inverse pair mass, $1/M_{\mathrm{B}}$.
Unlike in previous work \cite{Iskin2018,Torma2018} where the pair mass was also found to depend on quantum geometry, 
here $M_{\mathrm{B}}$ incorporates the self consistently determined pairing gap. 
Because $M_{\mathrm{B}}$ enters
the excitation spectrum of the pairs, 
the effect of geometry must be present in a host of general characteristics beyond
the superfluid stiffness including transport 
and thermodynamics\cite{Chen2000}, persisting even
into 
the pseudogap phase.
Here the ``pseudogap phase" refers to the non-superconducting
state with preformed pairs at $T_{\mathrm{BKT}} < T < T^*$. 
We reserve the term ``normal state"  for
a non-interacting system without pairing.

To physically understand the relation between the pair mass and
geometry, note that an increased magnitude of the quantum metric
reflects an increased spatial extent of the normal
state Wannier orbitals~\cite{Marzari1997, Marzari2012}. 
This increase
leads to larger pairs, which have a bigger overlap, leading to higher pair mobility (smaller
$M_{\mathrm{B}}$).  Nontrivial normal state band topology enhances
these effects, which become most apparent
in the so-called ``isolated flat band limit" \cite{Liang2017}, where
the conventional contributions to the pair mobility are negligible.
In analogy with earlier findings ~\cite{Peotta2015,Liang2017} we
demonstrate that a nontrivial band topology provides a lower bound for
$1/M_\text{B}$ in this limit.

Finally, it is important to determine the size of the geometric
contributions using experimentally accessible quantities.  We find that
the ratio of the pairing onset temperature, $T^*$, and
$T_{\mathrm{BKT}}$ allows quantification of the geometric
contributions and characterization of a given 2D superconductor more
generally.  We demonstrate how both temperatures can be determined from
the same voltage-current measurements \cite{Zhao2013}.

%NEW
The rest of the paper is organized as follows. Section~\ref{sec:
  framework} introduces the theoretical approach for deriving the BKT
transition temperature. This includes the introduction of the
topological band model, our pairing fluctuation theory and a 
procedure for calculating the transition temperature as approached
from the non-superconducting state.  We also present a discussion of
the isolated flat band limit where we derive a lower bound for
$n_{\mathrm{B}}/M_{\mathrm{B}}$ associated with band topology.
Here $n_{\mathrm{B}}$ is the areal density of the
preformed pairs. 
The corresponding numerical results for $T_{\mathrm{BKT}}$ and $T^*$
are presented in Sec.~\ref{sec: main-numerical}.  Based on our
numerical results and an experimental estimate of
$T^*/T_{\mathrm{BKT}}$, we speculate that magic angle TBLG is in the
BCS-BEC crossover regime, although it has not passed into the BEC, and
that the geometric contribution to $T_{\mathrm{BKT}}$ is significant.

Sec.~\ref{sec: comparison} contains a comparison of our results both
to numerical Monte-Carlo calculations and to other approaches.
Sec.~\ref{sec: conclusion} presents our conclusions.  Detailed
descriptions of the tight-binding model, derivations of our
multi-orbital pairing fluctuation theory, discussions of the relation
between quantum geometry and pair mass, and equations used for the
mean field superfluid stiffness and $T_{\mathrm{BKT}}$ can be found in
the appendices.

\section{Theoretical Framework} \label{sec: framework}

\subsection{Band model} 
\label{sec: model}

Our tight-binding model~\cite{Hofmann2019,Neupert2011} is defined on
a square lattice, which splits into two sublattices, $\{A,B\}$, due to a
staggered $\pi$ magnetic flux \cite{Appendix}.  The flux is opposite for opposite
spins with preserved time reversal symmetry.  This symmetry and
the absence of spin-orbit coupling reduces the four band pairing
problem, including sublattices and spin, to a two-band system with
sub-lattices only and we henceforth drop the spin.
Here we consider zero center-of-mass momentum and spin singlet pairing. 

As a result we have a simple normal state Hamiltonian ~\cite{Hofmann2019, Appendix} 
in $\vec{k}$ space,
\begin{gather}
H_{\mathrm{N}}(\vec{k}) = h_0(\vec{k}) +\vec{h}(\vec{k}) \cdot \vec{s} -\mu_\text{F},
\label{eq: HK2}
\end{gather}
written in the basis
$(c_{A}^\dagger (\vec{k}), \; c^\dagger_{B}(\vec{k}) )$.  Here
$\vec{s} =(s_x, s_y,s_z)$ are Pauli matrices defined for the
sublattice space,
$ h_0
%(\vec{k})
= -2 t_5  [ \cos 2(k_x+ k_y) +\cos 2(k_x
-k_y)]$, 
$ h_z
%(\vec{k})
= - 2 t_2  [ \cos(k_x + k_y) - \cos(k_x-k_y)
]$, 
$ h_x
% (\vec{k},\phi)
+ i \, h_y
%(\vec{k},\phi)
= -2 t  [e^{ i (- \phi
  - k_y)} \cos k_y + e^{i (\phi -k_y)} \cos k_x] $, with
$\phi= \pi/4$,  and $\mu_\text{F}$ is the fermionic chemical potential. We set the lattice constant $a_\text{L}^{}=1$.
Diagonalizing $H_{\mathrm{N}}(\vec{k})$ gives two energy bands,
$\xi_{\pm}(\vec{k}) = h_0 (\vec{k}) \pm |\vec{h}(\vec{k}) | -\mu_\text{F}$, with
a nonzero Chern number $C = \mp 1$.

For definiteness, following Ref.~\onlinecite{Hofmann2019} we consider two sets of
hopping parameters: (1)
$(t, t_2, t_5) = (1, 1/\sqrt{2}, (1-\sqrt{2}))/4$ and (2)
$(t, t_2, t_5)=(1, 1/\sqrt{2}, 0)$, corresponding, respectively, to a lower band width $W\approx 0.035 t$ and $0.83t$,
and to a band flatness (ratio) $\mathcal{F} \equiv W/E_g \approx 0.01$ and  0.2. Both sets have a band gap $E_g= 4 t$.
Throughout the paper we consider electron density $n=0.3$ per
square lattice site so that the lower band is only partially filled.

\subsection{Pairing fluctuation theory for $T\ge T_{\text{BKT}}$}
\label{sec: main-theory}

Our approach is based on a finite temperature formalism built on the
BCS ground state, which can readily be extended to include stronger
pairing correlations \cite{Leggett1980}. It was derived using an
equation of motion approach \cite{Chen2,Maly1999}, following Kadanoff
and Martin \cite{KadanoffMartin}, and extended to address pairing
(fluctuations) at an arbitrary strength in the context of BCS-BEC
crossover \cite{Chen2005}. Compared to other pairing fluctuation
theories \cite{Levin2010}, this formalism is consistent
with a BCS-like gap equation and simultaneously a
gapless
Anderson-Bogoliubov mode in the superfluid phase. This approach has
been used to address pairing and pseudogap phenomena in Fermi gases
and the cuprates \cite{Chen2005,Chen1999,Maly1999} as well as the
effects of spin-orbit coupling on ultracold Fermi gases \cite{He2013,
  Zhang2014, Fu2013, Zheng2014, Wu2015a}, and most recently to address
the two dimensional BKT transition \cite{Wu2015,Wang2020} in several
simple cases.  In 2D, the natural energy scale parameter,
$n_{\mathrm{B}}/M_{\mathrm{B}}$, enters to describe
$T_{\mathrm{BKT}}$.

To determine $n_\text{B}$ and $M_\text{B}$ we begin with the pair susceptibility
$\chi(Q)$.  We presume that $\chi(Q)$ assumes a special form
(involving one dressed and one bare Green's function) such that the
$Q=0$ pole of the many body T-matrix $t_\text{pg}$ \cite{Chen2005},
\begin{equation}
t_{ \mathrm{pg} }(Q) = \frac{-U}{1 - U \chi(Q)}\,,
\label{eq: tpgdef}
\end{equation}
yields the usual BCS gap equation for the pairing gap
$\Delta_{\mathrm{pg}}$  in the fermionic excitation energy
spectrum,
$E_{\pm}(\vec{k}) =\sqrt{\xi_\pm(\vec{k})^2 + \Delta_{\mathrm{pg}}^2
}$.
This $\Delta_{\mathrm{pg}}$ is to be distinguished
from the superconducting order parameter $\Delta_{\mathrm{sc}}$, which
vanishes at any finite $T$ in 2D.  Here $U>0$ is the
strength of a local attractive Hubbard interaction.
$Q\equiv (i\Omega_m,\vec{q})$ with $\Omega_m= 2 m \pi T$ the bosonic
Matsubara frequency \cite{Appendix}.  
%Expressions for $\chi(Q), t_{\mathrm{pg}}(Q)$,
%and details of the following derivations can be found in
%Appendix~\ref{app: theory}. 

Within ``the pseudogap approximation"~\cite{He2013,Zhang2014,Wu2015a},
it is presumed that
$t_{ \mathrm{pg} }(Q)$ is sharply peaked near $Q=0$, close to an
instability, so that \cite{Chen2005}
\begin{gather}
\Delta_{ \mathrm{pg} }^2 \equiv - T \sum_{Q \ne 0} t_{ \mathrm{pg} }(Q). \label{eq: DeltaPg2}
\end{gather}
Following Refs.~\onlinecite{Chen2005, Wu2015, Wang2020}, for small $Q$, we Taylor-expand $t_\text{pg}^{-1}(Q) =  {\mathcal{Z}}^{-1}(i \Omega_m -\vec{q}^2/(2 M_{\text{B}}) + \mu_{\text{B}})$, where
\begin{equation}
  \frac{ \mu_{\text{B}}}{ {\mathcal{Z}}}= -\frac{1}{U}+\chi(0) = -\frac{1}{U}+ \sum_{\vec{k} \in \mathrm{RBZ} } \sum_{ \alpha =\pm}  \frac{\tanh ( \beta E_\alpha/2) }{2 E_\alpha}.
  \label{eq:gap}
\end{equation}
For brevity, we have suppressed the $\vec{k}$ dependence
on the r.h.s.  ``RBZ" stands for reduced Brillouin zone. 
Here $\mu_{\mathrm{B}}$ is the bosonic pair chemical potential.  When $\mu_{\text{B}}$ is zero
Eq.~\eqref{eq:gap} is recognized as the BCS gap equation, but for the
present purposes we must include non-vanishing $\mu_{\text{B}}$.
Note that $t_{\mathrm{pg}}(Q)$ can be roughly viewed as a propagator for the preformed pairs with an energy $E_\text{B}=\vec{q}^2/2M_{\text{B}} -\mu_\text{B}$. 
Both expressions for ${\mathcal{Z}}$ and  $1/M_\text{B}$ are obtained as functions of $\{ \Delta_{\mathrm{pg}}, \mu_{\mathrm{F}} \}$ from the Taylor expansion.

In 2D, with a simple parabolic pair dispersion, Eq.~\eqref{eq: DeltaPg2} yields \cite{Wu2015, Wang2020}
\begin{align}
n_{\text{B}}\equiv \sum_{\vec{q}} f_{\text{B}}(E_{\text{B}}) = \mathcal{Z}^{-1}\Delta_{ \mathrm{pg} }^2    
  & = -\frac{M_{\text{B}}}{2\pi \beta}  \ln( 1- e^{\beta \mu_{\text{B}}^{}}), \label{eq: nBeqn2}
\end{align}
where $\beta=1/T$, and $f_{\text{B}}(x)=1/(e^{\beta x} -1)$. 
Then we have
\begin{equation}
n_\text{B} /M_{\text{B}} =  \Delta_{\mathrm{pg}}^2 /(M_\text{B} {\mathcal{Z}})
  =2  \;  \Delta_{\mathrm{pg}}^2 \big( T_{\mathrm{conv}} + T_{\mathrm{geom}} \big),
\label{eq: DBdef2}
\end{equation}
where we have split the contributions to the inverse pair mass into
two terms: $T_{\mathrm{conv}}$ is the conventional contribution that
only depends on the normal state dispersion
while $T_{\mathrm{geom}}$ is the geometric contribution that carries
information about the normal state wavefunction.  Here we present an
expression for $T_{\mathrm{geom}}$ with details discussed
elsewhere \cite{Appendix}.

\begin{align} 
& T_{\mathrm{geom} }
 = \sum_{\vec{k} \in \mathrm{RBZ} }  \sum_{ \{ \alpha, \alpha^\prime, \eta \} =\pm} \frac{ 1}{4} \bigg[1+ \eta \frac{\xi_\alpha}{ E_\alpha}  \bigg] \times
 \nonumber \\
& \hspace*{2em}
\frac{n_\text{F}(\eta E_\alpha) - n_\text{F}(-\xi_{ \alpha^\prime })}{\eta E_\alpha + \xi_{\alpha^\prime}} 
 (- \alpha \alpha^\prime)
\frac{1}{4} \sum_{\mu=x,y}  \partial_\mu \hat{h}\cdot \partial_\mu \hat{h},   \label{eq: Tgeom} 
\end{align}
where
$n_\text{F}(x)= 1/(e^{\beta x} +1)$ is the Fermi-Dirac distribution, and
$\hat{h}(\vec{k}) \equiv \vec{h}(\vec{k})/| \vec{h}(\vec{k})|$.
Interestingly, we see that $T_{\text{geom} }$ contains both intra- and
inter-band terms.

Quantum geometry enters into $T_{\text{geom}}$, or equivalently
$n_\text{B}/M_\text{B}$, through the diagonal components of the
quantum metric tensor, $g_{\mu\nu} (\vec{k})$:
\begin{align}
  g_{\mu\nu} (\vec{k})    & = \frac{1}{2} \partial_\mu \hat{h}(\vec{k}) \cdot \partial_\nu \hat{h}(\vec{k}),
\label{eq: gmunu2}
\end{align}
where $\{\mu,\nu\}=\{x,y\}$.  $g_{\mu\nu}$ is a measure of the
distance between two Bloch states in the projective normal state Hilbert space
\cite{Provost1980}.
In the BEC regime, where $n_\text{B}=n/2$,
$g_{\mu\nu}$ is directly connected to the inverse pair mass
$1/M_\text{B}$ \cite{Appendix}.
We stress that in contrast to other work
\cite{Iskin2018,Tovmasyan2016} here $1/M_\text{B}$ depends on the self
consistently determined pairing gap.

Finally, the electrons are subject to the number constraint
\cite{Chen2005, Wu2015, Wang2020},
\begin{equation}
\label{eq: num}
 n = \sum_{\vec{k} \in \mathrm{RBZ}} \sum_{\alpha =\pm} \big[ 1-
\frac{\xi_{\alpha}}{E_\alpha} \tanh ( \frac{\beta E_\alpha}{2}) \big]
.
\end{equation}

Equations \eqref{eq:gap}, \eqref{eq: nBeqn2}, and \eqref{eq: num} form a closed set that can be solved for
$\Delta_{\mathrm{pg}}$ and $\mu_\text{F}$, for given ($T$, $n$, $U$),
which also detemines the important
ratio $n_\text{B}/M_\text{B}$.

\subsection{BKT criterion} 
\label{sec: BKT}

It was initially proposed in
Ref.~\onlinecite{Ries2015} based on experiments in Fermi gases, that
the 2D BKT superconducting transition can be re-interpreted as a
``quasi-condensation" of preformed Cooper pairs.  The onset of
quasi-condensation provides a normal state access to the BKT
instability. Here the transition is approached from above, which is
complementary to the superfluid phase stiffness based approach (from
below).  The quasi-condensation onset is quantified through the
parameter $n_\text{B} /M_{\text{B}}$ which provides a natural 2D
energy scale.
More specifically, this approach to the BKT transition builds on
a Monte-Carlo study of weakly interacting bosons~\cite{Prokofev2002} where
it was found that at the onset of quasi-condensation,
i. e. $T=T_{\mathrm{BKT}}$, one has:
\begin{equation}
\frac{n_\text{B}(T)} { M_\text{B}(T)} =\frac{\mathcal{D}_\text{B}^\text{crit} } {2\pi}  T. 
\label{eq: BKT}
\end{equation}
Here $\mathcal{D}_{\mathrm{B}}^{\mathrm{crit}}$ is the critical value of the phase space density,
$\mathcal{D}_\text{B} (T) \equiv n_\text{B} \lambda_\text{B}^2 $
with $\lambda_\text{B} = \sqrt{2\pi/M_\text{B}T}$ the bosonic thermal de-Broglie wavelength (setting $\hbar=k_{\text{B}}=1$). 
This BKT criterion has been supported by experimental studies on atomic Bose gases~\cite{Jose2013, Tung2010, Clade2009}.

In general $\mathcal{D}_{\mathrm{B}}^{\mathrm{crit}}$ depends on the 
non-universal boson-boson
interaction strength $g_\text{B}$.  In the most general
case, $g_\text{B}$ is unknown
for a fermionic superconductor where Cooper pairs are the emergent composite bosons. However,
a small value of $g_{\mathrm{B}}$ appears consistent with the BCS
ground state, as the bosonic degrees of freedom enter this wavefunction in a
quasi-ideal manner.
Notably the dependence of $\mathcal{D}_{\mathrm{B}}^{\mathrm{crit}}$ on $g_\text{B}$ 
is logarithmic and therefore weak~\cite{Prokofev2002}.
Estimates for $\mathcal{D}_{\text{B}}^{\mathrm{crit}}$ for fermionic superfluids range from 4.9 to 6.45~\cite{Ries2015, Murthy2015}. 
We choose $\mathcal{D}_{\text{B}}^{\mathrm{crit}} =4.9$ that best fits the data
on Fermi gases \cite{Murthy2015}.

\subsection{Isolated flat band limit} 
\label{sec: Isolate}

It is useful to arrive at some
analytical insights on how $n_\text{B}/M_\text{B}$ depends on the
normal state band topology.  This can be done in the isolated flat
band limit, corresponding to $W\ll U \ll E_g$ 
(which is in the BEC regime). 
In this limit, superconductivity is restricted to
the lower flat band while the upper band is inactive, and
Eq.~\eqref{eq: DBdef2} simplifies to
\begin{align}
\frac{n_\text{B}}{M_\text{B}}  & \approx \Delta_{\mathrm{pg}}^2 \sum_{\vec{k}   \in \mathrm{RBZ} } \frac{ \tanh(\beta E_-(\vec{k}) /2 )  }{2 E_{-}(\vec{k})} \frac{1}{2} \sum_{\mu=x,y} g_{\mu\mu}(\vec{k}). 
\label{eq: isoflatband2}
\end{align}
Using an inequality between the quantum metric tensor and the normal state band Berry curvature, one obtains \cite{Appendix}
\begin{align}
\frac{n_\text{B}}{M_\text{B}}  
& \ge \Delta_{\mathrm{pg}}^2                   \frac{\tanh(\beta E_- /2 )}{4 E_{-} }        				\frac{|C|}{ \pi}, 
\label{eq: bound2}
\end{align}
which sets a lower bound for $n_{\text{B}}/M_{\text{B}} $ when
$C\ne 0 $, i.e. when the system is topologically nontrivial. Here
$E_-$ is $\vec{k}$ independent and $C=1$ is the normal
state conduction band Chern number.  Interestingly, this lower bound
is almost identical to the one derived for the MF superfluid phase
stiffness in Ref.~\onlinecite{Liang2017}, provided one replaces 
$\Delta_{\mathrm{pg}}$ with the MF superconducting order parameter.

\section{Numerical Results} 
\label{sec: main-numerical}

\begin{figure}%[h]
\centering
\includegraphics[width=\linewidth]{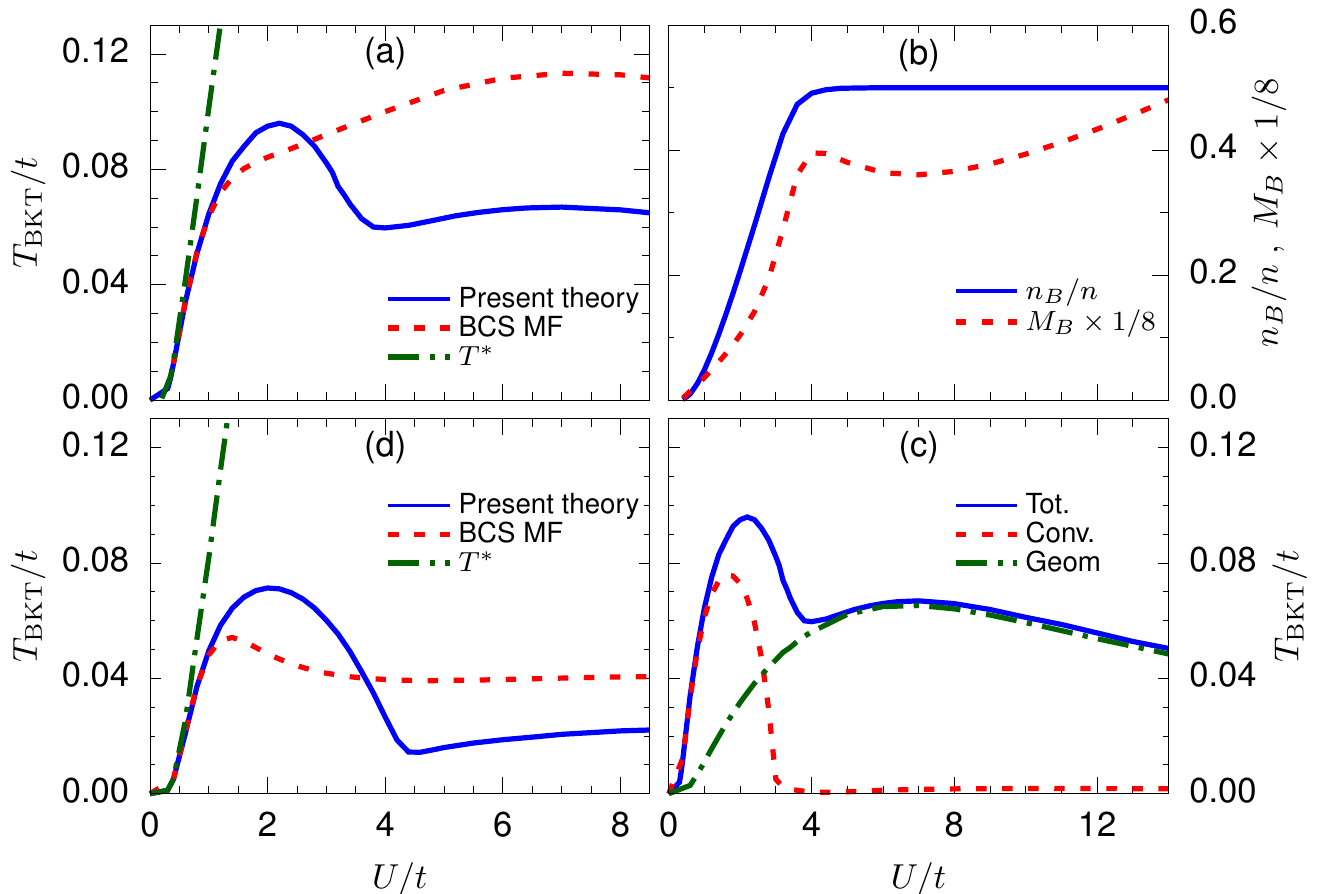}
\vspace{-0.5cm}
\caption{Behavior of calculated (a) $T_{\mathrm{BKT}}$ (labeled ``Present theory") and  (b) $\{ n_\text{B}/n, M_\text{B}\}$,   (c) decomposition of  $T_{\mathrm{BKT}}$ (``Tot'') into conventional (``Conv") and geometric contributions (``Geom") for topological bands, and  (d)  $T_{\mathrm{BKT}}$  for a non-topological system, as a function of $U/t$, all with  $\mathcal{F} = 0.2$. In comparison, also plotted in (a) and (d) are $T^*$ and  $T_{\mathrm{BKT}}$ (``BCS MF'') calculated using the MF  phase stiffness.
}
\label{fig: Fig1}
\end{figure}

In Fig.~\ref{fig: Fig1}(a) we compare the
calculated $T_{\mathrm{BKT}}$ from our pairing fluctuation theory with
that using the BCS MF superfluid phase stiffness $D_{s}$ for the
case of a flatness
parameter $\mathcal{F}=0.2$.  Also
plotted is the pairing onset temperature, $T^*$, well approximated by the
mean field transition temperature.  In the weak-coupling BCS limit,
all three temperatures converge.  However, in the strong coupling
regime, pairing flucuations become important and our
$T_{\mathrm{BKT}}$ is significantly reduced relative to its MF
counterpart, as a consequence of an additional bosonic excitation
channel.  Unlike the single band theory, where there is a more
dramatic $T_{\mathrm{BKT}}$ downturn near $U/t\approx 3$, in
this multi-orbital model the geometric contribution prevents the
expected strong decrease ~\cite{Wang2020}.

These features can be traced to the behavior of the pair mass,
$M_{\mathrm{B}}$, which is plotted along with $n_{\mathrm{B}}$ in
Fig.~\ref{fig: Fig1}(b).  In single band theories with conventional
contributions only, due to a large suppression of pair
hopping~\cite{Nozieres1985} and an increase of pair-pair repulsion with pair density~\cite{Micnas1990}, 
pairs tend to be localized near $U/t \approx 3$, corresponding to
$M_{\mathrm{B}} \rightarrow \infty$.
The presence of geometric terms prevents this pair mass divergence. 
Figures \ref{fig: Fig1}(a) and (b) reveal that, while the small $U$
behavior of $T_{\mathrm{BKT}}$ derives from variations in both
$M_{\mathrm{B}}$ and $n_{\mathrm{B}}$,
the behavior of $T_{\mathrm{BKT}}$ in the BEC regime reflects that of
 $1/M_{\mathrm{B}}$ only.

To see the importance of the geometric contributions more clearly,
in Fig.~\ref{fig: Fig1}(c) we present a decomposition of
$T_{\mathrm{BKT}}$ in terms of the conventional and geometric components,
by separating the total $n_\text{B}/M_\text{B}$ into two terms,
$ (n_\text{B}/M_\text{B})^{\text{conv}} \equiv 2 \Delta_{\text{pg}}^2
T_{\text{conv}}$ and
$ (n_\text{B}/M_\text{B})^{\text{geom}} \equiv 2 \Delta_{\text{pg}}^2
T_{\text{geom}}$.  We then apply the BKT criterion in Eq.~\eqref{eq:
  BKT} to each of
$\{n_\text{B}/M_\text{B} , (n_\text{B}/M_\text{B})^{\text{conv}} ,
(n_\text{B}/M_\text{B})^{\text{geom}} \} $ to arrive at the three
curves in Fig. \ref{fig: Fig1}(c). Here we see that $T_{\mathrm{BKT}}$
is almost completely geometric at $U/t \gtrsim 3$.
The conventional contribution in Fig. \ref{fig: Fig1}(c) exhibits a
dome-like dependence on $U$ with a maximum at $U \sim W$.  Its
contribution to  $T_{\mathrm{BKT}}$  in the pairing
fluctuation theory falls precipitously to almost zero at
$U/t \approx 3 $ and remains extremely small at larger $U$,
resulting from a cancellation between pair hopping and inter-pair repulsion
effects \cite{Appendix}.

\begin{figure}
\centering
\includegraphics[width=\linewidth]{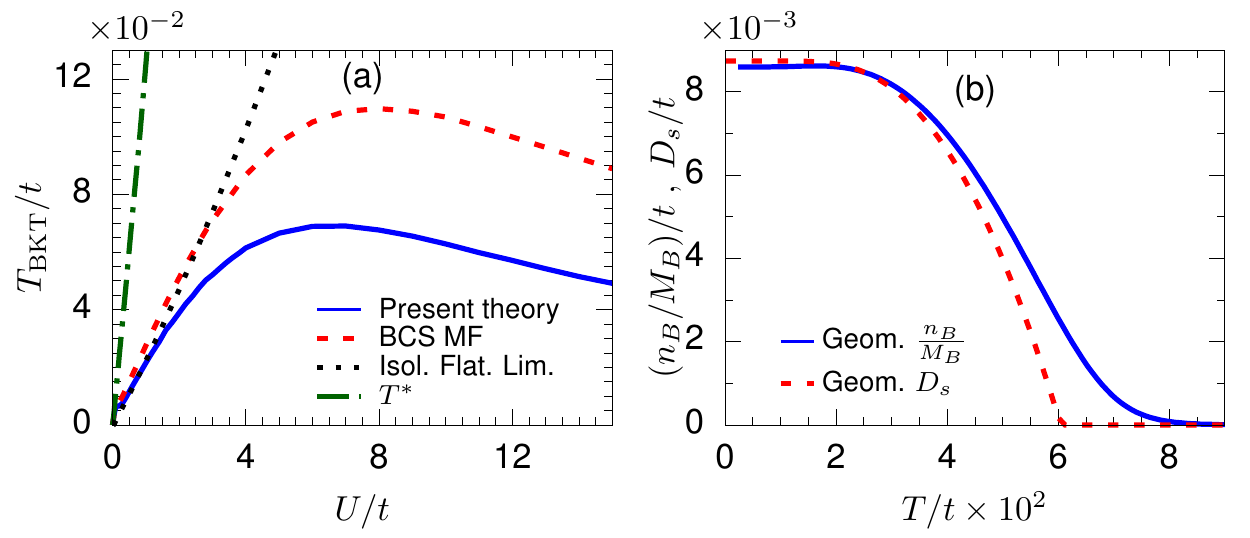}
\vspace{-0.5cm}
\caption{(a) Characteristic temperatures for the topological
  $\mathcal{F}=0.01$ superconductor, and comparison with lower bound
  of $T_{\mathrm{BKT}}$ in the isolated flat band limit
  (``Isol. Flat. Lim"), obtained using Eqs.~\eqref{eq: BKT} and
  \eqref{eq: bound2}.  This bound nearly coincides with the
  calculated $T_{\mathrm{BKT}}$ for the range
  $0.4 \lesssim U/t \lesssim 2$, where the system is in the BEC regime and $T_{\mathrm{BKT}}$ is nearly
  completely geometric.
  (b) Comparison between the $T$ dependence of $n_\text{B}/M_\text{B}$
  and that of BCS MF $D_s$ at $U/t=0.5$.  For the sake of clarity,
  only geometric contributions are included.  }
\label{fig: Fig2}
\end{figure}

It is instructive to compare with
a non-topological superconductor, as shown in Fig.~\ref{fig: Fig1}(d).
Our non-topological bands are constructed by adding a staggered
on-site potential to the topologically nontrivial Hamiltonian
$H_{\mathrm{N}}$ in Eq.~\eqref{eq: HK2}~\cite{Appendix}.  For a meaningful comparison
the trivial band structure is so chosen that both its conduction band
width $W$ and band gap $E_g$ are comparable to the nontrivial
$\mathcal{F}=0.2$ case. 
This ensures that the conventional
contributions to $T_{\mathrm{BKT}}$, as well as the $U$ dependence of
$\Delta_\text{pg}$ and $\mu_\text{F}$, are more or less the same in
both cases.  Comparison of  $T_{\mathrm{BKT}}$ in Fig.~\ref{fig:
  Fig1}(d) and Fig.~\ref{fig: Fig1}(a) at $U/t \gtrsim 4 $,
where the geometric component dominates, demonstrates that the geometric
contribution to $T_{\mathrm{BKT}}$ is significantly enhanced in the
non-trivial case.

In Fig.~\ref{fig: Fig2}(a) we present a comparison between the MF and
present theory for a nearly flat conduction band, with
$\mathcal{F} \approx 0.01$.  Just as in Fig.~\ref{fig: Fig1}(a),
pairing fluctuations suppress significantly the transition temperature
relative to the mean field result.  Also important is the absence of
the conventional $T_{\mathrm{BKT}}$ peak, seen in Fig.~\ref{fig:
  Fig1}(a).  There is a small residual feature at $U \sim W =0.035 t$
from the conventional term, which, however, is invisible in the plot.
In this nearly flat band limit, $T_{\mathrm{BKT}}$ is essentially
purely geometric for the entire range of $U/t$ displayed. Notably,
even a very
small attraction ($U/t \approx 0.3$) puts the system in the BEC
regime \cite{Appendix}, where $n_\text{B}/n$ reaches $1/2$.

Also plotted in Fig.~\ref{fig: Fig2}(a) are the pairing onset
temperature $T^*$ (dot-dashed) along with 
the lower bound of $T_{\mathrm{BKT}}$ in the
isolated flat band limit (black dotted line), which is obtained by applying
the BKT criterion in Eq.~\eqref{eq: BKT} to the r.h.s. of
Eq.~\eqref{eq: bound2}.  Interestingly this bound is
almost saturated by our calculated $T_{\mathrm{BKT}}$ when
$0.4\lesssim U/t \lesssim 2$.

Even with the reduction of $T_{\mathrm{BKT}}$ relative to the BCS MF
result, in the isolated flat band limit, $n_\text{B}/M_\text{B}$ is
essentially equal to its BCS MF counterpart $D_s$ at
$T=T_{\mathrm{BKT}}$ and even for higher temperatures, provided
$T \ll T^*$.  This can be seen through the comparison in
Fig.~\ref{fig: Fig2}(b) between our $n_\text{B}/M_\text{B}$ in
Eq.~\eqref{eq: isoflatband2} and that of the MF $D_s$, where
for clarity we have dropped the small but nonzero conventional term \cite{Appendix}.

\begin{figure}
\centering
\includegraphics[width=\linewidth]{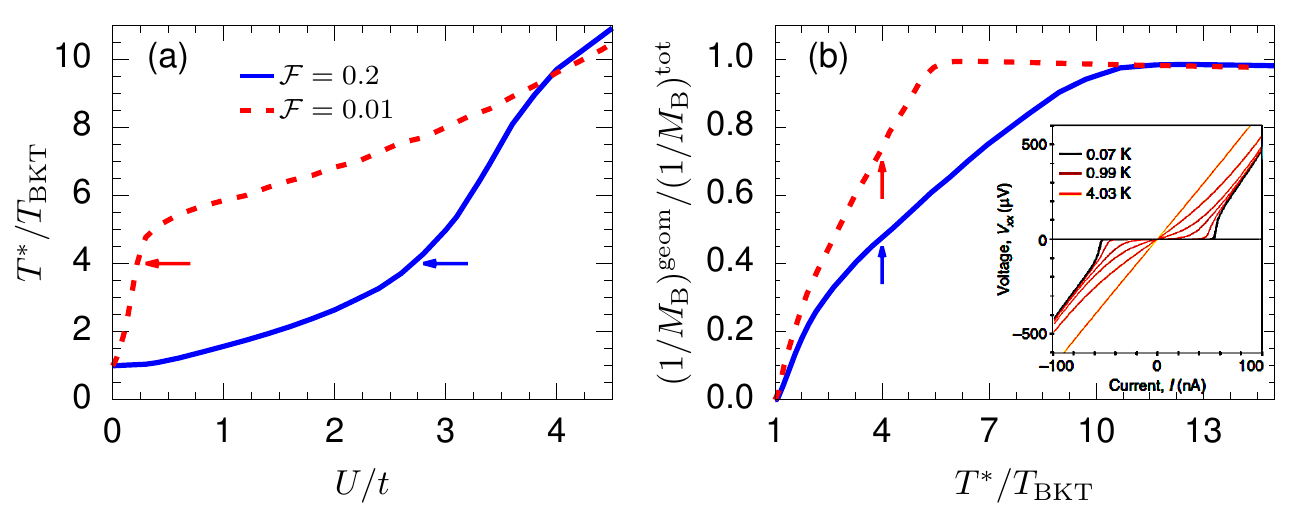}
\vspace{-0.5cm}
\caption{ (a) Calculated $T^*/T_{\mathrm{BKT}}$ as a function of $U$
  and (b) relative magnitude of the geometric terms plotted as
  $(n_\text{B}/M_\text{B})^{\mathrm{geom}}$ over
  $(n_\text{B}/M_\text{B})^{\mathrm{tot}}$ as a function of
  $T^*/T_{\mathrm{BKT}}$, for the topological $\mathcal{F}=0.2$
  and $\mathcal{F}=0.01$  cases.  Arrows
  correspond to where $T^*/T_{\mathrm{BKT}}=4$, deduced from the
  experiments in the inset of (b). (Inset) $V-I$  curves measured at
  different $T$  for magic-angle TBLG by Cao
  et. al.~\cite{Cao2018a}  }
\label{fig: Fig3}
\end{figure}

We turn finally to the physical implications of our calculations for a
given 2D superconductor.  We quantify the relative size of the
geometric terms by use of the dimensionless ratio
$T^*/T_{\mathrm{BKT}}$ which, importantly, has
been shown to be measurable in voltage current
($V -I $) experiments \cite{Zhao2013} with consistency checks from STM
data.  As shown in Figure \ref{fig: Fig3}(a),
$T^*/T_{\mathrm{BKT}}$
increases monotonically with interaction strength $U$ for both the
topological $\mathcal{F}=0.2$ and $\mathcal{F}=0.01$ cases, with an
even more rapid increase as the system approaches the BEC regime.
The fractional contribution
of the geometric terms,
$(n_\text{B}/M_\text{B})^{\mathrm{geom}}/
(n_\text{B}/M_\text{B})^{\mathrm{tot}}$,
is plotted in Fig.~\ref{fig: Fig3}(b). Once 
in the BEC regime, $T_{\mathrm{BKT}}$ is dominantly geometric.

To connect to experiments on TBLG, we present the experimental $V-I$
curves for an optimal example \cite{Cao2018a}, in the inset of
\ref{fig: Fig3}(b).  At $T=T_{\mathrm{BKT}}$ the $V-I$ curve follows a
power law, $V\propto R_{\mathrm{N}} I_c (I/I_c)^\alpha$ with
$\alpha=3$; $I_c$ is the critical current and $R_{\mathrm{N}}$ is the
normal state resistance \cite{Halperin1979, Epstein1981, Resnick1981,
  Hebard1983, Fiory1983, Kadin1983}.  Importantly, when $T$ reaches
$T^*$ the $V-I$ curve fully recovers its normal state Ohmic behavior,
$V \propto R_{\mathrm{N}} I$.

From the $V-I$ characteristics by Cao et al. \cite{Cao2018a}, we
estimate $T^*\approx 4 \mathrm{K}$ and
$T_{\mathrm{BKT}} \approx 1 \mathrm{K}$ ~\cite{Cao2018a}, which yield
$T^*/T_{\mathrm{BKT}} =4$.  At this ratio, the normalized geometric
contribution is about $70\%$ and 50\% for $\mathcal{F}=0.01$ and 0.2,
respectively, in Fig.~\ref{fig: Fig3}(b).  Which band flatness ratio
is more appropriate for magic angle TBLG depends on one's estimate of
the effective bandwidth $W$ and bandgap $E_g$.  If we take
$W\approx 3 \sim 5 \mathrm{meV}$, which is the energy range where the
bare flat band density of states is appreciable, and
$E_g\approx 20 \mathrm{meV}$~\cite{Cao2018a}, then
$\mathcal{F} \approx 0.15 \sim 0.25$.  At face value, this suggests
that the $\mathcal{F}=0.2$ case is more relevant to magic angle TBLG.
However, one should keep in mind that the estimated $W$ here only
provides an upper bound, as the superconductivity in TBLG may be
associated with a renormalized and therefore smaller effective band
width $W$.  In any case, the geometric contribution to
$T_{\mathrm{BKT}}$ is significant, ($\gtrsim 50\%$), and the system is
in the BCS-BEC crossover regime, although it has not yet passed into
the BEC.

\subsection{Further experimental estimates of $T^*$}
\label{sec: T*}

We stress that the $T^*/T_{\mathrm{BKT}}$ ratio
inferred from the
$V-I$ characteristics
 can be quite different from observations in other experiments and with different
samples
~\cite{Lu2019,Stepanov2020,Cao2020a}.
For example, for one superconductor studied in Ref.~\onlinecite{Stepanov2020}, the
ratio is only about $1.4$,
 with $T_{\mathrm{BKT}}=710 \mathrm{mK}$ and $T^* \approx 1 \mathrm{K}$.
This puts the corresponding system in the BCS weak-coupling regime in
Fig.~\ref{fig: Fig3}(a), and
consequently the corresponding geometric contribution to $T_{\mathrm{BKT}}$ from
Fig.~\ref{fig: Fig3}(b) is only about $10\sim 20\%$.
However, one should also take note that the $T^*$ read off from all the
existing $V-I$ curves is subject to
uncertainty since none of
the measurements provides a continuous sweep over closely separated
temperature intervals.

When the $V-I$ measurements are not available, it appears that
$T^*$ can be roughly estimated from dc transport.
This is based on a temperature feature in the longitudinal resistivity $\rho(T)$, which
corresponds to
the point where $\rho(T)$ begins to drop below its normal state extrapolation
\footnote{
Similar dc transport signatures of $T^*$ have been observed previously
in cuprates~\cite{Timusk1999}, although the pseudogap there can have a
completely different origin from preformed Cooper pairs.
}.
For example, in transport experiments on a TBLG sample with $T_{\mathrm{BKT}}=1\mathrm{K}$ in Ref.~\onlinecite{Cao2018a},
this transport signature yields $T^* = 4 \sim 5 \mathrm{K}$, roughly consistent with
the value obtained from $V-I$ measurements.
While $T^*$ identified in this way is necessarily greater than
or equal to $T_{\mathrm{BKT}}$, depending on the carrier density
and twist angle, it can be substantially larger.
As seen from transport studies in Fig.1 of Ref.~\onlinecite{Cao2020a},
the $T^*/T_{\mathrm{BKT}}$ ratio varies from a number close to $1$
to a number much larger than $10$ as the carrier density is tuned from one side of the superconducting dome
to the other in a given sample
\footnote{
Here we ignore the intervening correlated insulating phase at half filling of the lower and upper flat band,
and view the two superconducting domes flanking the insulating phase as one.
}.

One can speculate that this wide variation of $T^*/T_{\mathrm{BKT}}$  obtained
from transport, is unlikely to be due to disorder given that the measurements are
on the same
sample, though with different carrier density.
Instead, variations in Coulomb screening, which crucially depends on the carrier density may play a key role~\cite{Stepanov2020, Saito2020, Liu2020, Cao2020a}.

Because of the sensitivity of
the effective pairing interaction
to band filling and Coulomb screening,
determining whether superconducting magic angle TBLG is a
weak-coupling
or strong-coupling superconductor remains an open question. 
To firmly settle the issue, further $V-I$ experiments over finely
separated temperature intervals
in order to establish the temperature for the Ohmic recovery are much
needed. As in
Ref.
\onlinecite{Zhao2013},
for corroboration, these should ultimately be
combined with STM
measurements of the local pairing gap.
STM experiments~\cite{Jiang2019, Wong2020, Xie2019, Saito2020} on magic angle TBLG to date tend to be limited to the
normal state and have not yet reported signatures of the pairing
gap or of $T^*$.

\section{Comparison with theoretical literature }
\label{sec: comparison}

\begin{figure}
\centering
\includegraphics[width=0.8\linewidth]{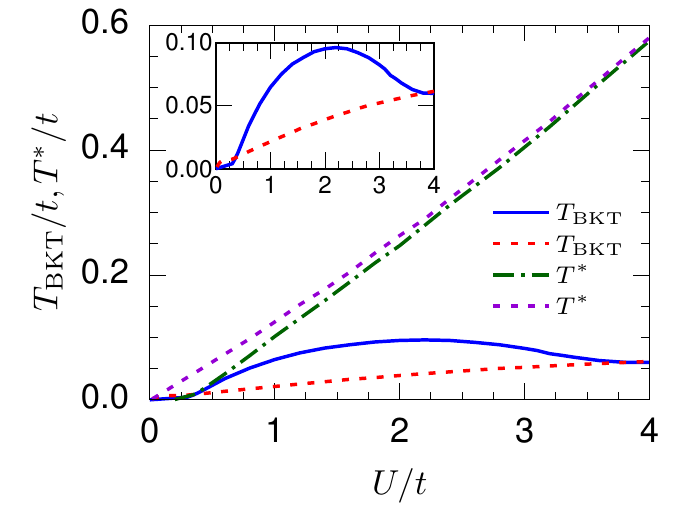}
\caption{
Results from the pairing fluctuation theory for $T_{\mathrm{BKT}}$ and $T^*$, for both the topological $\mathcal{F}=0.2$ band (solid blue and dark-green dash-dotted lines)
and $\mathcal{F}=0.01$ bands  (red and violet dashed lines). For a better comparison to the Monte-Carlo results~\cite{Hofmann2019}, only
data for $U/t$ up to about $4$ are shown.
(Inset) Zoomed view of $T_{\mathrm{BKT}}$.
 }
\label{fig: FigS4}
\end{figure}

Fig.~\ref{fig: FigS4} makes possible a comparison between our numerical results for $T^*$ and $T_{\mathrm{BKT}}$ and the Monte-Carlo (MC) calculations in Ref.~\onlinecite{Hofmann2019}.
Interestingly, our results are quite similar to the MC results, both qualitatively and even quantitatively.
The main difference is a small peak in $T_{\mathrm{BKT}}$ at $U/t\approx 2$ for $\mathcal{F}=0.2$,
which derives from the conventional terms and is absent in the MC results. Instead, the MC $T_{\mathrm{BKT}}$ for $\mathcal{F}=0.2$ has a $U$
dependence quite similar to that for $\mathcal{F}=0.01$, although the magnitude is larger in the former case (see Fig. 1 of Ref.~\onlinecite{Hofmann2019})
\footnote{We note that the results presented in Fig. 1 of Ref.~\onlinecite{Hofmann2019} are for electron density  $ n=0.5$ per site;
%, corresponding to lower band filling level $\nu=0.25$;
 while our results are for  $n=0.3$ per site. However, we do not expect the qualitative  $U$ dependence
of $T_{\mathrm{BKT}}$ and $T^*$ to change from $n=0.5$ to $n=0.3$.
 }.
Taken at face value, this suggests that our pairing fluctuation theory overestimates the size of the conventional contribution
to $T_{\mathrm{BKT}}$.
On the other hand, the MC results may suffer from finite size effects.
In any event, this comparison indicates that our pairing fluctuation theory
appears to have adequately accounted for the geometric contributions.

Prior to our work there have been studies of
the geometric contribution to the superfluid instability temperature
that were associated with beyond-mean-field Gaussian pairing fluctuations.
In a series of papers~\cite{Iskin2018, Iskin2018a, *Iskin2019a, *Iskin2020, *Iskin2020a}, M. Iskin called attention to the geometric contribution
in 2D and 3D spin-orbit coupled Fermi gases. Notably, for this specific
energy dispersion,
the geometric contribution
does not play a significant role and the conventional contribution dominates, due to the associated non-flat and unbounded band dispersion.
It should be noted that within a Gaussian fluctuation theory, which is most appropriate for
3D superfluids, there does appear an inter-band geometric contribution~\cite{Iskin2018}
similar to our Eq.~\eqref{eq: Tgeom}
\footnote{However, the intra-band contribution we identified, the term with $\alpha^\prime=\alpha$ in Eq.~\eqref{eq: Tgeom}, was missing.}.

Beyond mean
field effects and quantum geometry have also been discussed in Ref.~\onlinecite{Liang2017} in the
context of dynamical mean field theory (DMFT).
There it was similarly observed that the geometric contribution to the flat band
phase stiffness survives,
though reduced in magnitude.
These DMFT calculations were shown
to agree qualitatively with
the results of strict mean field theory, not in the
BCS regime but in the more strongly correlated BEC limit,
where one might expect a mean field approach to be less appropriate.
Finally, we note that there are other more analytical approaches
which incorporate bosonic fluctuation effects
on the superfluid phase stiffness across the entire BCS-BEC crossover~\cite{Benfatto2004, Fukushima2007, Bighin2016}
\footnote{Unfortunately, these calculations lead to an unusual double valued
functional form for the superfluid density.}.
While the role of this additional ``collective mode" bosonic branch is
to degrade the superfluid phase stiffness, as we find here,
these schemes have not addressed quantum geometric effects.
Further investigations are needed to resolve these issues.

\section{Conclusions}
\label{sec: conclusion}

In summary, we have established the quantum geometric contribution to superfluidity in a pair-fluctuation theory, where these contributions modify the pair mass.
In general the geometric contribution dominates in the strong coupling BEC regime and prevents localization of Cooper pairs.
We further show how to quantify the magnitude of the geometric contributions in
a multi-orbital 2D superconductor in terms of the $T^*/T_{\mathrm{BKT}}$
ratio. Our analysis was based on important experimental observations
\cite{Zhao2013} which have shown that the two temperature scales
($T_{\mathrm{BKT}}$ and $T^*$)
can be extracted from $V-I$ plots.
Using estimates of $T^*/T_{\mathrm{BKT}}$ from experiments
we have presented 
speculations on magic angle
TBLG, concerning the size of the geometric terms and the location of
this exotic superconductor within the BCS-BEC crossover.

%NEW FIRST SENTENCE NOTE THIS PARAGRAPH WAS MOVED.  
The rough comparison between theory and experiment in this paper is
based on the assumption that the simple model we studied captures some
essential features of the band structure of TBLG. While this sets up
the general framework and identifies the issues, clearly, a
calculation using a realistic band structure is ultimately needed. In
our model, the band topology comes from a nonzero spin Chern number.
On the other hand, (in the absence of
the hBN encapsulating substrate) \footnote{As shown by recent
  experimental and theoretical studies~\cite{Sharpe2019, Serlin2020,
    Zhang2019, Bultinck2020}, a coupling of the TBLG to the hBN
  substrate can break certain symmetry that leads to occurrence of
  Chern bands and the resulted anomalous Hall effect. } , the relevant
topology for the bare flat bands of TBLG was argued to be different
and to correspond to a so called ``fragile topology" ~\cite{Zou2018,
  Po2018, Po2018a, Po2019, Ahn2019}.  Whether this topology is
associated with the normal state out of which the superconductivity
emerges is still unclear \footnote{In principle, the normal state
  bands relevant to the superconductivity can be different from the
  bare band, due to renormalizations from interactions.}.  However, as
demonstrated in a BCS mean field calculation~\cite{Xie2020}, this
fragile topology exhibits similar Wannier obstruction effects that can
prevent the localization of Cooper pairs and hence enhance 2D
superconductivity for a flat band system.  Overall, we expect
most of our qualitative findings to survive in a more realistic band
calculation with fragile topology included.  The pairing fluctuation
theory that we presented for our two band model can be easily
generalized to a more-than-two-band structure, which is more relevant
to TBLG. We leave that for future work.

\begin{acknowledgments}
We thank M. Levin for a useful discussion. This work was primarily
funded by the University of Chicago Materials Research Science and
Engineering Center, and the National Science Foundation under Grant
No. DMR-1420709.  Q.C. was supported by NSF of China (Grant
No.~11774309).
\end{acknowledgments}

\appendix

\section{Tight-binding model} \label{app: model}
\begin{figure}
\includegraphics[width=0.425\linewidth]{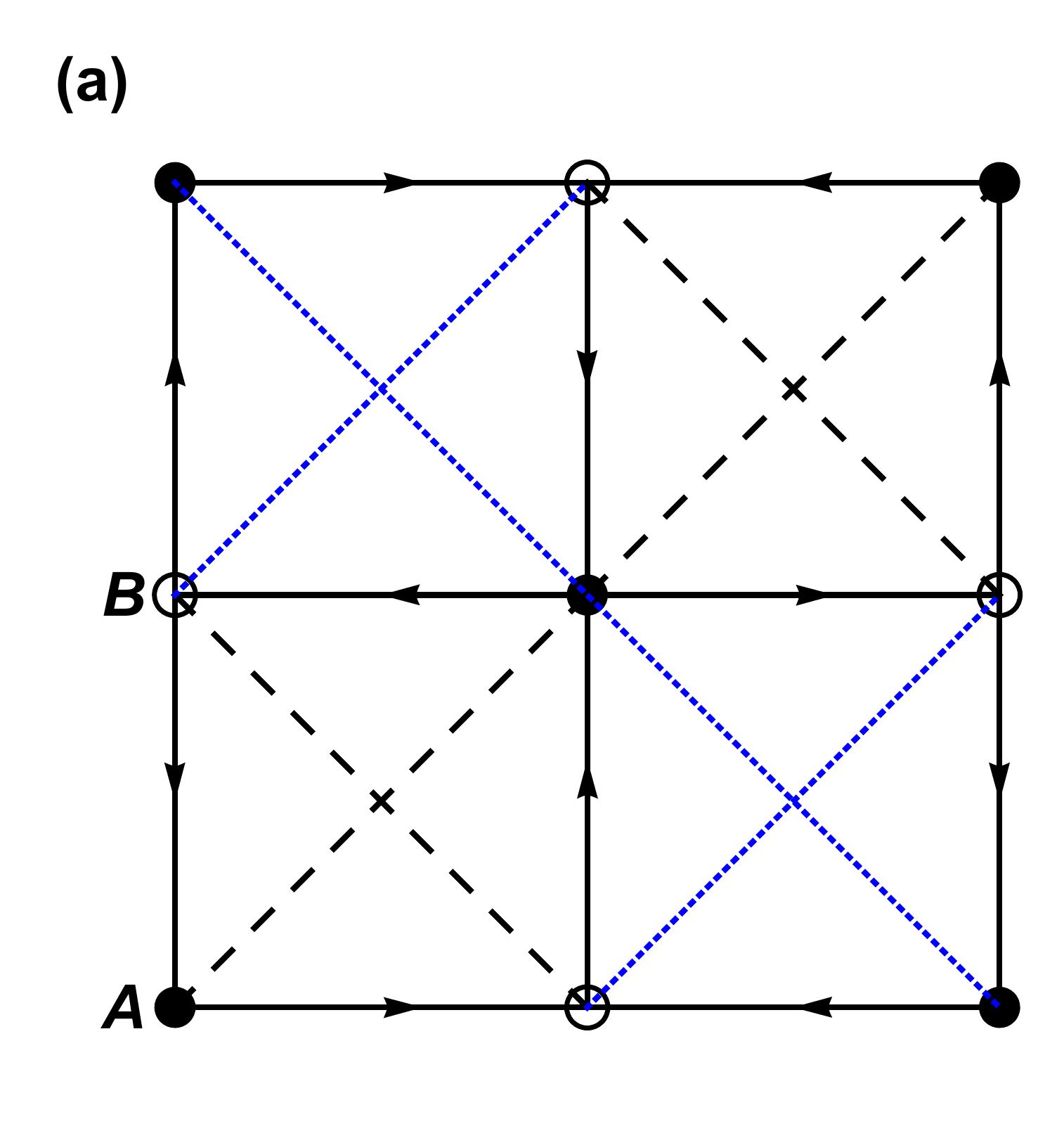}%
\hspace{1cm}
\includegraphics[width=0.41\linewidth]{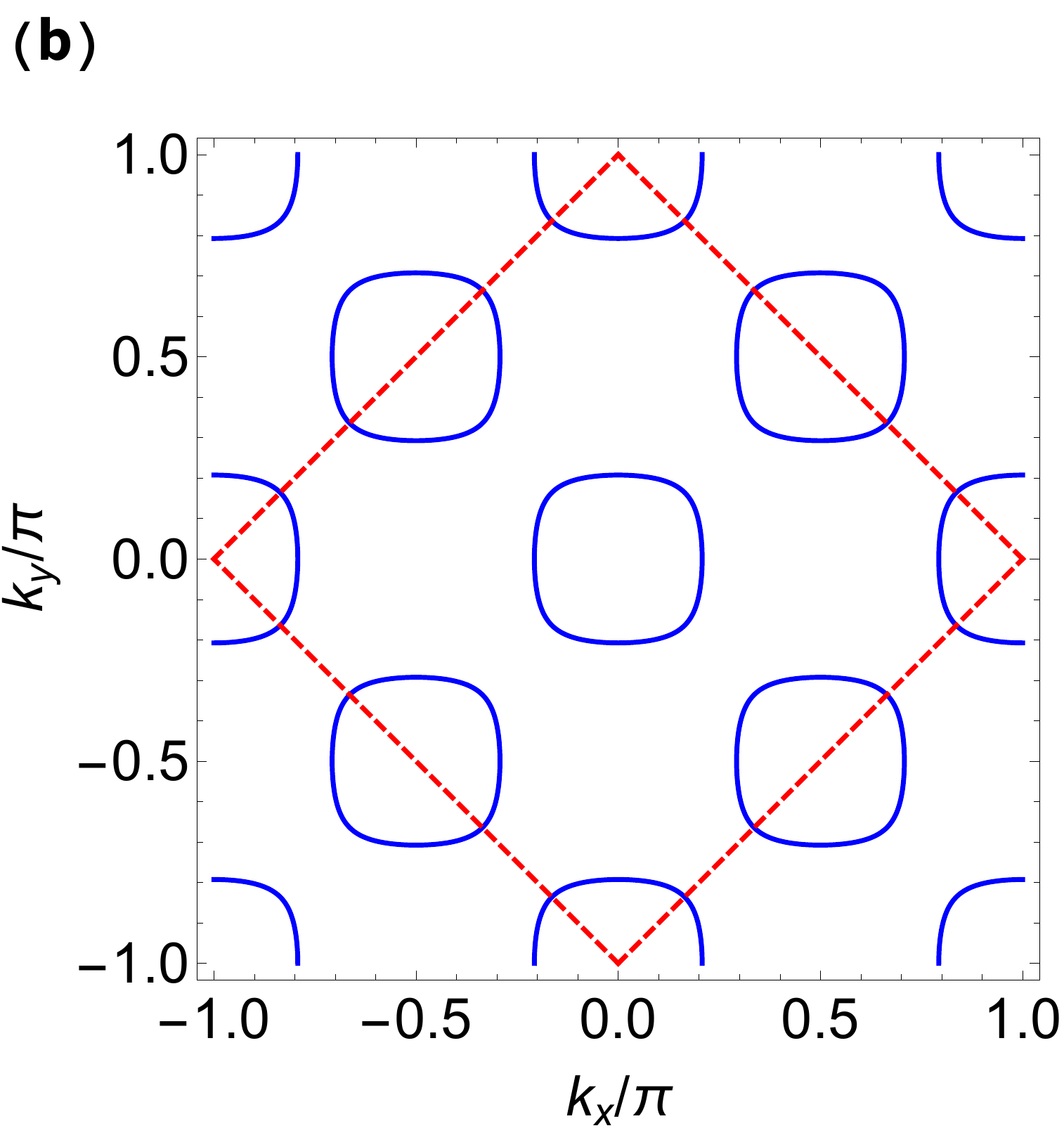} \\
\includegraphics[width=0.425\linewidth]{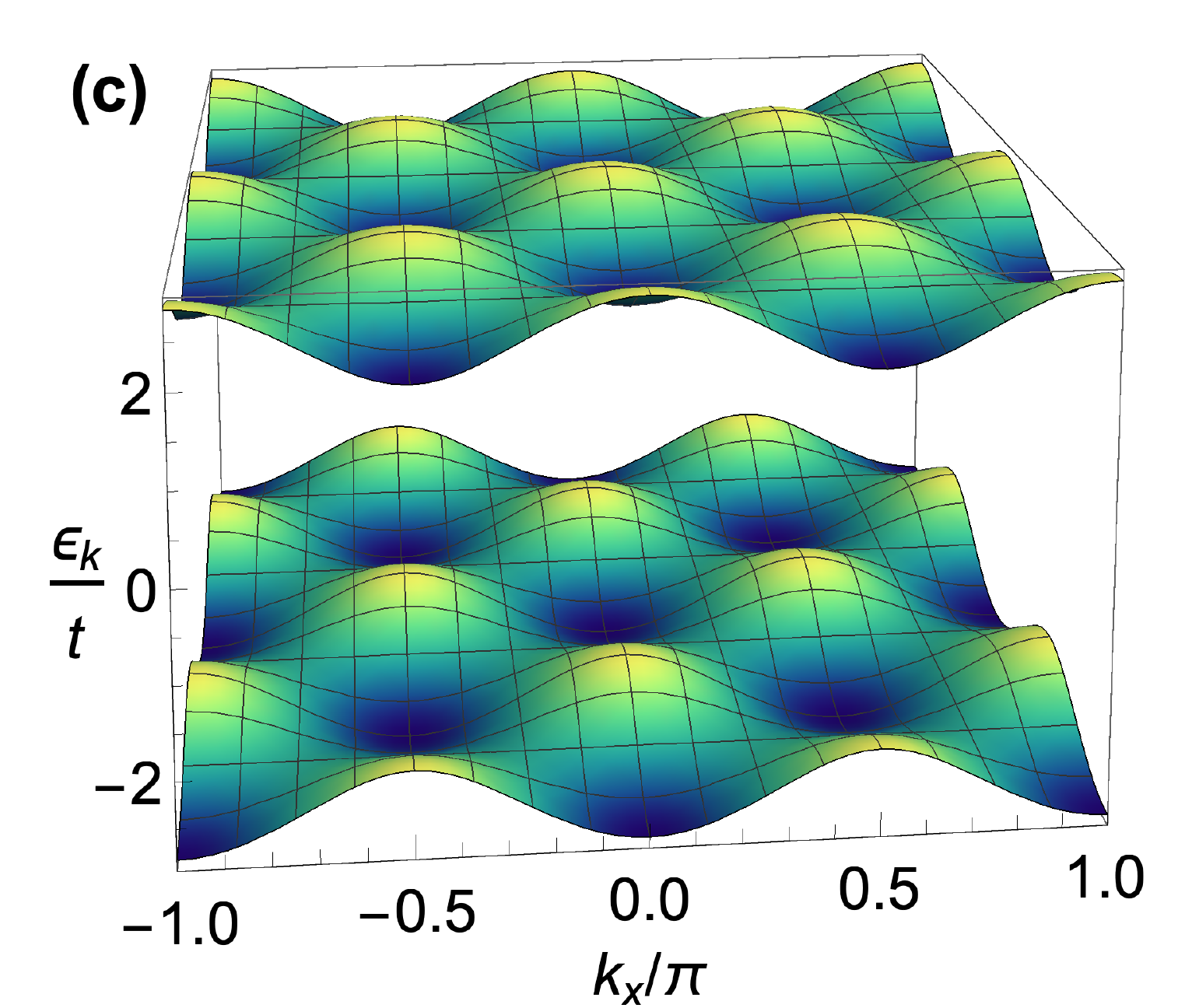}%
\hspace{1cm}
\includegraphics[width=0.425\linewidth]{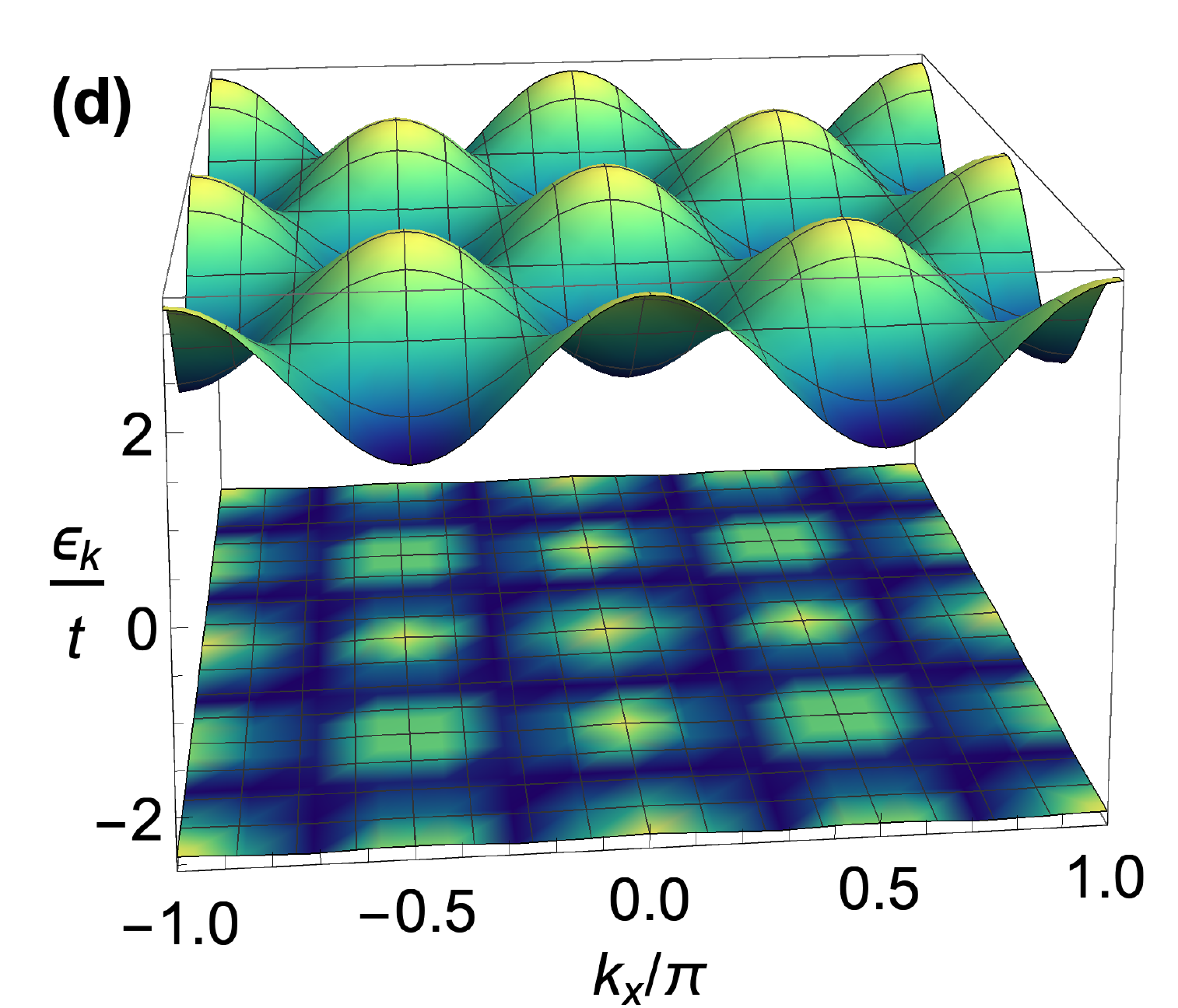} 
\caption{(a) The tight binding model for $H_{\mathrm{K}}$.
 $\{ A, B\}$ denote two different sub-lattices, resulting from a staggered $\pi$ flux.
 The NN hopping amplitudes are $t e^{i \pi/4}$ for spin $\uparrow$ along the direction depicted by the arrows. 
 Black dashed and blue dotted lines show the second NN bond with which the associated hopping amplitudes are $t_2$
 and $-t_2$, respectively. There is also a uniform hopping between the fifth NN sites, which is not shown for clarity. 
 (b) Fermi surfaces (FS), in blue, for the band flatness ratio $\mathcal{F}=0.2$ at electron density $n=0.3$ per site. 
The regime bounded by the red dashed lines defines the reduced Brillouin zone (RBZ).
(c) Corresponding band structure for $\mathcal{F}=0.2$. In the vertical axis,
$\epsilon_{\vec{k}} =h_0(\vec{k}) \pm |\vec{h}(\vec{k})|$. 
(d) Band structure for $\mathcal{F}=0.01$. 
}
\label{fig: model}
\end{figure}

Our model is defined on a square lattice, with a kinetic energy
contribution to the Hamiltonian,
$H_{\mathrm{K}}$,  given by
\begin{align}
H_{\mathrm{K}}   = &  \bigg\{ \bigg[  -t \sum_{\langle i, j\rangle } e^{i \phi^\sigma_{ij}} c^\dagger_{i,\sigma} c_{j,\sigma} 
               - t_2 \sum_{\langle i, j \rangle_2, \sigma} s_{\langle i, j \rangle_2} c^\dagger_{i,\sigma} c_{j,\sigma} \nonumber \\
            & -t_5 \sum_{\langle i, j \rangle_5, \sigma} c^\dagger_{i,\sigma} c_{j,\sigma} \bigg] + h.c. \bigg\} - \mu_{\mathrm{F}} \sum_i n_i. 
\end{align}
Here
$c^\dagger_{i,\sigma}$ ($c_{i,\sigma}$) are electron creation (annihilation) operators at site $i$ for spin $\sigma$. 
$(t, t_2, t_5)$ are the magnitudes of the hopping integrals defined for the nearest neighbor (NN), second NN , and the fifth NN bond
on the square lattice, respectively. 
 $\mu_{\mathrm{F}}$ is the fermionic chemical potential, and $n_i=\sum_{\sigma=\uparrow, \downarrow}  c^\dagger_{i,\sigma} c_{i,\sigma}$
 is the electron number at site $i$. 
 The NN hopping amplitude is modulated by the phase $e^{i \phi_{ij}^{\sigma}}$, where $\phi_{ij}^\sigma = s_{\sigma} \, (\pi/4) $ if the hopping is along the direction
 of the arrows depicted in Fig.~\ref{fig: model}.  $s_\sigma=+1$ ($-1$) for spin $\uparrow$ ($\downarrow$). 
 Because of $\phi^\sigma_{ij}$ there is a net $\pm \pi$ flux through each square plaquette for given spin.  
This flux is staggered from one plaquette to the next (see Fig.~\ref{fig: model}), which breaks the original lattice translational symmetry and leads to two different sublattices $\{A, B\}$. 
However, time reversal symmetry is still preserved,
because $\phi_{ij}^\sigma$ are opposite for opposite spin $\sigma$ so that the total flux through each plaquette is zero.
The sign of the second NN hopping amplitudes, $s_{\langle i,j \rangle_2}=\pm$, is also staggered, as shown in Fig.~\ref{fig: model}. 

Fourier transforming $H_{\mathrm{K}}$ to $\vec{k}$ space one finds the following block-diagonal Hamiltonian
\begin{gather}
H_{\mathrm{K}}(\vec{k}) =  
\begin{pmatrix}
H_{\uparrow} (\vec{k})   & 0 \\
         0                                 & H_{\downarrow}(\vec{k})
\end{pmatrix}, 
\label{eq: HK}
\end{gather}
in the basis $(c_{A,\uparrow}^\dagger (\vec{k}), \; c^\dagger_{B,\uparrow}(\vec{k}),  \; 
c_{A,\downarrow}^\dagger (\vec{k}),  \; c^\dagger_{B,\downarrow}(\vec{k}))$.
The diagonal block operating on the same spin is 
\begin{gather}
H_{\sigma}(\vec{k}) = h_0(\vec{k}) +\vec{h}(\vec{k},\phi_\sigma) \cdot \vec{s} -\mu_{\mathrm{F}},
\end{gather}
where $\vec{s} =(s_x, s_y,s_z)$ are the three Pauli matrices defined for the sublattice space
and 
\begin{subequations}
\begin{align}
& h_0(\vec{k})  = -2 t_5 \,  \big[ \cos 2(k_x+ k_y)  +\cos 2(k_x -k_y)\big],    \label{eq: h0} \\
 & h_z(\vec{k})   = - 2 t_2 \,  \big[ \cos(k_x + k_y) - \cos(k_x-k_y) \big],    \label{eq: hz} \\
 & h_x (\vec{k},\phi_\sigma) + i  \, h_y (\vec{k},\phi_\sigma)   \nonumber \\
&  \hspace{0.8cm} = -2 t \, e^{ i (- \phi_\sigma  - k_y)} \cos k_y     - 2 t \, e^{i (\phi_\sigma -k_y)} \cos k_x .   \label{eq: hxhy}
\end{align}
\end{subequations}
$\phi_\sigma = s_\sigma (\pi/4)$. Diagonalizing $H_{\mathrm{K}}(\vec{k})$ gives two energy bands, 
$\xi_{\pm}(\vec{k})  = h_0 (\vec{k})  \pm |\vec{h}(\vec{k},\phi_\sigma) | -\mu_{\mathrm{F}}$, 
 each of which are two-fold degenerate
due to the spin. The two bands have a nonzero spin dependent Chern number $C_{\alpha\sigma} = - \alpha s_\sigma$,
where $\alpha=\pm$. 

Although $H_{\sigma}(\vec{k})$ depends on spin due to $\phi_\sigma$, the final result of the time reversal invariant quantity, $n_{\text{B}}/M_{\text{B}}$
which determines the temperature, $T_{\text{BKT}}$, in our theory, is spin independent (see the following Sec.~\ref{app: theory}). 
Therefore, in the main text we drop the spin and keep only the spin $\uparrow$ block Hamiltonian, i. e. $H_{\text{N}} \equiv H_{\uparrow}$
in Eq.~\eqref{eq: HK2}.

\subsection{Non-topological model Hamiltonian}
In Fig.~\ref{fig: Fig1}(d) of the main text, we also considered a topologically trivial band structure with zero Chern number. 
The corresponding trivial Hamiltonian is obtained from $H_{\mathrm{K}}(\vec{k})$ by adding a staggered on-site potential term
\begin{gather}
H_{\mathrm{K}}^{ \mathrm{trivial} }(\vec{k}) = H_{\mathrm{K}}(\vec{k}) + m_z s_z \otimes \sigma_0,
\end{gather}
where $\sigma_0$ is the identity matrix in the spin space. 
The resultant bands from $H_{\mathrm{K}}^{ \mathrm{trivial} }(\vec{k})$ are trivial if $|m_z| > 4 t_2$. 
Using $(t, t_2, t_5, m_z) = (1, 0.02, 0, -3)$ gives a two-band model with $W \approx 1.2 \, t$ and $E_g \approx 5.8 \, t$,
corresponding to $\mathcal{F}=0.2$. 
$W$ and $E_g$ are comparable to those of the topological $\mathcal{F}=0.2$ band.  

\subsection{Attractive interaction}
For the interaction we choose a local attractive Hubbard model
\begin{gather}
V= -  U  \sum_i n_{i,\uparrow} n_{i,\downarrow},
\end{gather}
where $U>0$. We do not discuss the possible origin of this attractive interaction in TBLG, which is not important for our purposes.

\section{Multi-orbital BCS-based pairing fluctuation theory} \label{app: theory}

In the main text, we have sketched the derivation of our pairing fluctuation theory and outlined the main equations used. 
In this section we present the details. 
We first derive the expression for our pairing susceptibility
and the corresponding many-body T-matrix. 
From the two we then obtain 
the two central quantities for our calculation of $T_{\mathrm{BKT}}$,
 $n_{\mathrm{B}}$ and $M_{\mathrm{B}}$ of the preformed pairs. 

\subsection{Pairing susceptibility and many-body T-matrix $ t_{\mathrm{pg}}(Q)$}

Our pairing fluctuation theory is one type of the many BCS-BEC crossover theories. The central assumption behind most of these theories is that
even though the original BCS theory is a weak coupling one, the variational BCS ground state wavefunction 
has a wider applicability that goes beyond weak coupling~\cite{Leggett1980}.
Our theoretical framework is designed such that the $T=0$ ground state in this theory is identical to the BCS ground state and at the same time
it includes pairing fluctuation effects at finite $T$. Therefore, to derive such a theory for our
multi-orbital system, we first consider the corresponding BCS mean field problem. 

Within the BCS mean field, the Cooper pairing instability can be derived from the pairing vertex function $\Gamma(Q)$. 
Assuming a local s-wave singlet pairing order parameter $\hat{\Delta}_{\mathrm{sc}}(\vec{k}) =\Delta_{\mathrm{sc}} \;  i \sigma_y$, one can show that~\cite{Zhang2014}
\begin{align}
\frac{1}{\Gamma(Q)}   & = -\frac{1}{U} +\chi_0(Q), \\
\chi_0(Q)                  & = \frac{T}{2}  \sum_{K} 
  \mathrm{Tr} \big[ \mathcal{G}_0(K) i\sigma_y \widetilde{\mathcal{G}}_0(K-Q) (-i \sigma_y) \big].
\end{align}
$\chi_0(Q)$ is the bare pairing susceptibility. 
$K=(\omega_n,\vec{k})$ with $\omega_n=(2n+1) \pi T$ is the fermionic Matsubara frequency.
 The summation over $\vec{k}$ should
be restricted to the reduced Brillouin zone due to the unit cell doubling in real space. 
The trace is with respect to both sublattice and spin.
$\mathcal{G}_0(K)$ and $\widetilde{ \mathcal{G}}_0(K)$ are the normal state electronic and hole Green's function matrices,
whose definitions are
\begin{align}
\mathcal{G}_0(K)                      &  =          1/(i\omega_n - H_K(\vec{k})), \\
\widetilde{ \mathcal{G}}_0(K)    & \equiv    - [\mathcal{G}_0( - K)]^T. 
\end{align}
$1/\Gamma(Q=0)=0$ defines the BCS mean field $T_{c,\mathrm{BCS}}$, which will be taken as an estimate for
the pairing onset temperature
$T^*$ in our theory, i. e., $T^* =T_{c,\mathrm{BCS}}$. 

Correspondingly, the mean field BCS gap equation for $\Delta_{\mathrm{sc}}$ is given by
\begin{align}
 -\frac{1}{U} + \frac{T}{2}  \sum_{K} 
  \mathrm{Tr} \big[ \mathcal{G}(K) i\sigma_y \widetilde{\mathcal{G}}_0(K) (-i \sigma_y) \big] =0,
  \label{eq: BCSeq}
\end{align}
where $\mathcal{G}(K)$ is the electron Green's function with the superconducting pairing self energy $\Sigma_{\mathrm{sc}}(K)$ included 
\begin{align}
[\mathcal{G}(K)]^{-1}            &   = [\mathcal{G}_0(K)]^{-1}  -\Sigma_{\mathrm{sc}}(K),\\
\Sigma_{\mathrm{sc}}(K)     &  = \Delta_{\mathrm{sc}}^2 \widetilde{ \mathcal{G} }_0(K).  \label{eq: SigmaSc}
\end{align}
The zero temperature solution of $\Delta_{\mathrm{sc}}$ to the above gap equation gives the BCS ground state. 

Now we construct the pairing fluctuation theory. 
To account for the effects of scattering from non-condensed pairs on fermions, we include another pairing self energy, $\Sigma_{\mathrm{ pg}}$, into 
the dressed electronic Green's function $\mathcal{G}$
\begin{gather}
[\mathcal{G}(K)]^{-1} = [\mathcal{G}_0(K)]^{-1}  -\Sigma_{\mathrm{pg}}(K).
\end{gather}
$\Sigma_{\mathrm{ pg}}$ results from scatterings of electrons from non-condensed pairs, to be distinguished from $\Sigma_{\mathrm{sc}}$
which represents a true condensate. 
In three dimension (3D) we should include $\Sigma_{\mathrm{sc}}$ as in the BCS mean field theory. 
In 2D and at finite temperature, which is what we focus on, $\Sigma_{\mathrm{sc}} \equiv 0$ since
there is no true long range superconducting order parameter.

$\Sigma_{\mathrm{pg}}(K)$ is related to the many-body T-matrix $ t_{\mathrm{pg}}(Q)$ by
\begin{gather}
\Sigma_{\mathrm{pg}}(K) =  - T \sum_{Q \ne 0 } t_{\mathrm{pg}}(Q) \widetilde{\mathcal{G}}_0(K-Q). \label{eq: SigmaPg}
\end{gather}
All pair scattering effects are encapsulated in $ t_{\mathrm{pg}}(Q)$. 
Under the T-matrix approximation that has been widely used to understand BCS-BEC crossovers~\cite{Nozieres1985, Chen2005}
\begin{gather}
\frac{1}{t_{\mathrm{pg}}(Q)} = -  \frac{1}{U}  +\chi(Q), 
\label{eq: tpg}
\end{gather}
where
\begin{gather}
\chi(Q) =  \frac{T}{2}  \sum_{K} 
  \mathrm{Tr} \big[ \mathcal{G}(K) i\sigma_y \widetilde{\mathcal{G}}_0(K -Q) (-i \sigma_y) \big].  \label{eq: chidef}
\end{gather}
In the course of the developments of BCS-BEC crossover theories, there was a debate on whether the two Green's functions
used in the expression of $\chi(Q)$ should be $ \mathcal{G}_0 \widetilde{\mathcal{G}}_0$,  or $ \mathcal{G} \widetilde{\mathcal{G}}$, or  $ \mathcal{G} \widetilde{\mathcal{G}}_0$. We choose the asymmetric form,  $ \mathcal{G} \widetilde{\mathcal{G}}_0$,
so that in 3D, when the superconducting transition is interpreted as a BEC of Cooper pairs, the ground state of this pairing fluctuation theory is given by the
BCS wavefunction~\cite{Chen2005}.
This is reflected in the pole structure of the T-matrix, determined by $1/t_{\mathrm{pg}}(0)=0$
which yields the usual BCS gap equation for $\Delta_{\mathrm{pg}}$. 
It should be noted that the asymmetric form  $ \mathcal{G} \widetilde{\mathcal{G}}_0$ can in fact be derived within the equation of motion approach \cite{KadanoffMartin,ChenPhD}.

To proceed further, we note that for small pair chemical
potential we may approximate $t_{\mathrm{pg}}(Q)$, noting that it
is sharply peaked near $Q=0$ so that Eq.~\eqref{eq: SigmaPg}
can be written as
\begin{gather}
\Sigma_{\mathrm{pg}}(K) \approx  \Delta_{ \mathrm{pg}  }^2 \widetilde{\mathcal{G}}_0(K),  \label{eq: SigmaPg2}\\
\Delta_{\mathrm{pg}}^2 \equiv - T \sum_{Q \ne 0 } t_{ \mathrm{pg}  }(Q).  \label{eq: DeltaPg}
\end{gather}
We refer to this as the ``pg approximation", which (near the
superconducting instability) is supported by
numerical evidence \cite{Maly1999}. 
Eq.~\eqref{eq: SigmaPg2} is an analog to the BCS pairing self energy given in Eq.~\eqref{eq: SigmaSc}. 
Just as in the BCS mean field theory, the above form of $\Sigma_{\mathrm{pg}}(K)$ leads to a pseudogap $\Delta_{\mathrm{pg}}$ in the fermionic excitation energy spectrum
$E_{\pm}(\vec{k}) =\sqrt{\xi_{\pm}(\vec{k})^2 + \Delta_{ \mathrm{pg} }^2}$, which reflects the binding strength of non-condensed Cooper pairs.

For the Hamiltonian that is block diagonal in Eq.~\eqref{eq: HK}, we can carry out the spin trace in Eq.~\eqref{eq: chidef} and write 
\begin{align}
\chi(Q) & =  \frac{1}{2} \big[ \chi_{\uparrow \downarrow}(Q) + \chi_{\downarrow \uparrow}(Q) \big],  \label{eq: chitot}
\end{align}
where
\begin{align}
\chi_{\sigma \bar{\sigma}}(Q) =  T \sum_{K}  \mathrm{Tr} \big[ \mathcal{G}_{\sigma }(K)\widetilde{\mathcal{G}}_{0,\bar{\sigma}}(K-Q)  \big]. 
\end{align}
$\bar{\sigma}=\uparrow (\downarrow)$ if $\sigma=\downarrow (\uparrow)$. 
$\mathcal{G}_{\sigma}$ and $\widetilde{\mathcal{G}}_{0,\bar{\sigma}}$ are the spin $\sigma$ block of $\mathcal{G}$ and the
spin $\bar{\sigma}$ block of $\widetilde{\mathcal{G}}_{0}$, respectively. 
Substituting the definitions of $\mathcal{G}_{\sigma}(K) $ and $\mathcal{G}_{0,\bar{\sigma}}(K) $ into the expression of 
$\chi_{\sigma \bar{\sigma}}(Q)$ and completing the fermionic Matsubara sum, one gets
\begin{widetext}
\begin{align}
\chi_{\sigma \bar{\sigma}}(Q) 
 & =   \sum_{\vec{k} \in \mathrm{RBZ}} \sum_{ \{ \alpha, \alpha^\prime ,\eta \} =\pm}  
 \frac{1}{2} \big[ 1 + \eta \frac{\xi_{\alpha}(\vec{k})}{E_{\alpha}(\vec{k})} \big]
\frac{  n_F( \eta E_{\alpha}(\vec{k})) -n_F( - \xi_{\alpha^\prime} (\vec{k} -\vec{q}))  }{ i\Omega_m - \eta E_{\alpha}(\vec{k}) -\xi_{\alpha^\prime}(\vec{k} -\vec{q})}
 \mathrm{Tr} \big[   \hat{P}_{ \alpha,\sigma}(\vec{k}) \hat{P}_{\alpha^\prime, \sigma}(\vec{k} -\vec{q}) \big],  
 \label{eq: chispin}
\end{align}
\end{widetext}
where $n_{\mathrm{F}}(x)=1/(e^{\beta x} +1)$ with $\beta=1/T$ is the Fermi-Dirac distribution function and $\mathrm{Tr}[\cdots]$ is with respect to the sublattice subspace. 
\begin{gather}
\hat{P}_{\alpha,\sigma}(\vec{k}) \equiv \frac{1}{2}\big[ 1+ \alpha \; \hat{h}(\vec{k},\phi_\sigma)\cdot \vec{s} \big]
\label{eq: projector}
\end{gather}
is  the projection operator defined for the normal state band with energy $\xi_{\alpha}(\vec{k})$ and spin $\sigma$. 
$\hat{h}(\vec{k},\phi_\sigma) \equiv \vec{h}(\vec{k}, \phi_\sigma)/| \vec{h}(\vec{k},\phi_\sigma)|$.
Carrying out the trace in Eq.~\eqref{eq: chispin} leads to
\begin{gather}
   \mathrm{Tr} \big[   \hat{P}_{ \alpha,\sigma}(\vec{k}) \hat{P}_{\alpha^\prime, \sigma}(\vec{k} -\vec{q}) \big]
   =  \frac{  1+ \alpha \alpha^\prime \hat{h}(\vec{k},\phi_\sigma) \cdot \hat{h}(\vec{k}-\vec{q},\phi_\sigma) }{2}.
   \label{eq: projector2}
\end{gather}

\subsection{Small $Q$ expansion of $\chi(Q)$}
 
Within the ``pg approximation"
one can make the 
the following small $Q$ expansion for $\chi(Q)$~\cite{Chen2005, Wu2015, Wang2020},
\begin{align}
\chi(Q)        & \approx \chi(0) + b\,i\Omega_m -  c \; \vec{q}^2, \label{eq: chiSmallQ} 
\end{align}
where $\Omega_m = 2m \pi T$ is the bosonic Matsubara frequency, and
\begin{subequations} \label{eq: coeffabc}
\begin{align}
\chi(0)   & =  \sum_{\vec{k} \in \mathrm{RBZ} } \sum_{ \alpha =\pm}  \frac{1- 2 n_F(E_\alpha)}{2 E_\alpha},   \label{eq: coeffa}\\
b  & = -\sum_{\vec{k} \in \mathrm{ RBZ} } \sum_{ \{ \alpha ,\eta \} =\pm} \frac{\eta}{2 E_\alpha} \frac{n_F(\eta E_\alpha) - n_F(-\xi_\alpha)}{\eta E_\alpha + \xi_\alpha },  \label{eq: coeffb}\\
c  & = - \frac{1}{2}\frac{\partial^2}{\partial q_x^2} \chi(Q) \bigg|_{Q=0} \equiv  T_{\mathrm{conv}} + T_{\mathrm{geom}}.  \label{eq: coeffc}
\end{align}
\end{subequations}
Here to determine the coefficient $c$, we use only the $q_x^2$ component of the $\chi(Q)$ expansion, since the system possesses a $C_4$ rotational symmetry.

For our later discussion on quantum geometry we have broken up $c$ into two separate terms, 
\begin{widetext}
\begin{subequations} \label{eq: T1T2T3}
\begin{align}
 T_{\mathrm{conv}}      &  =  \sum_{\vec{k} \in \mathrm{RBZ}} \sum_{ \{ \alpha ,\eta \}=\pm} \frac{\eta}{4 E_\alpha} 
\bigg\{
  (\partial_{x} \xi_\alpha)^2  \; 2 \big[ \frac{n_F(\eta E_\alpha ) - n_F(-\xi_\alpha)}{(\eta E_\alpha + \xi_\alpha)^2} + \frac{\beta n_F(\xi_\alpha) n_F(-\xi_\alpha)}{\eta E_\alpha + \xi_\alpha} \big]  - \partial^2_{x} \xi_\alpha \frac{n_F(\eta E_\alpha ) - n_F(-\xi_\alpha)}{\eta E_\alpha + \xi_\alpha}
  \bigg\},   \label{eq: T1}\\
 T_{\mathrm{geom}}    &=    \sum_{\vec{k} \in \mathrm{RBZ}}  \sum_{ \{ \alpha, \alpha^\prime \eta \} =\pm} \frac{1}{4} \bigg[1+ \eta \frac{\xi_\alpha}{ E_\alpha}  \bigg]
\frac{n_F(\eta E_\alpha) - n_F(-\xi_{\alpha^\prime})}{\eta E_\alpha + \xi_{\alpha^\prime}} (- \alpha \alpha^\prime) \frac{1}{2} \partial_x \hat{h}\cdot \partial_x \hat{h},  \label{eq: T3}
\end{align}
\end{subequations}
\end{widetext}
where $\partial_x \equiv \partial_{k_x}$, and,
for brevity, we have suppressed the $\vec{k}$ dependence. 
The conventional term, $T_{\mathrm{conv}}$, is derived from the $q_x$ derivative
of the factors other than $\mathrm{Tr}[\cdots]$ in Eq~\eqref{eq: chispin};
while the geometric term, $T_{\mathrm{geom}}$, comes solely from that of the trace factor,
\begin{align}
& \qquad 
  \partial_{q_x}^2 \mathrm{Tr}\big[ \hat{P}_{\alpha,\sigma} (\vec{k})\hat{P}_{\alpha^\prime,\sigma} (\vec{k} - \vec{q})\big] |_{\vec{q}=0}  \nonumber \\
 &   =  (- \alpha \alpha^\prime)  \frac{1}{2} \partial_{k_x} \hat{h}(\vec{k},\phi_\sigma)\cdot \partial_{k_x} \hat{h}(\vec{k},\phi_\sigma). 
    \label{eq: gxxorigin}
\end{align}
$T_{\mathrm{geom}}$ depends on not only the normal state energy dispersion but also its wavefunctions,
through the projection operators in the trace factor.  This is
in sharp contrast to $T_{\mathrm{conv}}$. 
The scalar product, $\frac{1}{2} \partial_x \hat{h}\cdot \partial_x \hat{h}$, can be identified with the $xx-$component
of the quantum metric tensor which will be defined and discussed in detail in Sec.~\ref{app: geometry}.

We note that although $\hat{h}(\vec{k},\phi_\sigma)$ depends on spin due to $\phi_\sigma$, $\frac{1}{2} \partial_x \hat{h}\cdot \partial_x \hat{h}$
does not because it is even in the sign of $\phi_\sigma$. 
As a result, $\{\chi(0), b, c\}$ are all spin independent. 
So are the characteristic parameters for the non-condensed bosons such as $n_{\mathrm{B}}$ and $M_{\mathrm{B}}$. 

\subsection{$n_{\mathrm{B}}$ and $M_{\mathrm{B}}$}

Next we calculate $n_{\mathrm{B}}$ and $M_{\mathrm{B}}$ from $\{\chi(0), b,c\}$. 
Substituting Eq.~\eqref{eq: chiSmallQ} into 
Eq.~\eqref{eq: tpg} leads to
\begin{gather}
t_{ \mathrm{pg} }(Q)     \approx \frac{\mathcal{Z}}{i \Omega_m -\vec{q}^2/(2 M_{\text{B}}) + \mu_{\text{B}}},  \label{eq: tpgSmallQ} 
\end{gather}
with 
\begin{subequations} \label{eq: ZmuBMB}
\begin{align}
\mathcal{Z}  &=1/b,    \label{eq: Z}\\
\mu_{\text{B}}       & =\frac{-1/U + \chi(0)}{b},  \label{eq: muB} \\
M_{\text{B}}    & =b/(2c).  \label{eq: MB}
\end{align}
\end{subequations}
The quantity $t_{ \mathrm{pg} }(Q) $ in Eq.~\eqref{eq: tpgSmallQ} can be interpreted as the propagator for non-condensed pairs with an energy dispersion $E_{\text{B}}= \vec{q}^2/2M_{\text{B}}-\mu_{\mathrm{B}}$,
with $M_{\text{B}}$ the effective pair mass and $\mu_{\text{B}}$ the corresponding bosonic chemical potential. 
Then from Eqs.~\eqref{eq: DeltaPg} and \eqref{eq: tpgSmallQ} one can relate the areal density of non-condensed pairs, $n_{\text{B}}$, to $\Delta_{ \mathrm{pg} }^2$
by
\begin{align}
n_{\text{B}}  \equiv \sum_{\vec{q}} f_{\text{B}}(E_{\text{B}}) = \frac{\Delta_{ \mathrm{pg} }^2} {\mathcal{Z}}    
  & = \frac{M_{\text{B}}}{2\pi \beta} \big\{ - \ln[ 1- e^{\beta \mu_{\text{B}}}] \big\}, \label{eq: nBeqn}
\end{align}
where $f_{\text{B}}(x)=1/(e^{\beta x} -1)$ is the Bose-Einstein distribution. 
To obtain the r.h.s. of the last equality we have neglected the upper bound
in the $\vec{q}$ summation which is associated with a lattice.
This is consistent with the pg approximation which implies,
near the instability, a fast decrease of $t_{\mathrm{pg}}(Q)$
at large $Q$.

Eqs.~\eqref{eq: ZmuBMB} and \eqref{eq: nBeqn} combined together yield one independent nonlinear equation for two unknowns,
$\Delta_{\mathrm{pg}}$ and $\mu_{\mathrm{F}}$, in terms of $\{T,n,U\} $. 
The other independent equation comes from the electron density constraint~\cite{Chen2005, Wu2015, Wang2020}
\begin{align}
n    &   = \sum_{\vec{k} \in \mathrm{RBZ}} \sum_{\alpha =\pm}  \left[ 1- \frac{\xi_{\alpha}(\vec{k})}{E_\alpha (\vec{k})} \tanh  \frac{\beta E_\alpha(\vec{k})}{2} \right] . \label{eq: nk}
\end{align}
Solving the combined Eqs.~\eqref{eq: ZmuBMB} to \eqref{eq: nk} for given $\{ T, n, U\}$ numerically we are able to compute $\Delta_{\mathrm{pg}}$ and $\mu_{\mathrm{F}}$,
from which $n_{\text{B}}$ and $M_{\text{B}}$ can be determined.  
We then 
apply the BKT criterion, $n_\text{B}(T)/ M_\text{B}(T) =(\mathcal{D}_\text{B}^\text{crit} /2\pi)  T$, to determine 
$T_{\mathrm{BKT}}$. 

Using Eq.~\eqref{eq: nk} one can also rewrite the bosonic density as~\cite{He2013}
\begin{gather}
n_{\mathrm{B}} = \frac{n}{2} -
\sum_{\vec{k} \in \mathrm{RBZ} } \sum_{ \alpha =\pm} n_{\mathrm{F}}(\xi_{\alpha}(\vec{k})).
\end{gather}
This equation shows that $n_{\text{B}}$ increases with $U$ for given
temperature since $\mu_{\mathrm{F}}$ decreases with $U$.  As $U$
increases, the zero temperature $\mu_{\mathrm{F}}$ becomes negative,
i. e. lower than the conduction band bottom, at a certain value of
$U$.  Beyond this value $n_{\text{B}}$ saturates to $n/2$ at $T=0$
since $n_{\mathrm{F}}(\xi_{\alpha}(\vec{k})) =0$ for all $\vec{k}$.
The saturation defines the entrance to the BEC regime.  However, we
notice that at finite $T$, the point where $\mu_{\mathrm{F}}$ becomes
negative and the saturation onset of $n_{\mathrm{B}}$ do not occur
concomitantly at the same $U$ as
$n_{\mathrm{F}}(\xi_{\alpha}(\vec{k})) \ne 0$ even if
$\mu_{\mathrm{F}}$ is negative.

To see clearly where the quantum metric enters the bosonic parameters
we combine $n_{\text{B}}$ and $M_{\text{B}}$ and write the ratio as
\begin{equation}
n_\text{B} /M_{\text{B}} =2 \;  \Delta_{\mathrm{pg}}^2 \; c
  =2  \;  \Delta_{\mathrm{pg}}^2 \big( T_{\mathrm{conv}} + T_{\mathrm{geom}}  \big), 
\label{eq: DBdef}
\end{equation}
where Eqs.~\eqref{eq: nBeqn}, \eqref{eq: ZmuBMB}, and \eqref{eq:
  coeffc} have been used.  This equation shows explicitly that the
quantum metric affects $T_{\mathrm{BKT}}$ through the ratio
$n_\text{B} /M_{\text{B}} $, or $1/M_{\text{B}}$ in the BEC regime,
where $n_{\text{B}}=n/2=\text{const.}$

\subsection{Large $U$ limit of $n_{\text{B}}/M_{\text{B}}$ at
  $T=T_{\mathrm{BKT}}$} \label{app: largeU} In the $U \gg E_g$ limit,
$T_{\mathrm{conv}} \ll T_{\mathrm{geom}}$ so that one can neglect
$T_{\mathrm{conv}}$ in Eq.~\eqref{eq: DBdef}.  Also in this case,
$\mu_{\mathrm{F}}$ is large negative and $T_{\mathrm{BKT}}$ is much
smaller than $|\xi_{\alpha}|$ and $E_\alpha$ so that one can take
$T\approx 0$ in evaluating $T_{\mathrm{geom}}$ in Eq.~\eqref{eq:
  T3}. All Fermi functions $n_F(x)$ become either $0$ or $1$, so that one can
simplify $T_{\mathrm{geom}}$ and rewrite Eq.~\eqref{eq: DBdef} as
\begin{align}
\frac{ n_{\text{B}} } {M_{\text{B}} }
 & \approx \left(\sum_{\vec{k}}  \frac{1}{2} \partial_x \hat{h}\cdot \partial_x \hat{h}\right) 
  \frac{ 2 |\mu_{\mathrm{F}}|  + |\mu_{\mathrm{F}}|^2/E_0  -E_0}{2 E_0^2 (E_0+ |\mu_{\mathrm{F}}|)^2}  \Delta_{\text{pg}}^2
E_g^2,   
\end{align}
where $E_0 \equiv \sqrt{\mu_{\mathrm{F}}^2+\Delta_{\mathrm{pg}}^2}$.
Because  both $\Delta_{\mathrm{pg}}$ and $|\mu_{\mathrm{F}}|$ are proportional to  $  U$ at $U \gg E_g$,
we conclude that in the large $U$ limit $n_{\text{B}}/M_{\text{B}} \propto E_g^2/U$.

\section{Quantum geometry and the  pair mass}
\label{app: geometry}

Eq.~\eqref{eq: DBdef} suggests that the quantum metric can play an important role in
determining $T_{\mathrm{BKT}}$ through the $T_{\mathrm{geom}}$ term in $n_{\text{B}}/M_{\text{B}}$,
if the conventional contribution is small. 
In this section we introduce the definition for the quantum metric,
discuss the physical picture behind its interplay with delocalization of  
non-condensed pairs, and elucidate the role of the normal state band topology
in such an interplay. 
The latter becomes most clear in the isolated flat band limit, where we show
$n_{\text{B}}/M_{\text{B}}$ is lower bounded by the nontrivial band topology.

The quantum metric tensor, 
 $g_{\mu \nu}^{\alpha \sigma}(\vec{k})$ with $\{\mu,\nu\}=\{x,y\}$,
is defined for each $\alpha\sigma$ normal state band. 
It represents a distance in the projective Hilbert space between two states $\psi_{\alpha\sigma}(\vec{k})$ and $\psi_{\alpha\sigma}(\vec{k} +d \vec{k})$:
$d s^2 \equiv 1- | \langle \psi_{\alpha \sigma}(\vec{k}) | \psi_{\alpha \sigma}(\vec{k}+d\vec{k})\rangle |^2= \frac{1}{2} g_{\mu \nu}^{\alpha \sigma}(\vec{k}) d k_\mu d k_\nu +\mathcal{O}((dk)^3) $
~\cite{Provost1980, Anandan1990}.
Here $\psi_{\alpha\sigma}(\vec{k})$ is an eigenstate of $H_{\mathrm{K}}(\vec{k})$ in Eq.~\eqref{eq: HK}
with the quantum number $\alpha=\pm$ and $\sigma=\{ \uparrow, \downarrow\}$.  
Note that $g_{\mu \nu}^{\alpha\sigma}(\vec{k})$ is independent of the arbitrary $U(1)$ phase of $\psi_{\alpha \sigma}(\vec{k})$, and is therefore gauge invariant. 
By definition it is also positive definite.

The quantum metric tensor can be combined with the Berry curvature, $\mathcal{F}_{\mu\nu}^{\alpha \sigma}$, to define a quantum geometric tensor $\mathcal{R}_{\mu \nu}^{\alpha\sigma}$~\cite{Provost1980}:
\begin{subequations}
\begin{align}
\mathcal{R}_{\mu \nu}^{\alpha \sigma} &  \equiv  2 \mathrm{Tr}\big[  \hat{P}_{\alpha,\sigma} \partial_{k_\mu}\hat{P}_{\alpha,\sigma}  \partial_{k_\nu} \hat{P}_{\alpha,\sigma} \big]      
  = g_{\mu \nu}^{\alpha \sigma} + i \mathcal{F}_{\mu \nu}^{\alpha\sigma}/2.     \label{eq: Rmunu3}
\end{align}
\end{subequations}
Both $g_{\mu \nu}^{\alpha\sigma}$ and $\mathcal{F}_{\mu \nu}^{\alpha\sigma}$ are real. 
Using the definition of $\hat{P}_{\alpha,\sigma}$ in Eq.~\eqref{eq: projector} one obtains
\begin{subequations}
\begin{align}
g_{\mu\nu}^{\alpha\sigma} (\vec{k})    & = \frac{1}{2} \partial_\mu \hat{h}(\vec{k},\phi_\sigma) \cdot \partial_\nu \hat{h}(\vec{k},\phi_\sigma), 
\label{eq: gmunu}
\\
\mathcal{F}_{\mu \nu}^{\alpha\sigma}  (\vec{k})  & = \alpha \; \epsilon_{\mu \nu} \hat{h}(\vec{k},\phi_\sigma) \cdot [\partial_\mu \hat{h}(\vec{k},\phi_\sigma)
 \times \partial_\nu \hat{h}(\vec{k},\phi_\sigma)], \label{eq: Fmunu}
\end{align}
\end{subequations}
where $\epsilon_{\mu \nu}=-\epsilon_{\nu\mu}$ is the Levi-Civita symbol. 
$g_{\mu\nu}^{\alpha \sigma}$ is even under time reversal, and therefore independent of the spin $\sigma$. In contrast,
 $\mathcal{F}_{\mu \nu}^{\alpha\sigma}$ is odd  under time reversal, and therefore opposite for opposite spin. 
As a result, $g_{\mu\nu}^{\alpha \sigma}$ in Eq.~\eqref{eq: gmunu} is independent of $\{\alpha\sigma\}$ for our model. 

From its definition one can prove that $\mathcal{R}_{\mu\nu}^{\alpha\sigma}$ is positive definite~\cite{Peotta2015, Xie2020},
resulting an inequality between $g_{\mu\nu}^{\alpha\sigma}$ and $\mathcal{F}_{\mu\nu}^{\alpha\sigma}$:
$g_{xx}^{\alpha\sigma} g_{yy}^{\alpha\sigma} \ge ( \mathcal{F}_{xy}^{\alpha\sigma} )^2/4$. 
The inequality implies
\begin{gather}
\mathrm{Tr}[g_{\mu\nu}^{\alpha \sigma}] \ge 2 \sqrt{g_{xx}^{\alpha\sigma} g_{yy}^{\alpha\sigma}} \; \ge  | \mathcal{F}_{xy}^{\alpha\sigma}|.
\label{eq: ggF}
\end{gather}
Here $\mathrm{Tr}$ is with respect to $\{\mu \nu\}$.
Eq.~\eqref{eq: ggF} shows that in general a nonzero Chern number, which necessarily implies a nonzero $| \mathcal{F}_{xy}^{\alpha\sigma}|$,
enhances the magnitude of the quantum metric tensor. 
The physics behind this can be understood in terms of ``Wannier obstruction". 
Normal state Wannier functions $|\psi_{\alpha\sigma}(\vec{R}) \rangle$ can be 
constructed from the Bloch wavefunction $| \psi_{\alpha\sigma}(\vec{k}) \rangle$.
$|\psi_{\alpha\sigma}(\vec{R}) \rangle$ is in general not gauge invariant
because of the $U(1)$ phase ambiguity in defining $| \psi_{\alpha\sigma}(\vec{k}) \rangle$. 
Consequently, the spatial spread of $|\psi_{\alpha\sigma}(\vec{R}) \rangle$ contains both
a gauge invariant and non-invariant part~\cite{Marzari1997,Marzari2012}.
Interestingly, the former is equal to
$\sum_{\vec{k}} \mathrm{Tr} [ g_{\mu\nu}^{\alpha\sigma}]$.
If the $\alpha\sigma$ band is topologically trivial, then an exponentially localized $|\psi_{\alpha\sigma}(\vec{R}) \rangle$ can be 
constructed by choosing  a proper gauge.
On the other hand, if the $\alpha\sigma$ band is nontrivial, then this is impossible.
This is known as the ``Wannier obstruction"~\cite{Marzari1997, Marzari2012},
which implies a larger Wannier function spread, and therefore a larger $\sum_{\vec{k}} \mathrm{Tr} [ g_{\mu\nu}^{\alpha\sigma}]$.

The enhancement of $\mathrm{Tr} [ g_{\mu\nu}^{\alpha\sigma}]$ due to nontrivial band topology
also affects the pairing state through $n_{\text{B}}/M_{\text{B}}$.
The latter reflects the degree of delocalization of 
the non-condensed pairs. 
Both a larger $n_{\text{B}}$ and smaller $M_{\text{B}}$ imply a larger overlap between individual pair wavefunctions,
and therefore more delocalized pairs. 
How delocalized the pairs are must be connected to how delocalized the
normal state Wannier orbitals are.  
Therefore, it is not surprising that $g_{\mu\nu}^{\alpha \sigma}$,
which provides a measure of how delocalized the normal states are,
enters the expression of $n_{\text{B}}/M_{\text{B}}$ through $T_{\mathrm{geom}}$
in Eq.~\eqref{eq: T3}. 
However, $g_{\mu\nu}^{\alpha \sigma}$ appears in a complicated way because both the two normal bands
can contribute, and because both intra- and inter-band processes matter. 
Interestingly, the inter- and intra-band contributions in Eq.~\eqref{eq: T3} carry opposite
signs; the former partially cancels the latter which is positive. 

The above qualitative discussion suggests that in general, a nontrivial band topology enhances the quantum metric,
which in turn increases $n_{\text{B}}/M_{\text{B}}$. 
This emerges most clearly
in the isolated flat band limit, which was also heavily discussed in the 
literature addressing the superfluid
phase stiffness $D_s$~\cite{Peotta2015, Liang2017, Julku2020, Xie2020}, where a lower bound for
the mean field $D_s$ was found.
In the following we show that a similar bound exists for $n_{\text{B}}/M_{\text{B}}$ in this limit.

\subsection{Isolated flat band limit} \label{app: isolatedFlatLimit}

The isolated flat band limit for the Hamiltonian in Sec.~\ref{app: theory} is defined at $U$ such that
$W\ll U \ll E_g$. 
This regime
corresponds to a BEC superconductor. 
In this limit, superconductivity mainly occurs in the lower flat energy band while the upper one is inactive. 
As a consequence, all terms involving the upper energy band in the 
equations for $\{ T_{\mathrm{conv}}, T_{\mathrm{geom}} \}$
drop out. 
Also, the lower flat band term in  $T_{\mathrm{conv}}$ can be neglected because the band is flat.  
The only remaining term comes from $T_{\mathrm{geom}}$ which involves the lower flat band.
Then from Eq.~\eqref{eq: DBdef}, one finds
\begin{align}
\frac{n_{\mathrm{B}}}{M_{\mathrm{B}}}  
& = \Delta_{\mathrm{pg}}^2 \sum_{\vec{k}   \in \mathrm{RBZ} } \frac{ \tanh(\beta E_-(\vec{k}) /2 )  }{2 E_{-}(\vec{k})} g_{xx}(\vec{k}),
\label{eq: isoflatband}
\end{align}
where we have left the band dependence of $g_{\mu \nu}^{\alpha \sigma}$ unspecified since it is the same for different bands. 

Interestingly, this expression for $n_{\text{B}} / M_{\text{B}}$ is almost identical to that of the BCS mean field $D_s$ in the same limit (see 
Eq.~\eqref{eq: DsIsoFlatBand} of Sec.~\ref{app: meanfield} and also Ref.~\onlinecite{Liang2017}).
The only difference is that the gap parameter in $n_{\text{B}} / M_{\text{B}}$ is the pseudogap $\Delta_{\text{pg}}$ while that in $D_s$ is the BCS mean field superconducting order parameter.

Using Eq.~\eqref{eq: ggF} and $g_{xx}=g_{yy}$, one can derive the following lower bound for $n_{\text{B}}/M_{\text{B}} $
\begin{align}
\frac{n_{\mathrm{B}}}{M_{\mathrm{B}}}  
& \ge
  \Delta_{\mathrm{pg}}^2       \frac{\tanh(\beta E_- /2 )}{4 E_{-} }                                             \sum_{\vec{k} \in \mathrm{RBZ}}    |\mathcal{F}_{xy}(\vec{k})|   \nonumber  \\
& \ge   \Delta_{\mathrm{pg}}^2             \frac{\tanh(\beta E_- /2 )}{4 E_{-} }      					   |\sum_{\vec{k} \in \mathrm{RBZ}} \mathcal{F}_{xy}(\vec{k}) |  \nonumber  \\
& = \Delta_{\mathrm{pg}}^2                   \frac{\tanh(\beta E_- /2 )}{4 E_{-} }        				\frac{|C|}{ \pi}. 
\label{eq: bound}
\end{align}
$E_-$ is $\vec{k}$ independent since the band is flat.
We dropped the band dependence of the Berry curvature $\mathcal{F}_{xy}^{\alpha \sigma}(\vec{k})$ and also that of the Chern number $C_{\alpha \sigma}$,
since their absolute values are the same for all bands. 
To obtain the last line we have used Eq.~\eqref{eq: Fmunu}. 
This line clearly shows that $n_{\text{B}}/M_{\text{B}}$ is bounded below
when the flat band has a nonzero Chern number, i. e. it is topologically nontrivial. 

\section{Mean field calculation of $D_s(T)$ and $T_{\mathrm{BKT}}$} \label{app: meanfield}

In Figs.~\ref{fig: Fig1} and \ref{fig: Fig2} of the
main text we have included the mean field results of $D_s$ and $T_{\mathrm{BKT}}$
for comparison.
This section gives a summary of the main equations used. 

We start with the BCS mean field gap equation
\begin{align}
\frac{1}{U}  & = \sum_{\vec{k} \in \mathrm{RBZ}}       \sum_{\alpha=\pm}      \frac{1}{2 E_\alpha(\vec{k})} \tanh ( \frac{\beta E_\alpha(\vec{k})}{2} ),
\end{align}
where $E_\alpha \equiv \sqrt{\xi_\alpha^2 + \Delta_{ \mathrm{sc} }^2}$ with $\Delta_{ \mathrm{sc} }$ the BCS
mean field superconducting gap. 
This equation is derived from Eq.~\eqref{eq: BCSeq}. 
The electron density equation is the same as in Eq.~\eqref{eq: nk}. 
Solving the two equations for given $T$ and $U$ one obtains $\Delta_{\mathrm{sc}}$ and $\mu_{\mathrm{F}}$. 

From the mean field $\Delta_{\mathrm{sc}}$ and $\mu_{\mathrm{F}}$ we calculate the mean field $D_s$ by (for derivations see Refs.~\onlinecite{Liang2017,Julku2020})
\begin{widetext}
\begin{gather} 
D_s =\frac{1}{4} \sum_{\vec{k} \in \mathrm{RBZ}} \sum_{\{i,j\} =\{1,2,3,4\}} \frac{n_F(\mathcal{E}_j) - n_F(\mathcal{E}_i)}{\mathcal{E}_i - \mathcal{E}_j } 
\bigg\{
\langle \Psi_i| \; \partial_x H_{\mathrm{BdG}}[\Delta_{ \mathrm{sc} }=0]   |\Psi_j \rangle 
\langle \Psi_j | \partial_x H_{\mathrm{BdG}}  | \Psi_i \rangle 
- 
\langle \Psi_i| \; j_x^\dagger \; |\Psi_j \rangle
 \langle \Psi_j |  \; j_x \;  | \Psi_i \rangle
\bigg\},
\label{eq: MFDs}
\end{gather} 
\end{widetext}
where $\mathcal{E}_i =\pm E_\pm$  and $|\Psi_i\rangle$ are eigen-energies and eigenvectors of the following $4\times 4$ mean field BdG Hamiltonian matrix
\begin{gather}
H_{ \mathrm{BdG} }(\vec{k}) = 
\begin{pmatrix}
H_{\uparrow}(\vec{k})   & \Delta_{\mathrm{sc}} s_0  \\
 - \Delta_{\mathrm{sc}} s_0 & - H_{\downarrow}^T(-\vec{k})
\end{pmatrix}. 
\end{gather}
In the curly brace in Eq.~\eqref{eq: MFDs}, the first term is diamagnetic, while the second term is paramagnetic.
$j_x(\vec{k}) = (\partial_x H_{\mathrm{BdG}}(\vec{k})) \tau_z$ is the electric current operator, where $\tau_z$ is the $z$-component Pauli matrix
defined for the Nambu space. Following Ref.~\onlinecite{Liang2017} one can separate $D_s$ into the
conventional and geometric contributions, 
$ D_s= D_s^{\mathrm{conv}} + D_s^{\mathrm{geom}}$. 
Their expressions are~\cite{Liang2017}
\begin{widetext}
\begin{align}
D_s^{\mathrm{conv}}  
& = 
 \frac{1}{4}   \sum_{\vec{k} \in \mathrm{RBZ} } \sum_{\alpha=\pm}
\bigg[ - \frac{\beta}{2 \cosh^2(\beta E_\alpha (\vec{k})/2 )}  + \frac{\tanh(\beta E_\alpha(\vec{k})/2)}{ E_\alpha(\vec{k})}\bigg] 
\frac{|\Delta_{\mathrm{sc}}|^2}{E_\alpha(\vec{k})^2}  (\frac{\partial \xi_\alpha(\vec{k})}{\partial k_x})^2, \\
D_s^{\mathrm{geom}} 
&=
\frac{1}{4} \sum_{\vec{k} \in \mathrm{RBZ} } \sum_{\alpha=\pm}
\bigg[  \frac{\tanh(\beta E_{\alpha}(\vec{k})/2)}{E_{\alpha}(\vec{k})}  - \frac{\tanh(\beta E_{ - \alpha}(\vec{k})/2)}{E_{ - \alpha}(\vec{k})}  \bigg]  
\frac{\xi_{-\alpha} (\vec{k})- \xi_\alpha(\vec{k})} {\xi_{-\alpha} (\vec{k})+ \xi_{\alpha}(\vec{k})}     \;     |\Delta^2_{\mathrm{sc}}|   \;    g_{xx} (\vec{k}).
\end{align}
\end{widetext}
The prefactor $1/4$ comes from our different 
definition of $D_s$ from the one used in Ref.~\onlinecite{Liang2017}:
for the London equation under the Coulomb gauge we use $\vec{J} = - 4 D_s \vec{A}$, instead of $\vec{J} = - D_s \vec{A}$. 

In the isolated flat band limit, $D_s^{\mathrm{conv}} \approx 0$.
Also, because $\xi_{+} \gg \xi_-$ and $E_+ \gg E_-$,
the geometric term becomes~\cite{Liang2017}
\begin{gather}
D_s^{\mathrm{geom}} 
\approx
|\Delta_{\mathrm{sc}}|^2
\sum_{\vec{k} \in \mathrm{RBZ} }
\frac{\tanh(\beta E_{-}(\vec{k})/2)}{2 \, E_{-}(\vec{k})}  g_{xx}(\vec{k}).
\label{eq: DsIsoFlatBand}
\end{gather}

From $D_s(T)$ we determine the mean field $T_{\mathrm{BKT}}$ using the universal relation
\begin{gather}
T_{\mathrm{BKT}}= \frac{\pi}{2} D_s(T_{\mathrm{BKT}}). 
\label{eq: MFBKT}
\end{gather}

\section{Additional Numerical Results} 
\label{app: numerical}

\subsection{$\mathcal{F}=0.2$}
\label{sec: numerical-dispersive}

%%%%%%%%%%%   Insert Fig.2  %%%%%%%%%%%%%%%%%%
\begin{figure}[ht]
\centering
\includegraphics[width=0.75\linewidth]{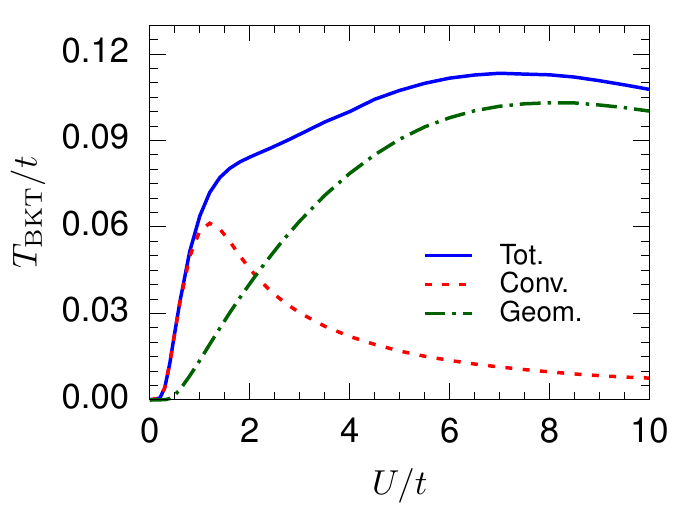}
\caption{
Decomposition of the BCS MF $T_{\mathrm{BKT}}$ into the conventional (``Conv" )
and geometric (``Geom") contributions for the topological $\mathcal{F}=0.2$ band. 
$n=0.3$. 
}
\label{fig: FigS1}
\end{figure}
%%%%%%%%%%%%%%%%%%%%%%%%%%%

In Fig.~\ref{fig: Fig1} of the main text we have decomposed our pairing fluctuation theory $T_{\mathrm{BKT}}$ 
into the conventional and geometric contributions. 
Here we make the same decomposition for the corresponding BCS MF theory $T_{\mathrm{BKT}}$
in Fig.~\ref{fig: FigS1}. 
Comparing the pairing fluctuation theory and MF results we see that the conventional term in both theories
has a dome shape dependence on $U$ with its maximum at $U \sim W$. 
However, the decrease of the mean field conventional $T_{\mathrm{BKT}}$ at large $U$
is much slower and follows a $t^2/U$ asymptote.
In contrast, the corresponding pairing fluctuation result falls precipitously to almost zero at $U/t \approx 3 $ and remains extremely small at larger $U$. 

The plummet of the pairing fluctuation theory $T_{\mathrm{BKT}}$ occurs near the point where $\mu_{\mathrm{F}}$ becomes negative.
It is associated with a rapid decrease of a term in $T_{\mathrm{conv}}$ in Eq.~\eqref{eq: T1},
(the second one in the square bracket),  
which is $\propto [ \partial_{\xi_{\alpha}} n_{\mathrm{F}}(\xi_{\alpha}) ]  (\partial_x \xi_{\alpha})^2$. 
This term vanishes at $T=0$ when $\mu_{\mathrm{F}}$ drops below the band bottom since
$[ \partial_{\xi_{\alpha}} n_{\mathrm{F}}(\xi_{\alpha}) ]  (\partial_x \xi_{\alpha})^2 =\delta(\xi_{\alpha}) ( \partial_x \xi_{\alpha})^2 \equiv 0$ for any $\vec{k}$.
The remaining two terms in Eq.~\eqref{eq: T1} cancel each other almost completely at $T=0$ 
when $\mu_{\mathrm{F}}$ is negative,
leading to the extremely small $T_{\mathrm{BKT}}$ at $U/t \gtrsim 3$. 
The near-complete cancellation does not occur when the electron density $n$ is small so that the conduction
band is much less than half-filled~\cite{Wang2020}, i. e. when the
preformed pairs in the BEC regime are dilute.
It suggests that the cancellation is a consequence of  a competition between pair hopping
and inter-site pair repulsion~\cite{Micnas1990}, the latter of which originates from Pauli exclusion that prevents two pairs from
occupying the same site.
The repulsion becomes more important as the density of the pairs, 
which is equal to $n/2$ in the BEC regime, increases, 
and it can severely restrict the motion of the pairs at high density~\cite{Micnas1990}, 
leading to almost zero $T_{\mathrm{BKT}}$.   
This effect of the repulsion is naturally not included in the calculated mean field $D_s$,
even when the pair density is high and when $U$ is very large~\cite{Denteneer1993}.  
To incorporate the inter-site pair repulsion effect into $D_s$ one needs to include
beyond mean field corrections~\cite{Dupuis2004, Benfatto2004}, in particular quantum fluctuation effects. 
On the other hand, numerical studies~\cite{Keller2001, Toschi2005, Paiva2010} on a simple 2D attractive (single-orbital) Hubbard model on a square lattice
do not seem to indicate a dramatic effect of the repulsion on $T_{\mathrm{BKT}}$. 
Of course, the numerical studies can be subject to finite size effects. 
At present, it is unclear if our calculated conventional $n_{\mathrm{B}}/M_{\mathrm{B}}$ has 
overestimated the pair repulsion effect or not.  
Further studies are needed to resolve this issue.

The geometric contribution behaves similarly in the two theories. At small $U$
it increases roughly linearly with $U$ except where
$U$ is very small. 
At $U/t \gtrsim 7$, it
begins to decrease, which comes from a cancellation
between the inter- and intra-band contributions to $T_{\mathrm{geom}}$ in Eq.~\eqref{eq: T3}.
The net result at large enough $U$  is $T_{\mathrm{BKT}}^{\mathrm{geom}} \propto (n_B/M_B)^{\text{geom}} \propto E_g^2/U$,
 as discussed in Sec.~\ref{app: largeU}.

%%%%%%%%%%%%%%%%%%%%%%%%%%%%%%%%%%%%%%%%%%%%%
\subsection{$\mathcal{F}=0.01$}

\begin{figure}[htp]
\centering
\includegraphics[width=1.\linewidth]{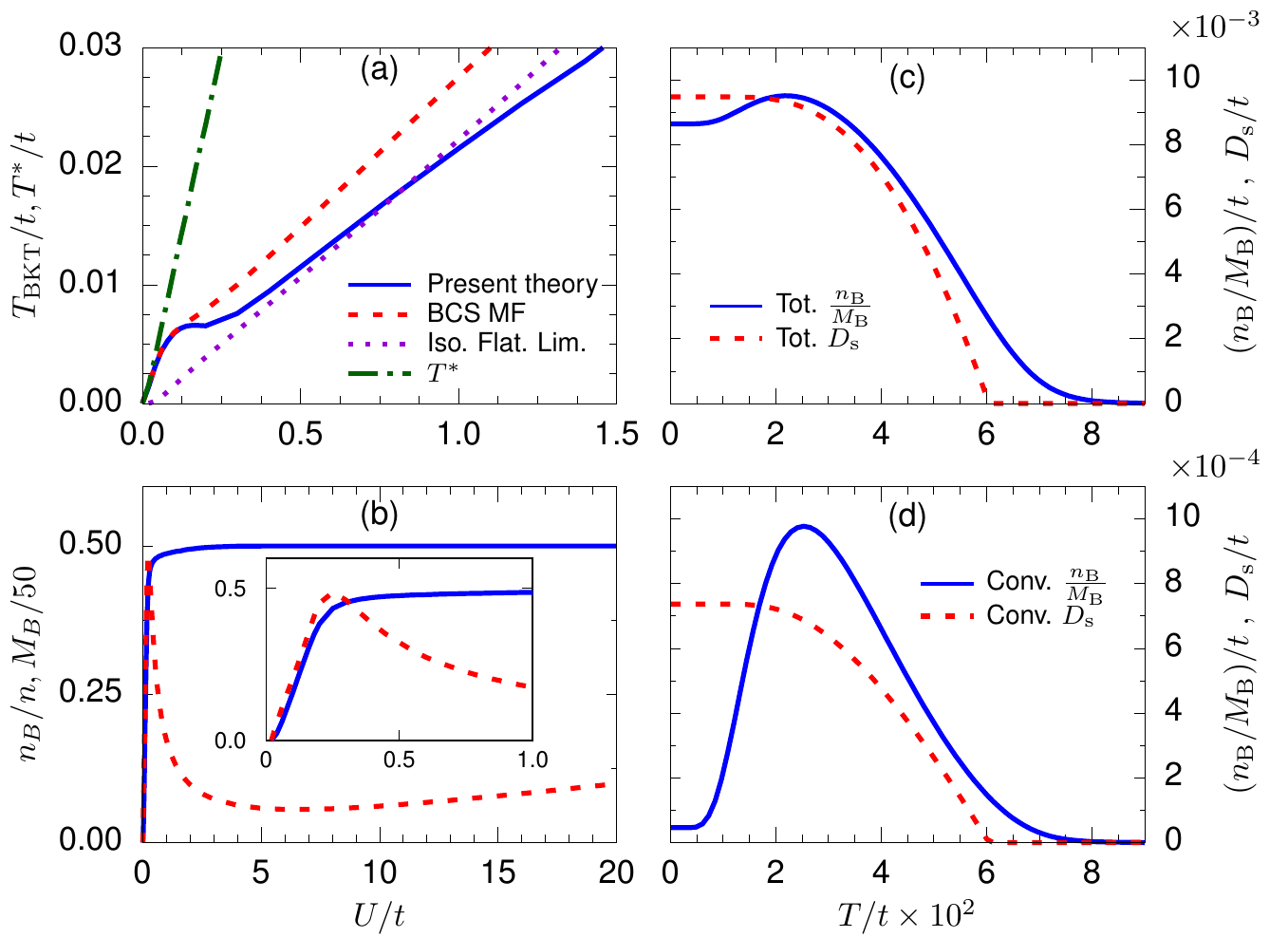} 
\caption{Results for the topological $\mathcal{F}=0.01$ band. 
(a) $\{ T_{\mathrm{BKT}},T^*\}$ and (b) $ \{n_{\mathrm{B}}/n, M_{\mathrm{B}}\}$
plotted as a function of $U/t$. 
``Iso. Flat. Lim." stands for the lower bound on $T_{\mathrm{BKT}}$ calculated
from the lower bound of $n_{\mathrm{B}}/M_{\mathrm{B}}$ in the isolated flat band limit,
given by the last line of Eq.~\eqref{eq: bound}. 
Inset in (b): zoomed view of  $ \{n_{\mathrm{B}}/n, M_{\mathrm{B}}\}$ at small $U/t$. 
$M_{\text{B}}$ is plotted in units of $t \, a_{\text{L}}^2$, where $a_{\text{L}}$ is the square lattice spacing. 
(c, d) $n_{\mathrm{B}}/M_{\mathrm{B}}$ and BCS MF $D_s$ plotted as a function of $T/t$ for $U/t=0.5$.  (c)  and (d) show the total  and conventional contributions,
respectively. 
 }
\label{fig: FigS3}
\end{figure}

Fig.~\ref{fig: FigS3} illustrates some additional numerical results for the $\mathcal{F}=0.01$ flat band. 
In Fig.~\ref{fig: FigS3}(a) we present a zoomed view of the $T_{\mathrm{BKT}}$ results
at small $U$. 
One sees that there is a remnant $T_{\mathrm{BKT}}$ peak
at $U/t \sim 0.1$, due to the small but still finite conventional contribution to $n_{\text{B}}/M_{\text{B}}$.
The latter comes from the fact that the conduction band is not completely flat. 
From Fig.~\ref{fig: FigS3}(a) one also sees that
the pairing fluctuation theory $T_{\mathrm{BKT}}$ almost saturates its lower bound
in the isolated flat band limit regime, i. e. at $W \ll U \ll E_g$
where $W\approx 0.035 t$ is the lower conduction band width. 
The near saturation comes from the fact that in the summation of the Berry curvature $\mathcal{F}_{xy}(\vec{k})$
in Eq.~\eqref{eq: bound},  $\mathcal{F}_{xy}(\vec{k})$ is dominated by one sign with large weight at most $\vec{k}$
so that the difference between $\sum_{\vec{k} } |\mathcal{F}_{xy}(\vec{k})|$ and $|\sum_{\vec{k}} \mathcal{F}_{xy}(\vec{k})|$
is roughly $10\%$. 
This suggests that, if the conduction band were trivial with zero Chern number,
 the resulting $n_{\text{B}}/M_{\text{B}}$ and $T_{\mathrm{BKT}}$ would be reduced by about $90\%$. 
Stated alternatively, for a flat band system, nontrivial band topology can significantly boost the two dimensional superconductivity
via the quantum metric effect. 

Fig.~\ref{fig: FigS3}(b) illustrates that, because of the extremely flat conduction band,
even a small attractive interaction $U/t \approx 0.3 $ already puts the system in the BEC regime where $n_{\mathrm{B}}/n$ saturates to $1/2$. 
Upon entering into the BEC regime,  $M_{\mathrm{B}}$ exhibits a sharp peak,
reflecting the strong localization tendency of the Cooper pairs due to the extremely small conventional contribution to $1/M_{\text{B}}$ in Eq.~\eqref{eq: T1}. 
Note that $M_{\text{B}}$ plotted in Fig.~\ref{fig: FigS3}(b) has been rescaled by a factor of $1/50$. 

In Fig.~\ref{fig: FigS3}(c) (\ref{fig: FigS3}(d)) we give a comparison between the total (conventional) $n_{\mathrm{B}}/M_{\mathrm{B}}$ and that of the BCS MF $D_s$
for $U/t=0.5$. 
 As shown in Fig.~\ref{fig: Fig2}(b) of the main text, the corresponding geometric 
contributions to $n_{\mathrm{B}}/M_{\mathrm{B}}$ and $D_s$ are almost identical at low temperatures,
even though the corresponding two $T_{\mathrm{BKT}}$ are different, as seen from Fig.~\ref{fig: FigS3}(a). 
Mathematically, the near coincidence derives from the fact that the expression for $n_{\mathrm{B}}/M_{\mathrm{B}}$ in this limit, given in Eq.~\eqref{eq: isoflatband},
is identical to that of $D_s$, given in Eq.~\eqref{eq: DsIsoFlatBand}, except that the gap parameters $\Delta$ in the two are different: $\Delta=\Delta_{\mathrm{pg}}$ in the former while $\Delta=\Delta_{\mathrm{sc}}$ in the latter case. 
However, at low temperatures,  $\Delta_{\mathrm{pg}}$ and $\Delta_{\mathrm{sc}}$ in the two approaches (which are based on the same
mean field equations) are essentially equal,
explaining why $n_{\mathrm{B}}/M_{\mathrm{B}}$ and $D_s$ are nearly the same.

On the other hand, the  conventional, as well as the total, contributions to $n_{\mathrm{B}}/M_{\mathrm{B}}$ and $D_s$ behave
quite differently. 
From Fig.~\ref{fig: FigS3}(c) we see that,
in contrast to the monotonic $D_s(T)$, the total $n_{\mathrm{B}}/M_{\mathrm{B}}$
has a small bump at $T \sim T^*/2$, which comes from the small
conventional $n_{\mathrm{B}}/M_{\mathrm{B}}$.
The latter depends on $T$ non-monotonically, as shown in Fig.~\ref{fig: FigS3}(d).
Interestingly, similar non-monotonic behavior has been observed in the phase stiffness of some 2D Josephson-junction
arrays where quantum fluctuations play an important role~\cite{Rojas1996, Capriotti2003}. 
The non-monotonicity comes from a competition between two physical processes. 
Near $T=0$, $(n_{\mathrm{B}}/M_{\mathrm{B}})^{\mathrm{conv}}$ is almost zero,
a consequence of
the competition between pair hopping and inter-pair repulsion,
as explained in Sec.~\ref{sec: numerical-dispersive}.
Increasing $T$ tends to enhance the pair hopping via an ionization process~\cite{Nozieres1985}, which becomes thermally more accessible. 
On the other hand, a large temperature also tends to dissociate
the Cooper pairs, leading to a decrease of $(n_{\mathrm{B}}/M_{\mathrm{B}})^{\mathrm{conv}}$ as $T$ increases towards $T^*$. 
Mathematically, the competition is between different temperature dependences of $n_{\mathrm{B}}$ and $(1/M_{\mathrm{B}})^{\mathrm{conv}}$. 
The net result is a peak of $(n_{\mathrm{B}}/M_{\mathrm{B}})^{\mathrm{conv}}$ near $T\sim T^*/2$. 
However, we should note that our results of $n_{\mathrm{B}}/M_{\mathrm{B}}$ become
unreliable at $T \sim T^*/2 \gg T_{\mathrm{BKT}}$ where the ``pg approximation" breaks down. 

\bibliography{Geometry,Review2}

%merlin.mbs apsrev4-1.bst 2010-07-25 4.21a (PWD, AO, DPC) hacked
%Control: key (0)
%Control: author (8) initials jnrlst
%Control: editor formatted (1) identically to author
%Control: production of article title (-1) disabled
%Control: page (0) single
%Control: year (1) truncated
%Control: production of eprint (0) enabled
\begin{thebibliography}{102}%
\makeatletter
\providecommand \@ifxundefined [1]{%
 \@ifx{#1\undefined}
}%
\providecommand \@ifnum [1]{%
 \ifnum #1\expandafter \@firstoftwo
 \else \expandafter \@secondoftwo
 \fi
}%
\providecommand \@ifx [1]{%
 \ifx #1\expandafter \@firstoftwo
 \else \expandafter \@secondoftwo
 \fi
}%
\providecommand \natexlab [1]{#1}%
\providecommand \enquote  [1]{``#1''}%
\providecommand \bibnamefont  [1]{#1}%
\providecommand \bibfnamefont [1]{#1}%
\providecommand \citenamefont [1]{#1}%
\providecommand \href@noop [0]{\@secondoftwo}%
\providecommand \href [0]{\begingroup \@sanitize@url \@href}%
\providecommand \@href[1]{\@@startlink{#1}\@@href}%
\providecommand \@@href[1]{\endgroup#1\@@endlink}%
\providecommand \@sanitize@url [0]{\catcode `\\12\catcode `\$12\catcode
  `\&12\catcode `\#12\catcode `\^12\catcode `\_12\catcode `\%12\relax}%
\providecommand \@@startlink[1]{}%
\providecommand \@@endlink[0]{}%
\providecommand \url  [0]{\begingroup\@sanitize@url \@url }%
\providecommand \@url [1]{\endgroup\@href {#1}{\urlprefix }}%
\providecommand \urlprefix  [0]{URL }%
\providecommand \Eprint [0]{\href }%
\providecommand \doibase [0]{http://dx.doi.org/}%
\providecommand \selectlanguage [0]{\@gobble}%
\providecommand \bibinfo  [0]{\@secondoftwo}%
\providecommand \bibfield  [0]{\@secondoftwo}%
\providecommand \translation [1]{[#1]}%
\providecommand \BibitemOpen [0]{}%
\providecommand \bibitemStop [0]{}%
\providecommand \bibitemNoStop [0]{.\EOS\space}%
\providecommand \EOS [0]{\spacefactor3000\relax}%
\providecommand \BibitemShut  [1]{\csname bibitem#1\endcsname}%
\let\auto@bib@innerbib\@empty
%</preamble>
\bibitem [{\citenamefont {Cao}\ \emph {et~al.}(2018{\natexlab{a}})\citenamefont
  {Cao}, \citenamefont {Fatemi}, \citenamefont {Demir}, \citenamefont {Fang},
  \citenamefont {Tomarken}, \citenamefont {Luo}, \citenamefont
  {Sanchez-Yamagishi}, \citenamefont {Watanabe}, \citenamefont {Taniguchi},
  \citenamefont {Kaxiras} \emph {et~al.}}]{Cao2018}%
  \BibitemOpen
  \bibfield  {author} {\bibinfo {author} {\bibfnamefont {Y.}~\bibnamefont
  {Cao}}, \bibinfo {author} {\bibfnamefont {V.}~\bibnamefont {Fatemi}},
  \bibinfo {author} {\bibfnamefont {A.}~\bibnamefont {Demir}}, \bibinfo
  {author} {\bibfnamefont {S.}~\bibnamefont {Fang}}, \bibinfo {author}
  {\bibfnamefont {S.~L.}\ \bibnamefont {Tomarken}}, \bibinfo {author}
  {\bibfnamefont {J.~Y.}\ \bibnamefont {Luo}}, \bibinfo {author} {\bibfnamefont
  {J.~D.}\ \bibnamefont {Sanchez-Yamagishi}}, \bibinfo {author} {\bibfnamefont
  {K.}~\bibnamefont {Watanabe}}, \bibinfo {author} {\bibfnamefont
  {T.}~\bibnamefont {Taniguchi}}, \bibinfo {author} {\bibfnamefont
  {E.}~\bibnamefont {Kaxiras}},  \emph {et~al.},\ }\href
  {https://www.nature.com/articles/nature26154} {\bibfield  {journal} {\bibinfo
   {journal} {Nature}\ }\textbf {\bibinfo {volume} {556}},\ \bibinfo {pages}
  {80} (\bibinfo {year} {2018}{\natexlab{a}})}\BibitemShut {NoStop}%
\bibitem [{\citenamefont {Cao}\ \emph {et~al.}(2018{\natexlab{b}})\citenamefont
  {Cao}, \citenamefont {Fatemi}, \citenamefont {Fang}, \citenamefont
  {Watanabe}, \citenamefont {Taniguchi}, \citenamefont {Kaxiras},\ and\
  \citenamefont {Jarillo-Herrero}}]{Cao2018a}%
  \BibitemOpen
  \bibfield  {author} {\bibinfo {author} {\bibfnamefont {Y.}~\bibnamefont
  {Cao}}, \bibinfo {author} {\bibfnamefont {V.}~\bibnamefont {Fatemi}},
  \bibinfo {author} {\bibfnamefont {S.}~\bibnamefont {Fang}}, \bibinfo {author}
  {\bibfnamefont {K.}~\bibnamefont {Watanabe}}, \bibinfo {author}
  {\bibfnamefont {T.}~\bibnamefont {Taniguchi}}, \bibinfo {author}
  {\bibfnamefont {E.}~\bibnamefont {Kaxiras}}, \ and\ \bibinfo {author}
  {\bibfnamefont {P.}~\bibnamefont {Jarillo-Herrero}},\ }\href
  {https://www.nature.com/articles/nature26160} {\bibfield  {journal} {\bibinfo
   {journal} {Nature}\ }\textbf {\bibinfo {volume} {556}},\ \bibinfo {pages}
  {43} (\bibinfo {year} {2018}{\natexlab{b}})}\BibitemShut {NoStop}%
\bibitem [{\citenamefont {Bistritzer}\ and\ \citenamefont
  {MacDonald}(2011)}]{Bistritzer2011}%
  \BibitemOpen
  \bibfield  {author} {\bibinfo {author} {\bibfnamefont {R.}~\bibnamefont
  {Bistritzer}}\ and\ \bibinfo {author} {\bibfnamefont {A.~H.}\ \bibnamefont
  {MacDonald}},\ }\href {https://www.pnas.org/content/108/30/12233} {\bibfield
  {journal} {\bibinfo  {journal} {Proceedings of the National Academy of
  Sciences}\ }\textbf {\bibinfo {volume} {108}},\ \bibinfo {pages} {12233}
  (\bibinfo {year} {2011})}\BibitemShut {NoStop}%
\bibitem [{\citenamefont {Wu}\ \emph {et~al.}(2018)\citenamefont {Wu},
  \citenamefont {MacDonald},\ and\ \citenamefont {Martin}}]{Wu2018}%
  \BibitemOpen
  \bibfield  {author} {\bibinfo {author} {\bibfnamefont {F.}~\bibnamefont
  {Wu}}, \bibinfo {author} {\bibfnamefont {A.~H.}\ \bibnamefont {MacDonald}}, \
  and\ \bibinfo {author} {\bibfnamefont {I.}~\bibnamefont {Martin}},\ }\href
  {\doibase 10.1103/PhysRevLett.121.257001} {\bibfield  {journal} {\bibinfo
  {journal} {Phys. Rev. Lett.}\ }\textbf {\bibinfo {volume} {121}},\ \bibinfo
  {pages} {257001} (\bibinfo {year} {2018})}\BibitemShut {NoStop}%
\bibitem [{\citenamefont {Dodaro}\ \emph {et~al.}(2018)\citenamefont {Dodaro},
  \citenamefont {Kivelson}, \citenamefont {Schattner}, \citenamefont {Sun},\
  and\ \citenamefont {Wang}}]{Dodaro2018}%
  \BibitemOpen
  \bibfield  {author} {\bibinfo {author} {\bibfnamefont {J.~F.}\ \bibnamefont
  {Dodaro}}, \bibinfo {author} {\bibfnamefont {S.~A.}\ \bibnamefont
  {Kivelson}}, \bibinfo {author} {\bibfnamefont {Y.}~\bibnamefont {Schattner}},
  \bibinfo {author} {\bibfnamefont {X.~Q.}\ \bibnamefont {Sun}}, \ and\
  \bibinfo {author} {\bibfnamefont {C.}~\bibnamefont {Wang}},\ }\href {\doibase
  10.1103/PhysRevB.98.075154} {\bibfield  {journal} {\bibinfo  {journal} {Phys.
  Rev. B}\ }\textbf {\bibinfo {volume} {98}},\ \bibinfo {pages} {075154}
  (\bibinfo {year} {2018})}\BibitemShut {NoStop}%
\bibitem [{\citenamefont {Kang}\ and\ \citenamefont {Vafek}(2018)}]{Kang2018}%
  \BibitemOpen
  \bibfield  {author} {\bibinfo {author} {\bibfnamefont {J.}~\bibnamefont
  {Kang}}\ and\ \bibinfo {author} {\bibfnamefont {O.}~\bibnamefont {Vafek}},\
  }\href {\doibase 10.1103/PhysRevX.8.031088} {\bibfield  {journal} {\bibinfo
  {journal} {Phys. Rev. X}\ }\textbf {\bibinfo {volume} {8}},\ \bibinfo {pages}
  {031088} (\bibinfo {year} {2018})}\BibitemShut {NoStop}%
\bibitem [{\citenamefont {Yankowitz}\ \emph {et~al.}(2019)\citenamefont
  {Yankowitz}, \citenamefont {Chen}, \citenamefont {Polshyn}, \citenamefont
  {Zhang}, \citenamefont {Watanabe}, \citenamefont {Taniguchi}, \citenamefont
  {Graf}, \citenamefont {Young},\ and\ \citenamefont {Dean}}]{Yankowitz2019}%
  \BibitemOpen
  \bibfield  {author} {\bibinfo {author} {\bibfnamefont {M.}~\bibnamefont
  {Yankowitz}}, \bibinfo {author} {\bibfnamefont {S.}~\bibnamefont {Chen}},
  \bibinfo {author} {\bibfnamefont {H.}~\bibnamefont {Polshyn}}, \bibinfo
  {author} {\bibfnamefont {Y.}~\bibnamefont {Zhang}}, \bibinfo {author}
  {\bibfnamefont {K.}~\bibnamefont {Watanabe}}, \bibinfo {author}
  {\bibfnamefont {T.}~\bibnamefont {Taniguchi}}, \bibinfo {author}
  {\bibfnamefont {D.}~\bibnamefont {Graf}}, \bibinfo {author} {\bibfnamefont
  {A.~F.}\ \bibnamefont {Young}}, \ and\ \bibinfo {author} {\bibfnamefont
  {C.~R.}\ \bibnamefont {Dean}},\ }\href
  {https://science.sciencemag.org/content/363/6431/1059} {\bibfield  {journal}
  {\bibinfo  {journal} {Science}\ }\textbf {\bibinfo {volume} {363}},\ \bibinfo
  {pages} {1059} (\bibinfo {year} {2019})}\BibitemShut {NoStop}%
\bibitem [{\citenamefont {Yuan}\ and\ \citenamefont {Fu}(2018)}]{Yuan2018}%
  \BibitemOpen
  \bibfield  {author} {\bibinfo {author} {\bibfnamefont {N.~F.~Q.}\
  \bibnamefont {Yuan}}\ and\ \bibinfo {author} {\bibfnamefont {L.}~\bibnamefont
  {Fu}},\ }\href {\doibase 10.1103/PhysRevB.98.045103} {\bibfield  {journal}
  {\bibinfo  {journal} {Phys. Rev. B}\ }\textbf {\bibinfo {volume} {98}},\
  \bibinfo {pages} {045103} (\bibinfo {year} {2018})}\BibitemShut {NoStop}%
\bibitem [{\citenamefont {Po}\ \emph {et~al.}(2018{\natexlab{a}})\citenamefont
  {Po}, \citenamefont {Zou}, \citenamefont {Vishwanath},\ and\ \citenamefont
  {Senthil}}]{Po2018}%
  \BibitemOpen
  \bibfield  {author} {\bibinfo {author} {\bibfnamefont {H.~C.}\ \bibnamefont
  {Po}}, \bibinfo {author} {\bibfnamefont {L.}~\bibnamefont {Zou}}, \bibinfo
  {author} {\bibfnamefont {A.}~\bibnamefont {Vishwanath}}, \ and\ \bibinfo
  {author} {\bibfnamefont {T.}~\bibnamefont {Senthil}},\ }\href {\doibase
  10.1103/PhysRevX.8.031089} {\bibfield  {journal} {\bibinfo  {journal} {Phys.
  Rev. X}\ }\textbf {\bibinfo {volume} {8}},\ \bibinfo {pages} {031089}
  (\bibinfo {year} {2018}{\natexlab{a}})}\BibitemShut {NoStop}%
\bibitem [{\citenamefont {Xu}\ and\ \citenamefont {Balents}(2018)}]{Xu2018}%
  \BibitemOpen
  \bibfield  {author} {\bibinfo {author} {\bibfnamefont {C.}~\bibnamefont
  {Xu}}\ and\ \bibinfo {author} {\bibfnamefont {L.}~\bibnamefont {Balents}},\
  }\href {\doibase 10.1103/PhysRevLett.121.087001} {\bibfield  {journal}
  {\bibinfo  {journal} {Phys. Rev. Lett.}\ }\textbf {\bibinfo {volume} {121}},\
  \bibinfo {pages} {087001} (\bibinfo {year} {2018})}\BibitemShut {NoStop}%
\bibitem [{\citenamefont {Roy}\ and\ \citenamefont {Juri\ifmmode \check{c}\else
  \v{c}\fi{}i\ifmmode~\acute{c}\else \'{c}\fi{}}(2019)}]{Roy2019}%
  \BibitemOpen
  \bibfield  {author} {\bibinfo {author} {\bibfnamefont {B.}~\bibnamefont
  {Roy}}\ and\ \bibinfo {author} {\bibfnamefont {V.}~\bibnamefont {Juri\ifmmode
  \check{c}\else \v{c}\fi{}i\ifmmode~\acute{c}\else \'{c}\fi{}}},\ }\href
  {\doibase 10.1103/PhysRevB.99.121407} {\bibfield  {journal} {\bibinfo
  {journal} {Phys. Rev. B}\ }\textbf {\bibinfo {volume} {99}},\ \bibinfo
  {pages} {121407(R)} (\bibinfo {year} {2019})}\BibitemShut {NoStop}%
\bibitem [{\citenamefont {Isobe}\ \emph {et~al.}(2018)\citenamefont {Isobe},
  \citenamefont {Yuan},\ and\ \citenamefont {Fu}}]{Isobe2018}%
  \BibitemOpen
  \bibfield  {author} {\bibinfo {author} {\bibfnamefont {H.}~\bibnamefont
  {Isobe}}, \bibinfo {author} {\bibfnamefont {N.~F.~Q.}\ \bibnamefont {Yuan}},
  \ and\ \bibinfo {author} {\bibfnamefont {L.}~\bibnamefont {Fu}},\ }\href
  {\doibase 10.1103/PhysRevX.8.041041} {\bibfield  {journal} {\bibinfo
  {journal} {Phys. Rev. X}\ }\textbf {\bibinfo {volume} {8}},\ \bibinfo {pages}
  {041041} (\bibinfo {year} {2018})}\BibitemShut {NoStop}%
\bibitem [{\citenamefont {Po}\ \emph {et~al.}(2019)\citenamefont {Po},
  \citenamefont {Zou}, \citenamefont {Senthil},\ and\ \citenamefont
  {Vishwanath}}]{Po2019}%
  \BibitemOpen
  \bibfield  {author} {\bibinfo {author} {\bibfnamefont {H.~C.}\ \bibnamefont
  {Po}}, \bibinfo {author} {\bibfnamefont {L.}~\bibnamefont {Zou}}, \bibinfo
  {author} {\bibfnamefont {T.}~\bibnamefont {Senthil}}, \ and\ \bibinfo
  {author} {\bibfnamefont {A.}~\bibnamefont {Vishwanath}},\ }\href {\doibase
  10.1103/PhysRevB.99.195455} {\bibfield  {journal} {\bibinfo  {journal} {Phys.
  Rev. B}\ }\textbf {\bibinfo {volume} {99}},\ \bibinfo {pages} {195455}
  (\bibinfo {year} {2019})}\BibitemShut {NoStop}%
\bibitem [{\citenamefont {Tarnopolsky}\ \emph {et~al.}(2019)\citenamefont
  {Tarnopolsky}, \citenamefont {Kruchkov},\ and\ \citenamefont
  {Vishwanath}}]{Tarnopolsky2019}%
  \BibitemOpen
  \bibfield  {author} {\bibinfo {author} {\bibfnamefont {G.}~\bibnamefont
  {Tarnopolsky}}, \bibinfo {author} {\bibfnamefont {A.~J.}\ \bibnamefont
  {Kruchkov}}, \ and\ \bibinfo {author} {\bibfnamefont {A.}~\bibnamefont
  {Vishwanath}},\ }\href {\doibase 10.1103/PhysRevLett.122.106405} {\bibfield
  {journal} {\bibinfo  {journal} {Phys. Rev. Lett.}\ }\textbf {\bibinfo
  {volume} {122}},\ \bibinfo {pages} {106405} (\bibinfo {year}
  {2019})}\BibitemShut {NoStop}%
\bibitem [{\citenamefont {Arora}\ \emph {et~al.}(2020)\citenamefont {Arora},
  \citenamefont {Polski}, \citenamefont {Zhang}, \citenamefont {Thomson},
  \citenamefont {Choi}, \citenamefont {Kim}, \citenamefont {Lin}, \citenamefont
  {Wilson}, \citenamefont {Xu}, \citenamefont {Chu} \emph
  {et~al.}}]{Arora2020}%
  \BibitemOpen
  \bibfield  {author} {\bibinfo {author} {\bibfnamefont {H.~S.}\ \bibnamefont
  {Arora}}, \bibinfo {author} {\bibfnamefont {R.}~\bibnamefont {Polski}},
  \bibinfo {author} {\bibfnamefont {Y.}~\bibnamefont {Zhang}}, \bibinfo
  {author} {\bibfnamefont {A.}~\bibnamefont {Thomson}}, \bibinfo {author}
  {\bibfnamefont {Y.}~\bibnamefont {Choi}}, \bibinfo {author} {\bibfnamefont
  {H.}~\bibnamefont {Kim}}, \bibinfo {author} {\bibfnamefont {Z.}~\bibnamefont
  {Lin}}, \bibinfo {author} {\bibfnamefont {I.~Z.}\ \bibnamefont {Wilson}},
  \bibinfo {author} {\bibfnamefont {X.}~\bibnamefont {Xu}}, \bibinfo {author}
  {\bibfnamefont {J.-H.}\ \bibnamefont {Chu}},  \emph {et~al.},\ }\href
  {https://www.nature.com/articles/s41586-020-2473-8} {\bibfield  {journal}
  {\bibinfo  {journal} {Nature}\ }\textbf {\bibinfo {volume} {583}},\ \bibinfo
  {pages} {379} (\bibinfo {year} {2020})}\BibitemShut {NoStop}%
\bibitem [{\citenamefont {Uemura}(2004)}]{Uemura2004}%
  \BibitemOpen
  \bibfield  {author} {\bibinfo {author} {\bibfnamefont {Y.~J.}\ \bibnamefont
  {Uemura}},\ }\href {\doibase 10.1088/0953-8984/16/40/007} {\bibfield
  {journal} {\bibinfo  {journal} {Journal of Physics: Condensed Matter}\
  }\textbf {\bibinfo {volume} {16}},\ \bibinfo {pages} {S4515} (\bibinfo {year}
  {2004})}\BibitemShut {NoStop}%
\bibitem [{\citenamefont {Wang}\ \emph {et~al.}(2020)\citenamefont {Wang},
  \citenamefont {Chen},\ and\ \citenamefont {Levin}}]{Wang2020}%
  \BibitemOpen
  \bibfield  {author} {\bibinfo {author} {\bibfnamefont {X.}~\bibnamefont
  {Wang}}, \bibinfo {author} {\bibfnamefont {Q.}~\bibnamefont {Chen}}, \ and\
  \bibinfo {author} {\bibfnamefont {K.}~\bibnamefont {Levin}},\ }\href
  {https://iopscience.iop.org/article/10.1088/1367-2630/ab890b} {\bibfield
  {journal} {\bibinfo  {journal} {New Journal of Physics}\ }\textbf {\bibinfo
  {volume} {22}},\ \bibinfo {pages} {063050} (\bibinfo {year}
  {2020})}\BibitemShut {NoStop}%
\bibitem [{\citenamefont {Berezinskii}(1971)}]{Berezinskii1971}%
  \BibitemOpen
  \bibfield  {author} {\bibinfo {author} {\bibfnamefont {V.}~\bibnamefont
  {Berezinskii}},\ }\href@noop {} {\bibfield  {journal} {\bibinfo  {journal}
  {Sov. Phys. JETP}\ }\textbf {\bibinfo {volume} {34}},\ \bibinfo {pages} {610}
  (\bibinfo {year} {1971})}\BibitemShut {NoStop}%
\bibitem [{\citenamefont {Kosterlitz}\ and\ \citenamefont
  {Thouless}(1973)}]{Kosterlitz1973}%
  \BibitemOpen
  \bibfield  {author} {\bibinfo {author} {\bibfnamefont {J.~M.}\ \bibnamefont
  {Kosterlitz}}\ and\ \bibinfo {author} {\bibfnamefont {D.~J.}\ \bibnamefont
  {Thouless}},\ }\href {\doibase 10.1088/0022-3719/6/7/010} {\bibfield
  {journal} {\bibinfo  {journal} {Journal of Physics C: Solid State Physics}\
  }\textbf {\bibinfo {volume} {6}},\ \bibinfo {pages} {1181} (\bibinfo {year}
  {1973})}\BibitemShut {NoStop}%
\bibitem [{\citenamefont {Benfatto}\ \emph {et~al.}(2007)\citenamefont
  {Benfatto}, \citenamefont {Castellani},\ and\ \citenamefont
  {Giamarchi}}]{Benfatto2007}%
  \BibitemOpen
  \bibfield  {author} {\bibinfo {author} {\bibfnamefont {L.}~\bibnamefont
  {Benfatto}}, \bibinfo {author} {\bibfnamefont {C.}~\bibnamefont
  {Castellani}}, \ and\ \bibinfo {author} {\bibfnamefont {T.}~\bibnamefont
  {Giamarchi}},\ }\href {\doibase 10.1103/PhysRevLett.98.117008} {\bibfield
  {journal} {\bibinfo  {journal} {Phys. Rev. Lett.}\ }\textbf {\bibinfo
  {volume} {98}},\ \bibinfo {pages} {117008} (\bibinfo {year}
  {2007})}\BibitemShut {NoStop}%
\bibitem [{\citenamefont {Peotta}\ and\ \citenamefont
  {T{\"o}rm{\"a}}(2015)}]{Peotta2015}%
  \BibitemOpen
  \bibfield  {author} {\bibinfo {author} {\bibfnamefont {S.}~\bibnamefont
  {Peotta}}\ and\ \bibinfo {author} {\bibfnamefont {P.}~\bibnamefont
  {T{\"o}rm{\"a}}},\ }\href {https://www.nature.com/articles/ncomms9944}
  {\bibfield  {journal} {\bibinfo  {journal} {Nature communications}\ }\textbf
  {\bibinfo {volume} {6}},\ \bibinfo {pages} {8944} (\bibinfo {year}
  {2015})}\BibitemShut {NoStop}%
\bibitem [{\citenamefont {Liang}\ \emph {et~al.}(2017)\citenamefont {Liang},
  \citenamefont {Vanhala}, \citenamefont {Peotta}, \citenamefont {Siro},
  \citenamefont {Harju},\ and\ \citenamefont {T\"orm\"a}}]{Liang2017}%
  \BibitemOpen
  \bibfield  {author} {\bibinfo {author} {\bibfnamefont {L.}~\bibnamefont
  {Liang}}, \bibinfo {author} {\bibfnamefont {T.~I.}\ \bibnamefont {Vanhala}},
  \bibinfo {author} {\bibfnamefont {S.}~\bibnamefont {Peotta}}, \bibinfo
  {author} {\bibfnamefont {T.}~\bibnamefont {Siro}}, \bibinfo {author}
  {\bibfnamefont {A.}~\bibnamefont {Harju}}, \ and\ \bibinfo {author}
  {\bibfnamefont {P.}~\bibnamefont {T\"orm\"a}},\ }\href {\doibase
  10.1103/PhysRevB.95.024515} {\bibfield  {journal} {\bibinfo  {journal} {Phys.
  Rev. B}\ }\textbf {\bibinfo {volume} {95}},\ \bibinfo {pages} {024515}
  (\bibinfo {year} {2017})}\BibitemShut {NoStop}%
\bibitem [{\citenamefont {Hu}\ \emph {et~al.}(2019)\citenamefont {Hu},
  \citenamefont {Hyart}, \citenamefont {Pikulin},\ and\ \citenamefont
  {Rossi}}]{Hu2019}%
  \BibitemOpen
  \bibfield  {author} {\bibinfo {author} {\bibfnamefont {X.}~\bibnamefont
  {Hu}}, \bibinfo {author} {\bibfnamefont {T.}~\bibnamefont {Hyart}}, \bibinfo
  {author} {\bibfnamefont {D.~I.}\ \bibnamefont {Pikulin}}, \ and\ \bibinfo
  {author} {\bibfnamefont {E.}~\bibnamefont {Rossi}},\ }\href {\doibase
  10.1103/PhysRevLett.123.237002} {\bibfield  {journal} {\bibinfo  {journal}
  {Phys. Rev. Lett.}\ }\textbf {\bibinfo {volume} {123}},\ \bibinfo {pages}
  {237002} (\bibinfo {year} {2019})}\BibitemShut {NoStop}%
\bibitem [{\citenamefont {Julku}\ \emph {et~al.}(2020)\citenamefont {Julku},
  \citenamefont {Peltonen}, \citenamefont {Liang}, \citenamefont {Heikkil\"a},\
  and\ \citenamefont {T\"orm\"a}}]{Julku2020}%
  \BibitemOpen
  \bibfield  {author} {\bibinfo {author} {\bibfnamefont {A.}~\bibnamefont
  {Julku}}, \bibinfo {author} {\bibfnamefont {T.~J.}\ \bibnamefont {Peltonen}},
  \bibinfo {author} {\bibfnamefont {L.}~\bibnamefont {Liang}}, \bibinfo
  {author} {\bibfnamefont {T.~T.}\ \bibnamefont {Heikkil\"a}}, \ and\ \bibinfo
  {author} {\bibfnamefont {P.}~\bibnamefont {T\"orm\"a}},\ }\href {\doibase
  10.1103/PhysRevB.101.060505} {\bibfield  {journal} {\bibinfo  {journal}
  {Phys. Rev. B}\ }\textbf {\bibinfo {volume} {101}},\ \bibinfo {pages}
  {060505(R)} (\bibinfo {year} {2020})}\BibitemShut {NoStop}%
\bibitem [{\citenamefont {Xie}\ \emph {et~al.}(2020)\citenamefont {Xie},
  \citenamefont {Song}, \citenamefont {Lian},\ and\ \citenamefont
  {Bernevig}}]{Xie2020}%
  \BibitemOpen
  \bibfield  {author} {\bibinfo {author} {\bibfnamefont {F.}~\bibnamefont
  {Xie}}, \bibinfo {author} {\bibfnamefont {Z.}~\bibnamefont {Song}}, \bibinfo
  {author} {\bibfnamefont {B.}~\bibnamefont {Lian}}, \ and\ \bibinfo {author}
  {\bibfnamefont {B.~A.}\ \bibnamefont {Bernevig}},\ }\href {\doibase
  10.1103/PhysRevLett.124.167002} {\bibfield  {journal} {\bibinfo  {journal}
  {Phys. Rev. Lett.}\ }\textbf {\bibinfo {volume} {124}},\ \bibinfo {pages}
  {167002} (\bibinfo {year} {2020})}\BibitemShut {NoStop}%
\bibitem [{\citenamefont {Hofmann}\ \emph {et~al.}(2019)\citenamefont
  {Hofmann}, \citenamefont {Berg},\ and\ \citenamefont
  {Chowdhury}}]{Hofmann2019}%
  \BibitemOpen
  \bibfield  {author} {\bibinfo {author} {\bibfnamefont {J.~S.}\ \bibnamefont
  {Hofmann}}, \bibinfo {author} {\bibfnamefont {E.}~\bibnamefont {Berg}}, \
  and\ \bibinfo {author} {\bibfnamefont {D.}~\bibnamefont {Chowdhury}},\ }\href
  {https://arxiv.org/abs/1912.08848} {\bibfield  {journal} {\bibinfo  {journal}
  {arXiv preprint arXiv:1912.08848}\ } (\bibinfo {year} {2019})}\BibitemShut
  {NoStop}%
\bibitem [{\citenamefont {Neupert}\ \emph {et~al.}(2011)\citenamefont
  {Neupert}, \citenamefont {Santos}, \citenamefont {Chamon},\ and\
  \citenamefont {Mudry}}]{Neupert2011}%
  \BibitemOpen
  \bibfield  {author} {\bibinfo {author} {\bibfnamefont {T.}~\bibnamefont
  {Neupert}}, \bibinfo {author} {\bibfnamefont {L.}~\bibnamefont {Santos}},
  \bibinfo {author} {\bibfnamefont {C.}~\bibnamefont {Chamon}}, \ and\ \bibinfo
  {author} {\bibfnamefont {C.}~\bibnamefont {Mudry}},\ }\href {\doibase
  10.1103/PhysRevLett.106.236804} {\bibfield  {journal} {\bibinfo  {journal}
  {Phys. Rev. Lett.}\ }\textbf {\bibinfo {volume} {106}},\ \bibinfo {pages}
  {236804} (\bibinfo {year} {2011})}\BibitemShut {NoStop}%
\bibitem [{\citenamefont {He}\ \emph {et~al.}(2013)\citenamefont {He},
  \citenamefont {Huang}, \citenamefont {Hu},\ and\ \citenamefont
  {Liu}}]{He2013}%
  \BibitemOpen
  \bibfield  {author} {\bibinfo {author} {\bibfnamefont {L.}~\bibnamefont
  {He}}, \bibinfo {author} {\bibfnamefont {X.-G.}\ \bibnamefont {Huang}},
  \bibinfo {author} {\bibfnamefont {H.}~\bibnamefont {Hu}}, \ and\ \bibinfo
  {author} {\bibfnamefont {X.-J.}\ \bibnamefont {Liu}},\ }\href {\doibase
  10.1103/PhysRevA.87.053616} {\bibfield  {journal} {\bibinfo  {journal} {Phys.
  Rev. A}\ }\textbf {\bibinfo {volume} {87}},\ \bibinfo {pages} {053616}
  (\bibinfo {year} {2013})}\BibitemShut {NoStop}%
\bibitem [{\citenamefont {Zhang}\ \emph {et~al.}(2014)\citenamefont {Zhang},
  \citenamefont {Hu}, \citenamefont {Liu},\ and\ \citenamefont
  {Pu}}]{Zhang2014}%
  \BibitemOpen
  \bibfield  {author} {\bibinfo {author} {\bibfnamefont {J.}~\bibnamefont
  {Zhang}}, \bibinfo {author} {\bibfnamefont {H.}~\bibnamefont {Hu}}, \bibinfo
  {author} {\bibfnamefont {X.-J.}\ \bibnamefont {Liu}}, \ and\ \bibinfo
  {author} {\bibfnamefont {H.}~\bibnamefont {Pu}},\ }\enquote {\bibinfo {title}
  {Annual review of cold atoms and molecules},}\ \ (\bibinfo  {publisher}
  {World Scientific, Singapore},\ \bibinfo {year} {2014})\ Chap.~\bibinfo
  {chapter} {2}, pp.\ \bibinfo {pages} {81--143}\BibitemShut {NoStop}%
\bibitem [{\citenamefont {Fu}\ \emph {et~al.}(2013)\citenamefont {Fu},
  \citenamefont {Huang}, \citenamefont {Meng}, \citenamefont {Wang},
  \citenamefont {Liu}, \citenamefont {Pu}, \citenamefont {Hu},\ and\
  \citenamefont {Zhang}}]{Fu2013}%
  \BibitemOpen
  \bibfield  {author} {\bibinfo {author} {\bibfnamefont {Z.}~\bibnamefont
  {Fu}}, \bibinfo {author} {\bibfnamefont {L.}~\bibnamefont {Huang}}, \bibinfo
  {author} {\bibfnamefont {Z.}~\bibnamefont {Meng}}, \bibinfo {author}
  {\bibfnamefont {P.}~\bibnamefont {Wang}}, \bibinfo {author} {\bibfnamefont
  {X.-J.}\ \bibnamefont {Liu}}, \bibinfo {author} {\bibfnamefont
  {H.}~\bibnamefont {Pu}}, \bibinfo {author} {\bibfnamefont {H.}~\bibnamefont
  {Hu}}, \ and\ \bibinfo {author} {\bibfnamefont {J.}~\bibnamefont {Zhang}},\
  }\href {\doibase 10.1103/PhysRevA.87.053619} {\bibfield  {journal} {\bibinfo
  {journal} {Phys. Rev. A}\ }\textbf {\bibinfo {volume} {87}},\ \bibinfo
  {pages} {053619} (\bibinfo {year} {2013})}\BibitemShut {NoStop}%
\bibitem [{\citenamefont {Zheng}\ \emph {et~al.}(2014)\citenamefont {Zheng},
  \citenamefont {Pu}, \citenamefont {Zou},\ and\ \citenamefont
  {Guo}}]{Zheng2014}%
  \BibitemOpen
  \bibfield  {author} {\bibinfo {author} {\bibfnamefont {Z.}~\bibnamefont
  {Zheng}}, \bibinfo {author} {\bibfnamefont {H.}~\bibnamefont {Pu}}, \bibinfo
  {author} {\bibfnamefont {X.}~\bibnamefont {Zou}}, \ and\ \bibinfo {author}
  {\bibfnamefont {G.}~\bibnamefont {Guo}},\ }\href {\doibase
  10.1103/PhysRevA.90.063623} {\bibfield  {journal} {\bibinfo  {journal} {Phys.
  Rev. A}\ }\textbf {\bibinfo {volume} {90}},\ \bibinfo {pages} {063623}
  (\bibinfo {year} {2014})}\BibitemShut {NoStop}%
\bibitem [{\citenamefont {Wu}\ \emph {et~al.}(2015{\natexlab{a}})\citenamefont
  {Wu}, \citenamefont {Anderson}, \citenamefont {Boyack},\ and\ \citenamefont
  {Levin}}]{Wu2015a}%
  \BibitemOpen
  \bibfield  {author} {\bibinfo {author} {\bibfnamefont {C.-T.}\ \bibnamefont
  {Wu}}, \bibinfo {author} {\bibfnamefont {B.~M.}\ \bibnamefont {Anderson}},
  \bibinfo {author} {\bibfnamefont {R.}~\bibnamefont {Boyack}}, \ and\ \bibinfo
  {author} {\bibfnamefont {K.}~\bibnamefont {Levin}},\ }\href {\doibase
  10.1103/PhysRevB.91.220504} {\bibfield  {journal} {\bibinfo  {journal} {Phys.
  Rev. B}\ }\textbf {\bibinfo {volume} {91}},\ \bibinfo {pages} {220504(R)}
  (\bibinfo {year} {2015}{\natexlab{a}})}\BibitemShut {NoStop}%
\bibitem [{\citenamefont {Leggett}(1980)}]{Leggett1980}%
  \BibitemOpen
  \bibfield  {author} {\bibinfo {author} {\bibfnamefont {A.~J.}\ \bibnamefont
  {Leggett}},\ }in\ \href@noop {} {\emph {\bibinfo {booktitle} {Modern trends
  in the theory of condensed matter}}}\ (\bibinfo  {publisher} {Springer,
  Berlin, Heidelberg},\ \bibinfo {year} {1980})\ pp.\ \bibinfo {pages}
  {13--27}\BibitemShut {NoStop}%
\bibitem [{\citenamefont {Iskin}(2018{\natexlab{a}})}]{Iskin2018}%
  \BibitemOpen
  \bibfield  {author} {\bibinfo {author} {\bibfnamefont {M.}~\bibnamefont
  {Iskin}},\ }\href {\doibase 10.1103/PhysRevA.97.033625} {\bibfield  {journal}
  {\bibinfo  {journal} {Phys. Rev. A}\ }\textbf {\bibinfo {volume} {97}},\
  \bibinfo {pages} {033625} (\bibinfo {year} {2018}{\natexlab{a}})}\BibitemShut
  {NoStop}%
\bibitem [{\citenamefont {T\"orm\"a}\ \emph {et~al.}(2018)\citenamefont
  {T\"orm\"a}, \citenamefont {Liang},\ and\ \citenamefont
  {Peotta}}]{Torma2018}%
  \BibitemOpen
  \bibfield  {author} {\bibinfo {author} {\bibfnamefont {P.}~\bibnamefont
  {T\"orm\"a}}, \bibinfo {author} {\bibfnamefont {L.}~\bibnamefont {Liang}}, \
  and\ \bibinfo {author} {\bibfnamefont {S.}~\bibnamefont {Peotta}},\ }\href
  {\doibase 10.1103/PhysRevB.98.220511} {\bibfield  {journal} {\bibinfo
  {journal} {Phys. Rev. B}\ }\textbf {\bibinfo {volume} {98}},\ \bibinfo
  {pages} {220511(R)} (\bibinfo {year} {2018})}\BibitemShut {NoStop}%
\bibitem [{\citenamefont {Chen}\ \emph {et~al.}(2000)\citenamefont {Chen},
  \citenamefont {Kosztin},\ and\ \citenamefont {Levin}}]{Chen2000}%
  \BibitemOpen
  \bibfield  {author} {\bibinfo {author} {\bibfnamefont {Q.}~\bibnamefont
  {Chen}}, \bibinfo {author} {\bibfnamefont {I.}~\bibnamefont {Kosztin}}, \
  and\ \bibinfo {author} {\bibfnamefont {K.}~\bibnamefont {Levin}},\ }\href
  {\doibase 10.1103/PhysRevLett.85.2801} {\bibfield  {journal} {\bibinfo
  {journal} {Phys. Rev. Lett.}\ }\textbf {\bibinfo {volume} {85}},\ \bibinfo
  {pages} {2801} (\bibinfo {year} {2000})}\BibitemShut {NoStop}%
\bibitem [{\citenamefont {Marzari}\ and\ \citenamefont
  {Vanderbilt}(1997)}]{Marzari1997}%
  \BibitemOpen
  \bibfield  {author} {\bibinfo {author} {\bibfnamefont {N.}~\bibnamefont
  {Marzari}}\ and\ \bibinfo {author} {\bibfnamefont {D.}~\bibnamefont
  {Vanderbilt}},\ }\href {\doibase 10.1103/PhysRevB.56.12847} {\bibfield
  {journal} {\bibinfo  {journal} {Phys. Rev. B}\ }\textbf {\bibinfo {volume}
  {56}},\ \bibinfo {pages} {12847} (\bibinfo {year} {1997})}\BibitemShut
  {NoStop}%
\bibitem [{\citenamefont {Marzari}\ \emph {et~al.}(2012)\citenamefont
  {Marzari}, \citenamefont {Mostofi}, \citenamefont {Yates}, \citenamefont
  {Souza},\ and\ \citenamefont {Vanderbilt}}]{Marzari2012}%
  \BibitemOpen
  \bibfield  {author} {\bibinfo {author} {\bibfnamefont {N.}~\bibnamefont
  {Marzari}}, \bibinfo {author} {\bibfnamefont {A.~A.}\ \bibnamefont
  {Mostofi}}, \bibinfo {author} {\bibfnamefont {J.~R.}\ \bibnamefont {Yates}},
  \bibinfo {author} {\bibfnamefont {I.}~\bibnamefont {Souza}}, \ and\ \bibinfo
  {author} {\bibfnamefont {D.}~\bibnamefont {Vanderbilt}},\ }\href {\doibase
  10.1103/RevModPhys.84.1419} {\bibfield  {journal} {\bibinfo  {journal} {Rev.
  Mod. Phys.}\ }\textbf {\bibinfo {volume} {84}},\ \bibinfo {pages} {1419}
  (\bibinfo {year} {2012})}\BibitemShut {NoStop}%
\bibitem [{\citenamefont {Zhao}\ \emph {et~al.}(2013)\citenamefont {Zhao},
  \citenamefont {Wang}, \citenamefont {Liu}, \citenamefont {Zhang},
  \citenamefont {Wang}, \citenamefont {Chen}, \citenamefont {Guo},
  \citenamefont {He}, \citenamefont {Chen}, \citenamefont {Wang}, \citenamefont
  {Wang}, \citenamefont {Xie}, \citenamefont {Niu}, \citenamefont {Wang},
  \citenamefont {Ma}, \citenamefont {Jain}, \citenamefont {Chan},\ and\
  \citenamefont {Xue}}]{Zhao2013}%
  \BibitemOpen
  \bibfield  {author} {\bibinfo {author} {\bibfnamefont {W.}~\bibnamefont
  {Zhao}}, \bibinfo {author} {\bibfnamefont {Q.}~\bibnamefont {Wang}}, \bibinfo
  {author} {\bibfnamefont {M.}~\bibnamefont {Liu}}, \bibinfo {author}
  {\bibfnamefont {W.}~\bibnamefont {Zhang}}, \bibinfo {author} {\bibfnamefont
  {Y.}~\bibnamefont {Wang}}, \bibinfo {author} {\bibfnamefont {M.}~\bibnamefont
  {Chen}}, \bibinfo {author} {\bibfnamefont {Y.}~\bibnamefont {Guo}}, \bibinfo
  {author} {\bibfnamefont {K.}~\bibnamefont {He}}, \bibinfo {author}
  {\bibfnamefont {X.}~\bibnamefont {Chen}}, \bibinfo {author} {\bibfnamefont
  {Y.}~\bibnamefont {Wang}}, \bibinfo {author} {\bibfnamefont {J.}~\bibnamefont
  {Wang}}, \bibinfo {author} {\bibfnamefont {X.}~\bibnamefont {Xie}}, \bibinfo
  {author} {\bibfnamefont {Q.}~\bibnamefont {Niu}}, \bibinfo {author}
  {\bibfnamefont {L.}~\bibnamefont {Wang}}, \bibinfo {author} {\bibfnamefont
  {X.}~\bibnamefont {Ma}}, \bibinfo {author} {\bibfnamefont {J.~K.}\
  \bibnamefont {Jain}}, \bibinfo {author} {\bibfnamefont {M.}~\bibnamefont
  {Chan}}, \ and\ \bibinfo {author} {\bibfnamefont {Q.-K.}\ \bibnamefont
  {Xue}},\ }\href {\doibase https://doi.org/10.1016/j.ssc.2013.04.025}
  {\bibfield  {journal} {\bibinfo  {journal} {Solid State Communications}\
  }\textbf {\bibinfo {volume} {165}},\ \bibinfo {pages} {59 } (\bibinfo {year}
  {2013})}\BibitemShut {NoStop}%
\bibitem [{App()}]{Appendix}%
  \BibitemOpen
  \href@noop {} {\emph {\bibinfo {title} {\textup{See the appendices for
  details }}}}\BibitemShut {NoStop}%
\bibitem [{\citenamefont {Chen}\ \emph {et~al.}(1998)\citenamefont {Chen},
  \citenamefont {Kosztin}, \citenamefont {Jank\'o},\ and\ \citenamefont
  {Levin}}]{Chen2}%
  \BibitemOpen
  \bibfield  {author} {\bibinfo {author} {\bibfnamefont {Q.}~\bibnamefont
  {Chen}}, \bibinfo {author} {\bibfnamefont {I.}~\bibnamefont {Kosztin}},
  \bibinfo {author} {\bibfnamefont {B.}~\bibnamefont {Jank\'o}}, \ and\
  \bibinfo {author} {\bibfnamefont {K.}~\bibnamefont {Levin}},\ }\href
  {\doibase 10.1103/PhysRevLett.81.4708} {\bibfield  {journal} {\bibinfo
  {journal} {Phys. Rev. Lett.}\ }\textbf {\bibinfo {volume} {81}},\ \bibinfo
  {pages} {4708} (\bibinfo {year} {1998})}\BibitemShut {NoStop}%
\bibitem [{\citenamefont {Maly}\ \emph {et~al.}(1999)\citenamefont {Maly},
  \citenamefont {Jank\'o},\ and\ \citenamefont {Levin}}]{Maly1999}%
  \BibitemOpen
  \bibfield  {author} {\bibinfo {author} {\bibfnamefont {J.}~\bibnamefont
  {Maly}}, \bibinfo {author} {\bibfnamefont {B.}~\bibnamefont {Jank\'o}}, \
  and\ \bibinfo {author} {\bibfnamefont {K.}~\bibnamefont {Levin}},\ }\href
  {\doibase https://doi.org/10.1016/S0921-4534(99)00326-3} {\bibfield
  {journal} {\bibinfo  {journal} {Physica C: Superconductivity}\ }\textbf
  {\bibinfo {volume} {321}},\ \bibinfo {pages} {113 } (\bibinfo {year}
  {1999})}\BibitemShut {NoStop}%
\bibitem [{\citenamefont {Kadanoff}\ and\ \citenamefont
  {Martin}(1961)}]{KadanoffMartin}%
  \BibitemOpen
  \bibfield  {author} {\bibinfo {author} {\bibfnamefont {L.~P.}\ \bibnamefont
  {Kadanoff}}\ and\ \bibinfo {author} {\bibfnamefont {P.~C.}\ \bibnamefont
  {Martin}},\ }\href {\doibase 10.1103/PhysRev.124.670} {\bibfield  {journal}
  {\bibinfo  {journal} {Phys. Rev.}\ }\textbf {\bibinfo {volume} {124}},\
  \bibinfo {pages} {670} (\bibinfo {year} {1961})}\BibitemShut {NoStop}%
\bibitem [{\citenamefont {Chen}\ \emph {et~al.}(2005)\citenamefont {Chen},
  \citenamefont {Stajic}, \citenamefont {Tan},\ and\ \citenamefont
  {Levin}}]{Chen2005}%
  \BibitemOpen
  \bibfield  {author} {\bibinfo {author} {\bibfnamefont {Q.}~\bibnamefont
  {Chen}}, \bibinfo {author} {\bibfnamefont {J.}~\bibnamefont {Stajic}},
  \bibinfo {author} {\bibfnamefont {S.}~\bibnamefont {Tan}}, \ and\ \bibinfo
  {author} {\bibfnamefont {K.}~\bibnamefont {Levin}},\ }\href {\doibase
  https://doi.org/10.1016/j.physrep.2005.02.005} {\bibfield  {journal}
  {\bibinfo  {journal} {Physics Reports}\ }\textbf {\bibinfo {volume} {412}},\
  \bibinfo {pages} {1 } (\bibinfo {year} {2005})}\BibitemShut {NoStop}%
\bibitem [{\citenamefont {Levin}\ \emph {et~al.}(2010)\citenamefont {Levin},
  \citenamefont {Chen}, \citenamefont {Chien},\ and\ \citenamefont
  {He}}]{Levin2010}%
  \BibitemOpen
  \bibfield  {author} {\bibinfo {author} {\bibfnamefont {K.}~\bibnamefont
  {Levin}}, \bibinfo {author} {\bibfnamefont {Q.}~\bibnamefont {Chen}},
  \bibinfo {author} {\bibfnamefont {C.-C.}\ \bibnamefont {Chien}}, \ and\
  \bibinfo {author} {\bibfnamefont {Y.}~\bibnamefont {He}},\ }\href {\doibase
  https://doi.org/10.1016/j.aop.2009.09.011} {\bibfield  {journal} {\bibinfo
  {journal} {Annals of Physics}\ }\textbf {\bibinfo {volume} {325}},\ \bibinfo
  {pages} {233 } (\bibinfo {year} {2010})}\BibitemShut {NoStop}%
\bibitem [{\citenamefont {Chen}\ \emph {et~al.}(1999)\citenamefont {Chen},
  \citenamefont {Kosztin}, \citenamefont {Jank\'o},\ and\ \citenamefont
  {Levin}}]{Chen1999}%
  \BibitemOpen
  \bibfield  {author} {\bibinfo {author} {\bibfnamefont {Q.}~\bibnamefont
  {Chen}}, \bibinfo {author} {\bibfnamefont {I.}~\bibnamefont {Kosztin}},
  \bibinfo {author} {\bibfnamefont {B.}~\bibnamefont {Jank\'o}}, \ and\
  \bibinfo {author} {\bibfnamefont {K.}~\bibnamefont {Levin}},\ }\href
  {\doibase 10.1103/PhysRevB.59.7083} {\bibfield  {journal} {\bibinfo
  {journal} {Phys. Rev. B}\ }\textbf {\bibinfo {volume} {59}},\ \bibinfo
  {pages} {7083} (\bibinfo {year} {1999})}\BibitemShut {NoStop}%
\bibitem [{\citenamefont {Wu}\ \emph {et~al.}(2015{\natexlab{b}})\citenamefont
  {Wu}, \citenamefont {Anderson}, \citenamefont {Boyack},\ and\ \citenamefont
  {Levin}}]{Wu2015}%
  \BibitemOpen
  \bibfield  {author} {\bibinfo {author} {\bibfnamefont {C.-T.}\ \bibnamefont
  {Wu}}, \bibinfo {author} {\bibfnamefont {B.~M.}\ \bibnamefont {Anderson}},
  \bibinfo {author} {\bibfnamefont {R.}~\bibnamefont {Boyack}}, \ and\ \bibinfo
  {author} {\bibfnamefont {K.}~\bibnamefont {Levin}},\ }\href {\doibase
  10.1103/PhysRevLett.115.240401} {\bibfield  {journal} {\bibinfo  {journal}
  {Phys. Rev. Lett.}\ }\textbf {\bibinfo {volume} {115}},\ \bibinfo {pages}
  {240401} (\bibinfo {year} {2015}{\natexlab{b}})}\BibitemShut {NoStop}%
\bibitem [{\citenamefont {Provost}\ and\ \citenamefont
  {Vallee}(1980)}]{Provost1980}%
  \BibitemOpen
  \bibfield  {author} {\bibinfo {author} {\bibfnamefont {J.~P.}\ \bibnamefont
  {Provost}}\ and\ \bibinfo {author} {\bibfnamefont {G.}~\bibnamefont
  {Vallee}},\ }\href {https://link.springer.com/article/10.1007/BF02193559}
  {\bibfield  {journal} {\bibinfo  {journal} {Communications in Mathematical
  Physics}\ }\textbf {\bibinfo {volume} {76}},\ \bibinfo {pages} {289}
  (\bibinfo {year} {1980})}\BibitemShut {NoStop}%
\bibitem [{\citenamefont {Tovmasyan}\ \emph {et~al.}(2016)\citenamefont
  {Tovmasyan}, \citenamefont {Peotta}, \citenamefont {T\"orm\"a},\ and\
  \citenamefont {Huber}}]{Tovmasyan2016}%
  \BibitemOpen
  \bibfield  {author} {\bibinfo {author} {\bibfnamefont {M.}~\bibnamefont
  {Tovmasyan}}, \bibinfo {author} {\bibfnamefont {S.}~\bibnamefont {Peotta}},
  \bibinfo {author} {\bibfnamefont {P.}~\bibnamefont {T\"orm\"a}}, \ and\
  \bibinfo {author} {\bibfnamefont {S.~D.}\ \bibnamefont {Huber}},\ }\href
  {\doibase 10.1103/PhysRevB.94.245149} {\bibfield  {journal} {\bibinfo
  {journal} {Phys. Rev. B}\ }\textbf {\bibinfo {volume} {94}},\ \bibinfo
  {pages} {245149} (\bibinfo {year} {2016})}\BibitemShut {NoStop}%
\bibitem [{\citenamefont {Ries}\ \emph {et~al.}(2015)\citenamefont {Ries},
  \citenamefont {Wenz}, \citenamefont {Z\"urn}, \citenamefont {Bayha},
  \citenamefont {Boettcher}, \citenamefont {Kedar}, \citenamefont {Murthy},
  \citenamefont {Neidig}, \citenamefont {Lompe},\ and\ \citenamefont
  {Jochim}}]{Ries2015}%
  \BibitemOpen
  \bibfield  {author} {\bibinfo {author} {\bibfnamefont {M.~G.}\ \bibnamefont
  {Ries}}, \bibinfo {author} {\bibfnamefont {A.~N.}\ \bibnamefont {Wenz}},
  \bibinfo {author} {\bibfnamefont {G.}~\bibnamefont {Z\"urn}}, \bibinfo
  {author} {\bibfnamefont {L.}~\bibnamefont {Bayha}}, \bibinfo {author}
  {\bibfnamefont {I.}~\bibnamefont {Boettcher}}, \bibinfo {author}
  {\bibfnamefont {D.}~\bibnamefont {Kedar}}, \bibinfo {author} {\bibfnamefont
  {P.~A.}\ \bibnamefont {Murthy}}, \bibinfo {author} {\bibfnamefont
  {M.}~\bibnamefont {Neidig}}, \bibinfo {author} {\bibfnamefont
  {T.}~\bibnamefont {Lompe}}, \ and\ \bibinfo {author} {\bibfnamefont
  {S.}~\bibnamefont {Jochim}},\ }\href {\doibase
  10.1103/PhysRevLett.114.230401} {\bibfield  {journal} {\bibinfo  {journal}
  {Phys. Rev. Lett.}\ }\textbf {\bibinfo {volume} {114}},\ \bibinfo {pages}
  {230401} (\bibinfo {year} {2015})}\BibitemShut {NoStop}%
\bibitem [{\citenamefont {Prokof'ev}\ and\ \citenamefont
  {Svistunov}(2002)}]{Prokofev2002}%
  \BibitemOpen
  \bibfield  {author} {\bibinfo {author} {\bibfnamefont {N.}~\bibnamefont
  {Prokof'ev}}\ and\ \bibinfo {author} {\bibfnamefont {B.}~\bibnamefont
  {Svistunov}},\ }\href {\doibase 10.1103/PhysRevA.66.043608} {\bibfield
  {journal} {\bibinfo  {journal} {Phys. Rev. A}\ }\textbf {\bibinfo {volume}
  {66}},\ \bibinfo {pages} {043608} (\bibinfo {year} {2002})}\BibitemShut
  {NoStop}%
\bibitem [{\citenamefont {Jose}(2013)}]{Jose2013}%
  \BibitemOpen
  \bibfield  {author} {\bibinfo {author} {\bibfnamefont {J.~V.}\ \bibnamefont
  {Jose}},\ }\href@noop {} {\emph {\bibinfo {title} {40 years of
  Berezinskii-Kosterlitz-Thouless theory}}}\ (\bibinfo  {publisher} {World
  Scientific, Singapore},\ \bibinfo {year} {2013})\BibitemShut {NoStop}%
\bibitem [{\citenamefont {Tung}\ \emph {et~al.}(2010)\citenamefont {Tung},
  \citenamefont {Lamporesi}, \citenamefont {Lobser}, \citenamefont {Xia},\ and\
  \citenamefont {Cornell}}]{Tung2010}%
  \BibitemOpen
  \bibfield  {author} {\bibinfo {author} {\bibfnamefont {S.}~\bibnamefont
  {Tung}}, \bibinfo {author} {\bibfnamefont {G.}~\bibnamefont {Lamporesi}},
  \bibinfo {author} {\bibfnamefont {D.}~\bibnamefont {Lobser}}, \bibinfo
  {author} {\bibfnamefont {L.}~\bibnamefont {Xia}}, \ and\ \bibinfo {author}
  {\bibfnamefont {E.~A.}\ \bibnamefont {Cornell}},\ }\href {\doibase
  10.1103/PhysRevLett.105.230408} {\bibfield  {journal} {\bibinfo  {journal}
  {Phys. Rev. Lett.}\ }\textbf {\bibinfo {volume} {105}},\ \bibinfo {pages}
  {230408} (\bibinfo {year} {2010})}\BibitemShut {NoStop}%
\bibitem [{\citenamefont {Clad\'e}\ \emph {et~al.}(2009)\citenamefont
  {Clad\'e}, \citenamefont {Ryu}, \citenamefont {Ramanathan}, \citenamefont
  {Helmerson},\ and\ \citenamefont {Phillips}}]{Clade2009}%
  \BibitemOpen
  \bibfield  {author} {\bibinfo {author} {\bibfnamefont {P.}~\bibnamefont
  {Clad\'e}}, \bibinfo {author} {\bibfnamefont {C.}~\bibnamefont {Ryu}},
  \bibinfo {author} {\bibfnamefont {A.}~\bibnamefont {Ramanathan}}, \bibinfo
  {author} {\bibfnamefont {K.}~\bibnamefont {Helmerson}}, \ and\ \bibinfo
  {author} {\bibfnamefont {W.~D.}\ \bibnamefont {Phillips}},\ }\href {\doibase
  10.1103/PhysRevLett.102.170401} {\bibfield  {journal} {\bibinfo  {journal}
  {Phys. Rev. Lett.}\ }\textbf {\bibinfo {volume} {102}},\ \bibinfo {pages}
  {170401} (\bibinfo {year} {2009})}\BibitemShut {NoStop}%
\bibitem [{\citenamefont {Murthy}\ \emph {et~al.}(2015)\citenamefont {Murthy},
  \citenamefont {Boettcher}, \citenamefont {Bayha}, \citenamefont {Holzmann},
  \citenamefont {Kedar}, \citenamefont {Neidig}, \citenamefont {Ries},
  \citenamefont {Wenz}, \citenamefont {Z\"urn},\ and\ \citenamefont
  {Jochim}}]{Murthy2015}%
  \BibitemOpen
  \bibfield  {author} {\bibinfo {author} {\bibfnamefont {P.~A.}\ \bibnamefont
  {Murthy}}, \bibinfo {author} {\bibfnamefont {I.}~\bibnamefont {Boettcher}},
  \bibinfo {author} {\bibfnamefont {L.}~\bibnamefont {Bayha}}, \bibinfo
  {author} {\bibfnamefont {M.}~\bibnamefont {Holzmann}}, \bibinfo {author}
  {\bibfnamefont {D.}~\bibnamefont {Kedar}}, \bibinfo {author} {\bibfnamefont
  {M.}~\bibnamefont {Neidig}}, \bibinfo {author} {\bibfnamefont {M.~G.}\
  \bibnamefont {Ries}}, \bibinfo {author} {\bibfnamefont {A.~N.}\ \bibnamefont
  {Wenz}}, \bibinfo {author} {\bibfnamefont {G.}~\bibnamefont {Z\"urn}}, \ and\
  \bibinfo {author} {\bibfnamefont {S.}~\bibnamefont {Jochim}},\ }\href
  {\doibase 10.1103/PhysRevLett.115.010401} {\bibfield  {journal} {\bibinfo
  {journal} {Phys. Rev. Lett.}\ }\textbf {\bibinfo {volume} {115}},\ \bibinfo
  {pages} {010401} (\bibinfo {year} {2015})}\BibitemShut {NoStop}%
\bibitem [{\citenamefont {Nozi{\`e}res}\ and\ \citenamefont
  {Schmitt-Rink}(1985)}]{Nozieres1985}%
  \BibitemOpen
  \bibfield  {author} {\bibinfo {author} {\bibfnamefont {P.}~\bibnamefont
  {Nozi{\`e}res}}\ and\ \bibinfo {author} {\bibfnamefont {S.}~\bibnamefont
  {Schmitt-Rink}},\ }\href
  {https://link.springer.com/article/10.1007/BF00683774} {\bibfield  {journal}
  {\bibinfo  {journal} {Journal of Low Temperature Physics}\ }\textbf {\bibinfo
  {volume} {59}},\ \bibinfo {pages} {195} (\bibinfo {year} {1985})}\BibitemShut
  {NoStop}%
\bibitem [{\citenamefont {Micnas}\ \emph {et~al.}(1990)\citenamefont {Micnas},
  \citenamefont {Ranninger},\ and\ \citenamefont {Robaszkiewicz}}]{Micnas1990}%
  \BibitemOpen
  \bibfield  {author} {\bibinfo {author} {\bibfnamefont {R.}~\bibnamefont
  {Micnas}}, \bibinfo {author} {\bibfnamefont {J.}~\bibnamefont {Ranninger}}, \
  and\ \bibinfo {author} {\bibfnamefont {S.}~\bibnamefont {Robaszkiewicz}},\
  }\href {\doibase 10.1103/RevModPhys.62.113} {\bibfield  {journal} {\bibinfo
  {journal} {Rev. Mod. Phys.}\ }\textbf {\bibinfo {volume} {62}},\ \bibinfo
  {pages} {113} (\bibinfo {year} {1990})}\BibitemShut {NoStop}%
\bibitem [{\citenamefont {Halperin}\ and\ \citenamefont
  {Nelson}(1979)}]{Halperin1979}%
  \BibitemOpen
  \bibfield  {author} {\bibinfo {author} {\bibfnamefont {B.}~\bibnamefont
  {Halperin}}\ and\ \bibinfo {author} {\bibfnamefont {D.~R.}\ \bibnamefont
  {Nelson}},\ }\href {https://link.springer.com/article/10.1007/BF00116988}
  {\bibfield  {journal} {\bibinfo  {journal} {Journal of low temperature
  physics}\ }\textbf {\bibinfo {volume} {36}},\ \bibinfo {pages} {599}
  (\bibinfo {year} {1979})}\BibitemShut {NoStop}%
\bibitem [{\citenamefont {Epstein}\ \emph {et~al.}(1981)\citenamefont
  {Epstein}, \citenamefont {Goldman},\ and\ \citenamefont
  {Kadin}}]{Epstein1981}%
  \BibitemOpen
  \bibfield  {author} {\bibinfo {author} {\bibfnamefont {K.}~\bibnamefont
  {Epstein}}, \bibinfo {author} {\bibfnamefont {A.~M.}\ \bibnamefont
  {Goldman}}, \ and\ \bibinfo {author} {\bibfnamefont {A.~M.}\ \bibnamefont
  {Kadin}},\ }\href {\doibase 10.1103/PhysRevLett.47.534} {\bibfield  {journal}
  {\bibinfo  {journal} {Phys. Rev. Lett.}\ }\textbf {\bibinfo {volume} {47}},\
  \bibinfo {pages} {534} (\bibinfo {year} {1981})}\BibitemShut {NoStop}%
\bibitem [{\citenamefont {Resnick}\ \emph {et~al.}(1981)\citenamefont
  {Resnick}, \citenamefont {Garland}, \citenamefont {Boyd}, \citenamefont
  {Shoemaker},\ and\ \citenamefont {Newrock}}]{Resnick1981}%
  \BibitemOpen
  \bibfield  {author} {\bibinfo {author} {\bibfnamefont {D.~J.}\ \bibnamefont
  {Resnick}}, \bibinfo {author} {\bibfnamefont {J.~C.}\ \bibnamefont
  {Garland}}, \bibinfo {author} {\bibfnamefont {J.~T.}\ \bibnamefont {Boyd}},
  \bibinfo {author} {\bibfnamefont {S.}~\bibnamefont {Shoemaker}}, \ and\
  \bibinfo {author} {\bibfnamefont {R.~S.}\ \bibnamefont {Newrock}},\ }\href
  {\doibase 10.1103/PhysRevLett.47.1542} {\bibfield  {journal} {\bibinfo
  {journal} {Phys. Rev. Lett.}\ }\textbf {\bibinfo {volume} {47}},\ \bibinfo
  {pages} {1542} (\bibinfo {year} {1981})}\BibitemShut {NoStop}%
\bibitem [{\citenamefont {Hebard}\ and\ \citenamefont
  {Fiory}(1983)}]{Hebard1983}%
  \BibitemOpen
  \bibfield  {author} {\bibinfo {author} {\bibfnamefont {A.~F.}\ \bibnamefont
  {Hebard}}\ and\ \bibinfo {author} {\bibfnamefont {A.~T.}\ \bibnamefont
  {Fiory}},\ }\href {\doibase 10.1103/PhysRevLett.50.1603} {\bibfield
  {journal} {\bibinfo  {journal} {Phys. Rev. Lett.}\ }\textbf {\bibinfo
  {volume} {50}},\ \bibinfo {pages} {1603} (\bibinfo {year}
  {1983})}\BibitemShut {NoStop}%
\bibitem [{\citenamefont {Fiory}\ \emph {et~al.}(1983)\citenamefont {Fiory},
  \citenamefont {Hebard},\ and\ \citenamefont {Glaberson}}]{Fiory1983}%
  \BibitemOpen
  \bibfield  {author} {\bibinfo {author} {\bibfnamefont {A.~T.}\ \bibnamefont
  {Fiory}}, \bibinfo {author} {\bibfnamefont {A.~F.}\ \bibnamefont {Hebard}}, \
  and\ \bibinfo {author} {\bibfnamefont {W.~I.}\ \bibnamefont {Glaberson}},\
  }\href {\doibase 10.1103/PhysRevB.28.5075} {\bibfield  {journal} {\bibinfo
  {journal} {Phys. Rev. B}\ }\textbf {\bibinfo {volume} {28}},\ \bibinfo
  {pages} {5075} (\bibinfo {year} {1983})}\BibitemShut {NoStop}%
\bibitem [{\citenamefont {Kadin}\ \emph {et~al.}(1983)\citenamefont {Kadin},
  \citenamefont {Epstein},\ and\ \citenamefont {Goldman}}]{Kadin1983}%
  \BibitemOpen
  \bibfield  {author} {\bibinfo {author} {\bibfnamefont {A.~M.}\ \bibnamefont
  {Kadin}}, \bibinfo {author} {\bibfnamefont {K.}~\bibnamefont {Epstein}}, \
  and\ \bibinfo {author} {\bibfnamefont {A.~M.}\ \bibnamefont {Goldman}},\
  }\href {\doibase 10.1103/PhysRevB.27.6691} {\bibfield  {journal} {\bibinfo
  {journal} {Phys. Rev. B}\ }\textbf {\bibinfo {volume} {27}},\ \bibinfo
  {pages} {6691} (\bibinfo {year} {1983})}\BibitemShut {NoStop}%
\bibitem [{\citenamefont {Lu}\ \emph {et~al.}(2019)\citenamefont {Lu},
  \citenamefont {Stepanov}, \citenamefont {Yang}, \citenamefont {Xie},
  \citenamefont {Aamir}, \citenamefont {Das}, \citenamefont {Urgell},
  \citenamefont {Watanabe}, \citenamefont {Taniguchi}, \citenamefont {Zhang},
  \citenamefont {Bachtold}, \citenamefont {MacDonald},\ and\ \citenamefont
  {Efetov}}]{Lu2019}%
  \BibitemOpen
  \bibfield  {author} {\bibinfo {author} {\bibfnamefont {X.}~\bibnamefont
  {Lu}}, \bibinfo {author} {\bibfnamefont {P.}~\bibnamefont {Stepanov}},
  \bibinfo {author} {\bibfnamefont {W.}~\bibnamefont {Yang}}, \bibinfo {author}
  {\bibfnamefont {M.}~\bibnamefont {Xie}}, \bibinfo {author} {\bibfnamefont
  {M.~A.}\ \bibnamefont {Aamir}}, \bibinfo {author} {\bibfnamefont
  {I.}~\bibnamefont {Das}}, \bibinfo {author} {\bibfnamefont {C.}~\bibnamefont
  {Urgell}}, \bibinfo {author} {\bibfnamefont {K.}~\bibnamefont {Watanabe}},
  \bibinfo {author} {\bibfnamefont {T.}~\bibnamefont {Taniguchi}}, \bibinfo
  {author} {\bibfnamefont {G.}~\bibnamefont {Zhang}}, \bibinfo {author}
  {\bibfnamefont {A.}~\bibnamefont {Bachtold}}, \bibinfo {author}
  {\bibfnamefont {A.~H.}\ \bibnamefont {MacDonald}}, \ and\ \bibinfo {author}
  {\bibfnamefont {D.~K.}\ \bibnamefont {Efetov}},\ }\href
  {https://www.nature.com/articles/s41586-019-1695-0} {\bibfield  {journal}
  {\bibinfo  {journal} {Nature}\ }\textbf {\bibinfo {volume} {574}},\ \bibinfo
  {pages} {653} (\bibinfo {year} {2019})}\BibitemShut {NoStop}%
\bibitem [{\citenamefont {Stepanov}\ \emph {et~al.}(2020)\citenamefont
  {Stepanov}, \citenamefont {Das}, \citenamefont {Lu}, \citenamefont
  {Fahimniya}, \citenamefont {Watanabe}, \citenamefont {Taniguchi},
  \citenamefont {Koppens}, \citenamefont {Lischner}, \citenamefont {Levitov},\
  and\ \citenamefont {Efetov}}]{Stepanov2020}%
  \BibitemOpen
  \bibfield  {author} {\bibinfo {author} {\bibfnamefont {P.}~\bibnamefont
  {Stepanov}}, \bibinfo {author} {\bibfnamefont {I.}~\bibnamefont {Das}},
  \bibinfo {author} {\bibfnamefont {X.}~\bibnamefont {Lu}}, \bibinfo {author}
  {\bibfnamefont {A.}~\bibnamefont {Fahimniya}}, \bibinfo {author}
  {\bibfnamefont {K.}~\bibnamefont {Watanabe}}, \bibinfo {author}
  {\bibfnamefont {T.}~\bibnamefont {Taniguchi}}, \bibinfo {author}
  {\bibfnamefont {F.~H.}\ \bibnamefont {Koppens}}, \bibinfo {author}
  {\bibfnamefont {J.}~\bibnamefont {Lischner}}, \bibinfo {author}
  {\bibfnamefont {L.}~\bibnamefont {Levitov}}, \ and\ \bibinfo {author}
  {\bibfnamefont {D.~K.}\ \bibnamefont {Efetov}},\ }\href
  {https://www.nature.com/articles/s41586-020-2459-6} {\bibfield  {journal}
  {\bibinfo  {journal} {Nature}\ }\textbf {\bibinfo {volume} {583}},\ \bibinfo
  {pages} {375} (\bibinfo {year} {2020})}\BibitemShut {NoStop}%
\bibitem [{\citenamefont {Cao}\ \emph {et~al.}(2020)\citenamefont {Cao},
  \citenamefont {Rodan-Legrain}, \citenamefont {Park}, \citenamefont {Yuan},
  \citenamefont {Watanabe}, \citenamefont {Taniguchi}, \citenamefont
  {Fernandes}, \citenamefont {Fu},\ and\ \citenamefont
  {Jarillo-Herrero}}]{Cao2020a}%
  \BibitemOpen
  \bibfield  {author} {\bibinfo {author} {\bibfnamefont {Y.}~\bibnamefont
  {Cao}}, \bibinfo {author} {\bibfnamefont {D.}~\bibnamefont {Rodan-Legrain}},
  \bibinfo {author} {\bibfnamefont {J.~M.}\ \bibnamefont {Park}}, \bibinfo
  {author} {\bibfnamefont {F.~N.}\ \bibnamefont {Yuan}}, \bibinfo {author}
  {\bibfnamefont {K.}~\bibnamefont {Watanabe}}, \bibinfo {author}
  {\bibfnamefont {T.}~\bibnamefont {Taniguchi}}, \bibinfo {author}
  {\bibfnamefont {R.~M.}\ \bibnamefont {Fernandes}}, \bibinfo {author}
  {\bibfnamefont {L.}~\bibnamefont {Fu}}, \ and\ \bibinfo {author}
  {\bibfnamefont {P.}~\bibnamefont {Jarillo-Herrero}},\ }\href
  {https://arxiv.org/abs/2004.04148} {\bibfield  {journal} {\bibinfo  {journal}
  {arXiv preprint arXiv:2004.04148}\ } (\bibinfo {year} {2020})}\BibitemShut
  {NoStop}%
\bibitem [{Note1()}]{Note1}%
  \BibitemOpen
  \bibinfo {note} {Similar dc transport signatures of $T^*$ have been observed
  previously in cuprates~\cite {Timusk1999}, although the pseudogap there can
  have a completely different origin from preformed Cooper pairs.}\BibitemShut
  {Stop}%
\bibitem [{Note2()}]{Note2}%
  \BibitemOpen
  \bibinfo {note} {Here we ignore the intervening correlated insulating phase
  at half filling of the lower and upper flat band, and view the two
  superconducting domes flanking the insulating phase as one.}\BibitemShut
  {Stop}%
\bibitem [{\citenamefont {Saito}\ \emph {et~al.}(2020)\citenamefont {Saito},
  \citenamefont {Ge}, \citenamefont {Watanabe}, \citenamefont {Taniguchi},\
  and\ \citenamefont {Young}}]{Saito2020}%
  \BibitemOpen
  \bibfield  {author} {\bibinfo {author} {\bibfnamefont {Y.}~\bibnamefont
  {Saito}}, \bibinfo {author} {\bibfnamefont {J.}~\bibnamefont {Ge}}, \bibinfo
  {author} {\bibfnamefont {K.}~\bibnamefont {Watanabe}}, \bibinfo {author}
  {\bibfnamefont {T.}~\bibnamefont {Taniguchi}}, \ and\ \bibinfo {author}
  {\bibfnamefont {A.~F.}\ \bibnamefont {Young}},\ }\href@noop {} {\bibfield
  {journal} {\bibinfo  {journal} {Nature Physics}\ }\textbf {\bibinfo {volume}
  {16}},\ \bibinfo {pages} {926} (\bibinfo {year} {2020})}\BibitemShut
  {NoStop}%
\bibitem [{\citenamefont {Liu}\ \emph {et~al.}(2020)\citenamefont {Liu},
  \citenamefont {Wang}, \citenamefont {Watanabe}, \citenamefont {Taniguchi},
  \citenamefont {Vafek},\ and\ \citenamefont {Li}}]{Liu2020}%
  \BibitemOpen
  \bibfield  {author} {\bibinfo {author} {\bibfnamefont {X.}~\bibnamefont
  {Liu}}, \bibinfo {author} {\bibfnamefont {Z.}~\bibnamefont {Wang}}, \bibinfo
  {author} {\bibfnamefont {K.}~\bibnamefont {Watanabe}}, \bibinfo {author}
  {\bibfnamefont {T.}~\bibnamefont {Taniguchi}}, \bibinfo {author}
  {\bibfnamefont {O.}~\bibnamefont {Vafek}}, \ and\ \bibinfo {author}
  {\bibfnamefont {J.}~\bibnamefont {Li}},\ }\href
  {https://arxiv.org/abs/2003.11072} {\bibfield  {journal} {\bibinfo  {journal}
  {arXiv preprint arXiv:2003.11072}\ } (\bibinfo {year} {2020})}\BibitemShut
  {NoStop}%
\bibitem [{\citenamefont {Jiang}\ \emph {et~al.}(2019)\citenamefont {Jiang},
  \citenamefont {Lai}, \citenamefont {Watanabe}, \citenamefont {Taniguchi},
  \citenamefont {Haule}, \citenamefont {Mao},\ and\ \citenamefont
  {Andrei}}]{Jiang2019}%
  \BibitemOpen
  \bibfield  {author} {\bibinfo {author} {\bibfnamefont {Y.}~\bibnamefont
  {Jiang}}, \bibinfo {author} {\bibfnamefont {X.}~\bibnamefont {Lai}}, \bibinfo
  {author} {\bibfnamefont {K.}~\bibnamefont {Watanabe}}, \bibinfo {author}
  {\bibfnamefont {T.}~\bibnamefont {Taniguchi}}, \bibinfo {author}
  {\bibfnamefont {K.}~\bibnamefont {Haule}}, \bibinfo {author} {\bibfnamefont
  {J.}~\bibnamefont {Mao}}, \ and\ \bibinfo {author} {\bibfnamefont {E.~Y.}\
  \bibnamefont {Andrei}},\ }\href
  {https://www.nature.com/articles/s41586-019-1460-4} {\bibfield  {journal}
  {\bibinfo  {journal} {Nature}\ }\textbf {\bibinfo {volume} {573}},\ \bibinfo
  {pages} {91} (\bibinfo {year} {2019})}\BibitemShut {NoStop}%
\bibitem [{\citenamefont {Wong}\ \emph {et~al.}(2020)\citenamefont {Wong},
  \citenamefont {Nuckolls}, \citenamefont {Oh}, \citenamefont {Lian},
  \citenamefont {Xie}, \citenamefont {Jeon}, \citenamefont {Watanabe},
  \citenamefont {Taniguchi}, \citenamefont {Bernevig},\ and\ \citenamefont
  {Yazdani}}]{Wong2020}%
  \BibitemOpen
  \bibfield  {author} {\bibinfo {author} {\bibfnamefont {D.}~\bibnamefont
  {Wong}}, \bibinfo {author} {\bibfnamefont {K.~P.}\ \bibnamefont {Nuckolls}},
  \bibinfo {author} {\bibfnamefont {M.}~\bibnamefont {Oh}}, \bibinfo {author}
  {\bibfnamefont {B.}~\bibnamefont {Lian}}, \bibinfo {author} {\bibfnamefont
  {Y.}~\bibnamefont {Xie}}, \bibinfo {author} {\bibfnamefont {S.}~\bibnamefont
  {Jeon}}, \bibinfo {author} {\bibfnamefont {K.}~\bibnamefont {Watanabe}},
  \bibinfo {author} {\bibfnamefont {T.}~\bibnamefont {Taniguchi}}, \bibinfo
  {author} {\bibfnamefont {B.~A.}\ \bibnamefont {Bernevig}}, \ and\ \bibinfo
  {author} {\bibfnamefont {A.}~\bibnamefont {Yazdani}},\ }\href
  {https://www.nature.com/articles/s41586-020-2339-0} {\bibfield  {journal}
  {\bibinfo  {journal} {Nature}\ }\textbf {\bibinfo {volume} {582}},\ \bibinfo
  {pages} {198} (\bibinfo {year} {2020})}\BibitemShut {NoStop}%
\bibitem [{\citenamefont {Xie}\ \emph {et~al.}(2019)\citenamefont {Xie},
  \citenamefont {Lian}, \citenamefont {J{\"a}ck}, \citenamefont {Liu},
  \citenamefont {Chiu}, \citenamefont {Watanabe}, \citenamefont {Taniguchi},
  \citenamefont {Bernevig},\ and\ \citenamefont {Yazdani}}]{Xie2019}%
  \BibitemOpen
  \bibfield  {author} {\bibinfo {author} {\bibfnamefont {Y.}~\bibnamefont
  {Xie}}, \bibinfo {author} {\bibfnamefont {B.}~\bibnamefont {Lian}}, \bibinfo
  {author} {\bibfnamefont {B.}~\bibnamefont {J{\"a}ck}}, \bibinfo {author}
  {\bibfnamefont {X.}~\bibnamefont {Liu}}, \bibinfo {author} {\bibfnamefont
  {C.-L.}\ \bibnamefont {Chiu}}, \bibinfo {author} {\bibfnamefont
  {K.}~\bibnamefont {Watanabe}}, \bibinfo {author} {\bibfnamefont
  {T.}~\bibnamefont {Taniguchi}}, \bibinfo {author} {\bibfnamefont {B.~A.}\
  \bibnamefont {Bernevig}}, \ and\ \bibinfo {author} {\bibfnamefont
  {A.}~\bibnamefont {Yazdani}},\ }\href
  {https://www.nature.com/articles/s41586-019-1422-x} {\bibfield  {journal}
  {\bibinfo  {journal} {Nature}\ }\textbf {\bibinfo {volume} {572}},\ \bibinfo
  {pages} {101} (\bibinfo {year} {2019})}\BibitemShut {NoStop}%
\bibitem [{Note3()}]{Note3}%
  \BibitemOpen
  \bibinfo {note} {We note that the results presented in Fig. 1 of
  Ref.~\protect \rev@citealpnum {Hofmann2019} are for electron density $ n=0.5$
  per site; while our results are for $n=0.3$ per site. However, we do not
  expect the qualitative $U$ dependence of $T_{\protect \text {BKT}}$ and $T^*$
  to change from $n=0.5$ to $n=0.3$.}\BibitemShut {Stop}%
\bibitem [{\citenamefont {Iskin}(2018{\natexlab{b}})}]{Iskin2018a}%
  \BibitemOpen
  \bibfield  {author} {\bibinfo {author} {\bibfnamefont {M.}~\bibnamefont
  {Iskin}},\ }\href {\doibase 10.1103/PhysRevA.97.063625} {\bibfield  {journal}
  {\bibinfo  {journal} {Phys. Rev. A}\ }\textbf {\bibinfo {volume} {97}},\
  \bibinfo {pages} {063625} (\bibinfo {year} {2018}{\natexlab{b}})}\BibitemShut
  {NoStop}%
\bibitem [{\citenamefont {Iskin}(2019)}]{Iskin2019a}%
  \BibitemOpen
  \bibfield  {author} {\bibinfo {author} {\bibfnamefont {M.}~\bibnamefont
  {Iskin}},\ }\href {\doibase 10.1103/PhysRevA.99.053603} {\bibfield  {journal}
  {\bibinfo  {journal} {Phys. Rev. A}\ }\textbf {\bibinfo {volume} {99}},\
  \bibinfo {pages} {053603} (\bibinfo {year} {2019})}\BibitemShut {NoStop}%
\bibitem [{\citenamefont {Iskin}(2020{\natexlab{a}})}]{Iskin2020}%
  \BibitemOpen
  \bibfield  {author} {\bibinfo {author} {\bibfnamefont {M.}~\bibnamefont
  {Iskin}},\ }\href {\doibase 10.1103/PhysRevA.101.053631} {\bibfield
  {journal} {\bibinfo  {journal} {Phys. Rev. A}\ }\textbf {\bibinfo {volume}
  {101}},\ \bibinfo {pages} {053631} (\bibinfo {year}
  {2020}{\natexlab{a}})}\BibitemShut {NoStop}%
\bibitem [{\citenamefont {Iskin}(2020{\natexlab{b}})}]{Iskin2020a}%
  \BibitemOpen
  \bibfield  {author} {\bibinfo {author} {\bibfnamefont {M.}~\bibnamefont
  {Iskin}},\ }\href {\doibase https://doi.org/10.1016/j.physb.2020.412260}
  {\bibfield  {journal} {\bibinfo  {journal} {Physica B: Condensed Matter}\
  }\textbf {\bibinfo {volume} {592}},\ \bibinfo {pages} {412260} (\bibinfo
  {year} {2020}{\natexlab{b}})}\BibitemShut {NoStop}%
\bibitem [{Note4()}]{Note4}%
  \BibitemOpen
  \bibinfo {note} {However, the intra-band contribution we identified, the term
  with $\alpha ^\prime =\alpha $ in Eq.~\protect \textup {\hbox {\mathsurround
  \z@ \protect \normalfont (\ignorespaces \ref {eq: Tgeom}\unskip \@@italiccorr
  )}}, was missing.}\BibitemShut {Stop}%
\bibitem [{\citenamefont {Benfatto}\ \emph {et~al.}(2004)\citenamefont
  {Benfatto}, \citenamefont {Toschi},\ and\ \citenamefont
  {Caprara}}]{Benfatto2004}%
  \BibitemOpen
  \bibfield  {author} {\bibinfo {author} {\bibfnamefont {L.}~\bibnamefont
  {Benfatto}}, \bibinfo {author} {\bibfnamefont {A.}~\bibnamefont {Toschi}}, \
  and\ \bibinfo {author} {\bibfnamefont {S.}~\bibnamefont {Caprara}},\ }\href
  {\doibase 10.1103/PhysRevB.69.184510} {\bibfield  {journal} {\bibinfo
  {journal} {Phys. Rev. B}\ }\textbf {\bibinfo {volume} {69}},\ \bibinfo
  {pages} {184510} (\bibinfo {year} {2004})}\BibitemShut {NoStop}%
\bibitem [{\citenamefont {Fukushima}\ \emph {et~al.}(2007)\citenamefont
  {Fukushima}, \citenamefont {Ohashi}, \citenamefont {Taylor},\ and\
  \citenamefont {Griffin}}]{Fukushima2007}%
  \BibitemOpen
  \bibfield  {author} {\bibinfo {author} {\bibfnamefont {N.}~\bibnamefont
  {Fukushima}}, \bibinfo {author} {\bibfnamefont {Y.}~\bibnamefont {Ohashi}},
  \bibinfo {author} {\bibfnamefont {E.}~\bibnamefont {Taylor}}, \ and\ \bibinfo
  {author} {\bibfnamefont {A.}~\bibnamefont {Griffin}},\ }\href {\doibase
  10.1103/PhysRevA.75.033609} {\bibfield  {journal} {\bibinfo  {journal} {Phys.
  Rev. A}\ }\textbf {\bibinfo {volume} {75}},\ \bibinfo {pages} {033609}
  (\bibinfo {year} {2007})}\BibitemShut {NoStop}%
\bibitem [{\citenamefont {Bighin}\ and\ \citenamefont
  {Salasnich}(2016)}]{Bighin2016}%
  \BibitemOpen
  \bibfield  {author} {\bibinfo {author} {\bibfnamefont {G.}~\bibnamefont
  {Bighin}}\ and\ \bibinfo {author} {\bibfnamefont {L.}~\bibnamefont
  {Salasnich}},\ }\href {\doibase 10.1103/PhysRevB.93.014519} {\bibfield
  {journal} {\bibinfo  {journal} {Phys. Rev. B}\ }\textbf {\bibinfo {volume}
  {93}},\ \bibinfo {pages} {014519} (\bibinfo {year} {2016})}\BibitemShut
  {NoStop}%
\bibitem [{Note5()}]{Note5}%
  \BibitemOpen
  \bibinfo {note} {Unfortunately, these calculations lead to an unusual double
  valued functional form for the superfluid density.}\BibitemShut {Stop}%
\bibitem [{Note6()}]{Note6}%
  \BibitemOpen
  \bibinfo {note} {As shown by recent experimental and theoretical
  studies~\cite {Sharpe2019, Serlin2020, Zhang2019, Bultinck2020}, a coupling
  of the TBLG to the hBN substrate can break certain symmetry that leads to
  occurrence of Chern bands and the resulted anomalous Hall
  effect.}\BibitemShut {Stop}%
\bibitem [{\citenamefont {Zou}\ \emph {et~al.}(2018)\citenamefont {Zou},
  \citenamefont {Po}, \citenamefont {Vishwanath},\ and\ \citenamefont
  {Senthil}}]{Zou2018}%
  \BibitemOpen
  \bibfield  {author} {\bibinfo {author} {\bibfnamefont {L.}~\bibnamefont
  {Zou}}, \bibinfo {author} {\bibfnamefont {H.~C.}\ \bibnamefont {Po}},
  \bibinfo {author} {\bibfnamefont {A.}~\bibnamefont {Vishwanath}}, \ and\
  \bibinfo {author} {\bibfnamefont {T.}~\bibnamefont {Senthil}},\ }\href
  {\doibase 10.1103/PhysRevB.98.085435} {\bibfield  {journal} {\bibinfo
  {journal} {Phys. Rev. B}\ }\textbf {\bibinfo {volume} {98}},\ \bibinfo
  {pages} {085435} (\bibinfo {year} {2018})}\BibitemShut {NoStop}%
\bibitem [{\citenamefont {Po}\ \emph {et~al.}(2018{\natexlab{b}})\citenamefont
  {Po}, \citenamefont {Watanabe},\ and\ \citenamefont {Vishwanath}}]{Po2018a}%
  \BibitemOpen
  \bibfield  {author} {\bibinfo {author} {\bibfnamefont {H.~C.}\ \bibnamefont
  {Po}}, \bibinfo {author} {\bibfnamefont {H.}~\bibnamefont {Watanabe}}, \ and\
  \bibinfo {author} {\bibfnamefont {A.}~\bibnamefont {Vishwanath}},\ }\href
  {\doibase 10.1103/PhysRevLett.121.126402} {\bibfield  {journal} {\bibinfo
  {journal} {Phys. Rev. Lett.}\ }\textbf {\bibinfo {volume} {121}},\ \bibinfo
  {pages} {126402} (\bibinfo {year} {2018}{\natexlab{b}})}\BibitemShut
  {NoStop}%
\bibitem [{\citenamefont {Ahn}\ \emph {et~al.}(2019)\citenamefont {Ahn},
  \citenamefont {Park},\ and\ \citenamefont {Yang}}]{Ahn2019}%
  \BibitemOpen
  \bibfield  {author} {\bibinfo {author} {\bibfnamefont {J.}~\bibnamefont
  {Ahn}}, \bibinfo {author} {\bibfnamefont {S.}~\bibnamefont {Park}}, \ and\
  \bibinfo {author} {\bibfnamefont {B.-J.}\ \bibnamefont {Yang}},\ }\href
  {\doibase 10.1103/PhysRevX.9.021013} {\bibfield  {journal} {\bibinfo
  {journal} {Phys. Rev. X}\ }\textbf {\bibinfo {volume} {9}},\ \bibinfo {pages}
  {021013} (\bibinfo {year} {2019})}\BibitemShut {NoStop}%
\bibitem [{Note7()}]{Note7}%
  \BibitemOpen
  \bibinfo {note} {In principle, the normal state bands relevant to the
  superconductivity can be different from the bare band, due to
  renormalizations from interactions.}\BibitemShut {Stop}%
\bibitem [{\citenamefont {Chen}(2000)}]{ChenPhD}%
  \BibitemOpen
  \bibfield  {author} {\bibinfo {author} {\bibfnamefont {Q.~J.}\ \bibnamefont
  {Chen}},\ }\emph {\bibinfo {title} {Generalization of BCS theory to short
  coherence length superconductors: A BCS--Bose-Einstein crossover
  scenario,}},\ \href {https://arxiv.org/abs/1801.06266} {Ph.D. thesis},\
  \bibinfo  {school} {University of Chicago} (\bibinfo {year} {2000}),\
  \bibinfo {note} {arXiv:1801.06266}\BibitemShut {NoStop}%
\bibitem [{\citenamefont {Anandan}\ and\ \citenamefont
  {Aharonov}(1990)}]{Anandan1990}%
  \BibitemOpen
  \bibfield  {author} {\bibinfo {author} {\bibfnamefont {J.}~\bibnamefont
  {Anandan}}\ and\ \bibinfo {author} {\bibfnamefont {Y.}~\bibnamefont
  {Aharonov}},\ }\href {\doibase 10.1103/PhysRevLett.65.1697} {\bibfield
  {journal} {\bibinfo  {journal} {Phys. Rev. Lett.}\ }\textbf {\bibinfo
  {volume} {65}},\ \bibinfo {pages} {1697} (\bibinfo {year}
  {1990})}\BibitemShut {NoStop}%
\bibitem [{\citenamefont {Denteneer}\ \emph {et~al.}(1993)\citenamefont
  {Denteneer}, \citenamefont {An},\ and\ \citenamefont {van
  Leeuwen}}]{Denteneer1993}%
  \BibitemOpen
  \bibfield  {author} {\bibinfo {author} {\bibfnamefont {P.~J.~H.}\
  \bibnamefont {Denteneer}}, \bibinfo {author} {\bibfnamefont {G.}~\bibnamefont
  {An}}, \ and\ \bibinfo {author} {\bibfnamefont {J.~M.~J.}\ \bibnamefont {van
  Leeuwen}},\ }\href {\doibase 10.1103/PhysRevB.47.6256} {\bibfield  {journal}
  {\bibinfo  {journal} {Phys. Rev. B}\ }\textbf {\bibinfo {volume} {47}},\
  \bibinfo {pages} {6256} (\bibinfo {year} {1993})}\BibitemShut {NoStop}%
\bibitem [{\citenamefont {Dupuis}(2004)}]{Dupuis2004}%
  \BibitemOpen
  \bibfield  {author} {\bibinfo {author} {\bibfnamefont {N.}~\bibnamefont
  {Dupuis}},\ }\href {\doibase 10.1103/PhysRevB.70.134502} {\bibfield
  {journal} {\bibinfo  {journal} {Phys. Rev. B}\ }\textbf {\bibinfo {volume}
  {70}},\ \bibinfo {pages} {134502} (\bibinfo {year} {2004})}\BibitemShut
  {NoStop}%
\bibitem [{\citenamefont {Keller}\ \emph {et~al.}(2001)\citenamefont {Keller},
  \citenamefont {Metzner},\ and\ \citenamefont {Schollw\"ock}}]{Keller2001}%
  \BibitemOpen
  \bibfield  {author} {\bibinfo {author} {\bibfnamefont {M.}~\bibnamefont
  {Keller}}, \bibinfo {author} {\bibfnamefont {W.}~\bibnamefont {Metzner}}, \
  and\ \bibinfo {author} {\bibfnamefont {U.}~\bibnamefont {Schollw\"ock}},\
  }\href {\doibase 10.1103/PhysRevLett.86.4612} {\bibfield  {journal} {\bibinfo
   {journal} {Phys. Rev. Lett.}\ }\textbf {\bibinfo {volume} {86}},\ \bibinfo
  {pages} {4612} (\bibinfo {year} {2001})}\BibitemShut {NoStop}%
\bibitem [{\citenamefont {Toschi}\ \emph {et~al.}(2005)\citenamefont {Toschi},
  \citenamefont {Barone}, \citenamefont {Capone},\ and\ \citenamefont
  {Castellani}}]{Toschi2005}%
  \BibitemOpen
  \bibfield  {author} {\bibinfo {author} {\bibfnamefont {A.}~\bibnamefont
  {Toschi}}, \bibinfo {author} {\bibfnamefont {P.}~\bibnamefont {Barone}},
  \bibinfo {author} {\bibfnamefont {M.}~\bibnamefont {Capone}}, \ and\ \bibinfo
  {author} {\bibfnamefont {C.}~\bibnamefont {Castellani}},\ }\href {\doibase
  10.1088/1367-2630/7/1/007} {\bibfield  {journal} {\bibinfo  {journal} {New
  Journal of Physics}\ }\textbf {\bibinfo {volume} {7}},\ \bibinfo {pages} {7}
  (\bibinfo {year} {2005})}\BibitemShut {NoStop}%
\bibitem [{\citenamefont {Paiva}\ \emph {et~al.}(2010)\citenamefont {Paiva},
  \citenamefont {Scalettar}, \citenamefont {Randeria},\ and\ \citenamefont
  {Trivedi}}]{Paiva2010}%
  \BibitemOpen
  \bibfield  {author} {\bibinfo {author} {\bibfnamefont {T.}~\bibnamefont
  {Paiva}}, \bibinfo {author} {\bibfnamefont {R.}~\bibnamefont {Scalettar}},
  \bibinfo {author} {\bibfnamefont {M.}~\bibnamefont {Randeria}}, \ and\
  \bibinfo {author} {\bibfnamefont {N.}~\bibnamefont {Trivedi}},\ }\href
  {\doibase 10.1103/PhysRevLett.104.066406} {\bibfield  {journal} {\bibinfo
  {journal} {Phys. Rev. Lett.}\ }\textbf {\bibinfo {volume} {104}},\ \bibinfo
  {pages} {066406} (\bibinfo {year} {2010})}\BibitemShut {NoStop}%
\bibitem [{\citenamefont {Rojas}\ and\ \citenamefont
  {Jos\'e}(1996)}]{Rojas1996}%
  \BibitemOpen
  \bibfield  {author} {\bibinfo {author} {\bibfnamefont {C.}~\bibnamefont
  {Rojas}}\ and\ \bibinfo {author} {\bibfnamefont {J.~V.}\ \bibnamefont
  {Jos\'e}},\ }\href {\doibase 10.1103/PhysRevB.54.12361} {\bibfield  {journal}
  {\bibinfo  {journal} {Phys. Rev. B}\ }\textbf {\bibinfo {volume} {54}},\
  \bibinfo {pages} {12361} (\bibinfo {year} {1996})}\BibitemShut {NoStop}%
\bibitem [{\citenamefont {Capriotti}\ \emph {et~al.}(2003)\citenamefont
  {Capriotti}, \citenamefont {Cuccoli}, \citenamefont {Fubini}, \citenamefont
  {Tognetti},\ and\ \citenamefont {Vaia}}]{Capriotti2003}%
  \BibitemOpen
  \bibfield  {author} {\bibinfo {author} {\bibfnamefont {L.}~\bibnamefont
  {Capriotti}}, \bibinfo {author} {\bibfnamefont {A.}~\bibnamefont {Cuccoli}},
  \bibinfo {author} {\bibfnamefont {A.}~\bibnamefont {Fubini}}, \bibinfo
  {author} {\bibfnamefont {V.}~\bibnamefont {Tognetti}}, \ and\ \bibinfo
  {author} {\bibfnamefont {R.}~\bibnamefont {Vaia}},\ }\href {\doibase
  10.1103/PhysRevLett.91.247004} {\bibfield  {journal} {\bibinfo  {journal}
  {Phys. Rev. Lett.}\ }\textbf {\bibinfo {volume} {91}},\ \bibinfo {pages}
  {247004} (\bibinfo {year} {2003})}\BibitemShut {NoStop}%
\bibitem [{\citenamefont {Timusk}\ and\ \citenamefont
  {Statt}(1999)}]{Timusk1999}%
  \BibitemOpen
  \bibfield  {author} {\bibinfo {author} {\bibfnamefont {T.}~\bibnamefont
  {Timusk}}\ and\ \bibinfo {author} {\bibfnamefont {B.}~\bibnamefont {Statt}},\
  }\href {\doibase 10.1088/0034-4885/62/1/002} {\bibfield  {journal} {\bibinfo
  {journal} {Reports on Progress in Physics}\ }\textbf {\bibinfo {volume}
  {62}},\ \bibinfo {pages} {61} (\bibinfo {year} {1999})}\BibitemShut {NoStop}%
\bibitem [{\citenamefont {Sharpe}\ \emph {et~al.}(2019)\citenamefont {Sharpe},
  \citenamefont {Fox}, \citenamefont {Barnard}, \citenamefont {Finney},
  \citenamefont {Watanabe}, \citenamefont {Taniguchi}, \citenamefont
  {Kastner},\ and\ \citenamefont {Goldhaber-Gordon}}]{Sharpe2019}%
  \BibitemOpen
  \bibfield  {author} {\bibinfo {author} {\bibfnamefont {A.~L.}\ \bibnamefont
  {Sharpe}}, \bibinfo {author} {\bibfnamefont {E.~J.}\ \bibnamefont {Fox}},
  \bibinfo {author} {\bibfnamefont {A.~W.}\ \bibnamefont {Barnard}}, \bibinfo
  {author} {\bibfnamefont {J.}~\bibnamefont {Finney}}, \bibinfo {author}
  {\bibfnamefont {K.}~\bibnamefont {Watanabe}}, \bibinfo {author}
  {\bibfnamefont {T.}~\bibnamefont {Taniguchi}}, \bibinfo {author}
  {\bibfnamefont {M.~A.}\ \bibnamefont {Kastner}}, \ and\ \bibinfo {author}
  {\bibfnamefont {D.}~\bibnamefont {Goldhaber-Gordon}},\ }\href
  {https://science.sciencemag.org/content/365/6453/605} {\bibfield  {journal}
  {\bibinfo  {journal} {Science}\ }\textbf {\bibinfo {volume} {365}},\ \bibinfo
  {pages} {605} (\bibinfo {year} {2019})}\BibitemShut {NoStop}%
\bibitem [{\citenamefont {Serlin}\ \emph {et~al.}(2020)\citenamefont {Serlin},
  \citenamefont {Tschirhart}, \citenamefont {Polshyn}, \citenamefont {Zhang},
  \citenamefont {Zhu}, \citenamefont {Watanabe}, \citenamefont {Taniguchi},
  \citenamefont {Balents},\ and\ \citenamefont {Young}}]{Serlin2020}%
  \BibitemOpen
  \bibfield  {author} {\bibinfo {author} {\bibfnamefont {M.}~\bibnamefont
  {Serlin}}, \bibinfo {author} {\bibfnamefont {C.~L.}\ \bibnamefont
  {Tschirhart}}, \bibinfo {author} {\bibfnamefont {H.}~\bibnamefont {Polshyn}},
  \bibinfo {author} {\bibfnamefont {Y.}~\bibnamefont {Zhang}}, \bibinfo
  {author} {\bibfnamefont {J.}~\bibnamefont {Zhu}}, \bibinfo {author}
  {\bibfnamefont {K.}~\bibnamefont {Watanabe}}, \bibinfo {author}
  {\bibfnamefont {T.}~\bibnamefont {Taniguchi}}, \bibinfo {author}
  {\bibfnamefont {L.}~\bibnamefont {Balents}}, \ and\ \bibinfo {author}
  {\bibfnamefont {A.~F.}\ \bibnamefont {Young}},\ }\href
  {https://science.sciencemag.org/content/367/6480/900} {\bibfield  {journal}
  {\bibinfo  {journal} {Science}\ }\textbf {\bibinfo {volume} {367}},\ \bibinfo
  {pages} {900} (\bibinfo {year} {2020})}\BibitemShut {NoStop}%
\bibitem [{\citenamefont {Zhang}\ \emph {et~al.}(2019)\citenamefont {Zhang},
  \citenamefont {Mao},\ and\ \citenamefont {Senthil}}]{Zhang2019}%
  \BibitemOpen
  \bibfield  {author} {\bibinfo {author} {\bibfnamefont {Y.-H.}\ \bibnamefont
  {Zhang}}, \bibinfo {author} {\bibfnamefont {D.}~\bibnamefont {Mao}}, \ and\
  \bibinfo {author} {\bibfnamefont {T.}~\bibnamefont {Senthil}},\ }\href
  {\doibase 10.1103/PhysRevResearch.1.033126} {\bibfield  {journal} {\bibinfo
  {journal} {Phys. Rev. Research}\ }\textbf {\bibinfo {volume} {1}},\ \bibinfo
  {pages} {033126} (\bibinfo {year} {2019})}\BibitemShut {NoStop}%
\bibitem [{\citenamefont {Bultinck}\ \emph {et~al.}(2020)\citenamefont
  {Bultinck}, \citenamefont {Chatterjee},\ and\ \citenamefont
  {Zaletel}}]{Bultinck2020}%
  \BibitemOpen
  \bibfield  {author} {\bibinfo {author} {\bibfnamefont {N.}~\bibnamefont
  {Bultinck}}, \bibinfo {author} {\bibfnamefont {S.}~\bibnamefont
  {Chatterjee}}, \ and\ \bibinfo {author} {\bibfnamefont {M.~P.}\ \bibnamefont
  {Zaletel}},\ }\href {\doibase 10.1103/PhysRevLett.124.166601} {\bibfield
  {journal} {\bibinfo  {journal} {Phys. Rev. Lett.}\ }\textbf {\bibinfo
  {volume} {124}},\ \bibinfo {pages} {166601} (\bibinfo {year}
  {2020})}\BibitemShut {NoStop}%
\end{thebibliography}%

\end{document}